\documentclass[11pt,twoside,a4paper,cmspaper,final,collab]{cms-tdr}
\def\svnVersion{c2a0f91}\def\svnDate{2026/12/01}\def\cmsCernNoTag{CERN-EP-2026-140}\def\cmsCernDate{\today}\def\cmsMessage{Published in the Journal of High Energy Physics as \href{http://dx.doi.org/10.1007/JHEP11(2026)244}{\doi{10.1007/JHEP11(2026)244}.}}
\begin{document}\cmsNoteHeader{TOP-24-008}

\newcommand{\pp}{\ensuremath{\Pp\Pp}\xspace}

\newcommand{\sigeta}{\ensuremath{\eta}\xspace}
\newcommand{\abseta}{\ensuremath{\abs{\sigeta}}\xspace}
\newcommand{\Deta}{\ensuremath{\Delta\sigeta}\xspace}
\newcommand{\Dphi}{\ensuremath{\Delta\phi}\xspace}
\newcommand{\PQj}{{\HepParticle{j}{}{}}\xspace}
\newcommand{\Njets}{\ensuremath{N_{\PQj}}\xspace}
\newcommand{\Nbjets}{\ensuremath{N_{\PQb}}\xspace}
\newcommand{\mll}{\ensuremath{m(\Pell\Pell)}\xspace}

\newcommand{\tttt}{\ensuremath{\ttbar\ttbar}\xspace}
\newcommand{\ttt}{\ensuremath{\PQt\PQt\PQt}\xspace}
\newcommand{\tttq}{\ensuremath{\PQt\PQt\PQt\PQq}\xspace}
\newcommand{\tttW}{\ensuremath{\PQt\PQt\PQt\PW}\xspace}
\newcommand{\ttZ}{\ensuremath{\ttbar\PZ}\xspace}
\newcommand{\ttW}{\ensuremath{\ttbar\PW}\xspace}
\newcommand{\ttH}{\ensuremath{\ttbar\PH}\xspace}
\newcommand{\ttX}{\ensuremath{\ttbar\PX}\xspace}
\newcommand{\tH}{\ensuremath{\PQt\PH}\xspace}
\newcommand{\WZ}{\ensuremath{\PW\PZ}\xspace}
\newcommand{\ZZ}{\ensuremath{\PZ\PZ}\xspace}
\newcommand{\Zjets}{\ensuremath{\PZ\text{+jets}}\xspace}
\newcommand{\VVV}{\ensuremath{\PV\PV(\PV)}\xspace}
\newcommand{\Xgamma}{\ensuremath{\PX\PGg}\xspace}
\newcommand{\ttttPLUSttt}{\ensuremath{\tttt\text{+}\ttt}\xspace}

\newcommand{\DeepJet}{\ensuremath{\textsc{DeepJet}}\xspace}
\newcommand{\MADSPIN}{\ensuremath{\textsc{madspin}}\xspace}

\newcommand{\twol}{\ensuremath{2\Pell}\xspace}
\newcommand{\threel}{\ensuremath{3\Pell}\xspace}
\newcommand{\fourl}{\ensuremath{4\Pell}\xspace}

\newcommand{\SRtwo}{\ensuremath{\text{SR-}\twol}\xspace}
\newcommand{\SRthree}{\ensuremath{\text{SR-}\threel}\xspace}
\newcommand{\SRfour}{\ensuremath{\text{SR-}\fourl}\xspace}
\newcommand{\CRtwoHigh}{\ensuremath{\text{CR-}\twol\text{-45j2\PQb}}\xspace}
\newcommand{\CRtwoLow}{\ensuremath{\text{CR-}\twol\text{-23j1\PQb}}\xspace}
\newcommand{\CRthreeLow}{\ensuremath{\text{CR-}\threel\text{-2j1\PQb}}\xspace}
\newcommand{\CRthreeZ}{\ensuremath{\text{CR-}\threel\text{-\PZ}}\xspace}
\newcommand{\CRfourZ}{\ensuremath{\text{CR-}\fourl\text{-\PZ}}\xspace}

\newcommand{\SRtwoTTTT}{\ensuremath{\text{SR-}\twol\text{-}\tttt}\xspace}
\newcommand{\SRtwoTTX}{\ensuremath{\text{SR-}\twol\text{-}\ttX}\xspace}
\newcommand{\SRtwoTT}{\ensuremath{\text{SR-}\twol\text{-}\ttbar}\xspace}
\newcommand{\SRthreeTTTT}{\ensuremath{\text{SR-}\threel\text{-}\tttt}\xspace}
\newcommand{\SRthreeTTX}{\ensuremath{\text{SR-}\threel\text{-}\ttX}\xspace}
\newcommand{\SRthreeTT}{\ensuremath{\text{SR-}\threel\text{-}\ttbar}\xspace}

\newcommand{\muR}{\ensuremath{\mu_{\mathrm{R}}}\xspace}
\newcommand{\muF}{\ensuremath{\mu_{\mathrm{F}}}\xspace}

\newcommand{\Lagrangian}{\ensuremath{\mathcal{L}}\xspace}
\newcommand{\LagrangianSM}{\ensuremath{\Lagrangian_{\text{SM}}}\xspace}
\newcommand{\Operator}{\ensuremath{\mathcal{Q}}\xspace}
\newcommand{\OperatorD}{\ensuremath{\Operator_i^{(d)}}\xspace}
\newcommand{\WilsonD}{\ensuremath{c_i^{(d)}}\xspace}
\newcommand{\MatElement}{\ensuremath{\mathcal{M}}\xspace}
\newcommand{\MatElementSM}{\ensuremath{\MatElement_{\text{SM}}}\xspace}
\newcommand{\MatElementEFT}[1][i]{\ensuremath{\MatElement_{#1}}\xspace}

\newcommand{\cQQone}{\ensuremath{\smash{c_{\PQQ\PQQ}^{(1)}}}\xspace}
\newcommand{\ctt}{\ensuremath{\smash{c_{\PQt\PQt}}}\xspace}
\newcommand{\cQtone}{\ensuremath{\smash{c_{\PQQ\PQt}^{(1)}}}\xspace}
\newcommand{\cQteight}{\ensuremath{\smash{c_{\PQQ\PQt}^{(8)}}}\xspace}
\newcommand{\ctH}{\ensuremath{c_{\PQt\PH}}\xspace}
\newcommand{\ctHRe}{\ensuremath{\smash{\ctH^{\mathrm{Re}}}}\xspace}
\newcommand{\ctHIm}{\ensuremath{\smash{\ctH^{\mathrm{Im}}}}\xspace}

\newcommand{\Sone}{{\HepParticle{S}{1}{}}\xspace}
\newcommand{\Seight}{{\HepParticle{S}{8}{}}\xspace}
\newcommand{\Pone}{{\HepParticle{P}{1}{}}\xspace}
\newcommand{\Peight}{{\HepParticle{P}{8}{}}\xspace}
\newcommand{\Vone}{{\HepParticle{V}{1}{}}\xspace}
\newcommand{\Veight}{{\HepParticle{V}{8}{}}\xspace}

\newcommand{\ySone}{\ensuremath{y_{1\HepParticle{S}{}{}}}\xspace}
\newcommand{\ySeight}{\ensuremath{y_{8\HepParticle{S}{}{}}}\xspace}
\newcommand{\yPone}{\ensuremath{y_{1\HepParticle{P}{}{}}}\xspace}
\newcommand{\yPeight}{\ensuremath{y_{8\HepParticle{P}{}{}}}\xspace}
\newcommand{\gVone}{\ensuremath{g_1}\xspace}
\newcommand{\gVoneL}{\ensuremath{g_{1\mathrm{L}}}\xspace}
\newcommand{\gVoneR}{\ensuremath{g_{1\mathrm{R}}}\xspace}
\newcommand{\gVeight}{\ensuremath{g_8}\xspace}
\newcommand{\gVeightL}{\ensuremath{g_{8\mathrm{L}}}\xspace}
\newcommand{\gVeightR}{\ensuremath{g_{8\mathrm{R}}}\xspace}

\newcommand{\mX}{\ensuremath{m_{\PX}}\xspace}
\newcommand{\mSone}{\ensuremath{m_{\HepParticle{S}{}{}1}}\xspace}
\newcommand{\mSeight}{\ensuremath{m_{\HepParticle{S}{}{}8}}\xspace}
\newcommand{\mPone}{\ensuremath{m_{\HepParticle{P}{}{}1}}\xspace}
\newcommand{\mPeight}{\ensuremath{m_{\HepParticle{P}{}{}8}}\xspace}
\newcommand{\mVone}{\ensuremath{m_{\HepParticle{V}{}{}1}}\xspace}
\newcommand{\mVeight}{\ensuremath{m_{\HepParticle{V}{}{}8}}\xspace}

\newcommand{\yukeven}{\ensuremath{\kappa_{\PQt}}\xspace}
\newcommand{\yukodd}{\ensuremath{\tilde{\kappa}_{\PQt}}\xspace}

\newcommand{\CP}{\ensuremath{CP}\xspace}
\newcommand{\NLLpr}{\ensuremath{\text{NLL}^\prime}\xspace}
\newcommand{\iu}{\ensuremath{\mathrm{i}\mkern1mu}\xspace}

\newcommand{\GammaH}{\ensuremath{\Gamma_{\PH}}\xspace}
\newcommand{\GammaX}{\ensuremath{\Gamma_{\PX}}\xspace}

\cmsNoteHeader{TOP-24-008}
\title{Search for physics beyond the standard model in four and three top quark production events using proton-proton collisions at \texorpdfstring{$\sqrt{s}=13\TeV$}{sqrt(s)=13 TeV}}

\author[cern]{The CMS Collaboration}
\date{\today}

\abstract{A search for physics beyond the standard model using four and three top quark production events is reported. The analyzed proton-proton collision data were recorded at 13\TeV with the CMS detector at the CERN LHC in 2016--2018 and correspond to an integrated luminosity of 138\fbinv. Events with two same-sign, three, or four leptons (electrons and/or muons) are selected. Constraints on six Wilson coefficients that modify interactions between four third-generation quarks or between top quarks and the Higgs boson in the standard model effective field theory framework are derived. The data are further used to exclude narrow topphilic heavy resonances in the mass ranges between 400\GeV and 1.6\TeV depending on their spin and color states. Finally, the top quark Yukawa coupling is extracted, considering both \CP-even and \CP-odd contributions.}

\hypersetup{%
pdfauthor={CMS Collaboration},%
pdftitle={Search for physics beyond the standard model in four and three top quark production events using proton-proton collisions at sqrt(s)=13 TeV},%
pdfsubject={CMS},%
pdfkeywords={CMS, top quark, 4top, 3top, BSM},%
}

\maketitle

\section{Introduction}

The top quark is the heaviest elementary particle in the standard model (SM) of particle physics.
As such, it plays an important role in understanding electroweak symmetry breaking via the Brout--Englert--Higgs mechanism in the SM or via alternative models beyond the SM (BSM).
Rare top quark production processes are important probes to test SM predictions and search for BSM physics.
This paper studies the simultaneous production of four top quarks (\tttt) or of three top quarks in association with either a \PW boson or a lighter quark (\tttW and \tttq, referred to together as \ttt in the following).
The \tttt production cross section is enhanced in many BSM theories with new heavy, strongly interacting particles that can decay to top quarks, such as supersymmetry~\cite{Nilles:1983ge, Farrar:1978xj, Toharia:2005gm, Plehn:2008ae, Calvet:2012rk, Beck:2015cga, Darme:2018dvz}, compositeness~\cite{Lillie:2007hd, Pomarol:2008bh, Kumar:2009vs, Cacciapaglia:2015eqa}, extra dimensions~\cite{Cacciapaglia:2011kz}, extended Higgs sectors~\cite{BhupalDev:2014bir, Dicus:1994bm, Craig:2015jba, Craig:2016ygr, Anisha:2023xmh}, dark matter models with heavy vector mediators~\cite{Ducu:2015fda}, and axion-like particles~\cite{Blasi:2023hvb}.
The sensitivity of \ttt production to BSM scenarios has been studied in the context of flavor-changing neutral currents~\cite{Kohda:2017fkn, Cao:2019qrb, Khanpour:2019qnw}, new heavy resonances~\cite{Iguro:2017ysu, Cho:2019stk}, and dark matter~\cite{Abasov:2024mwk}.

Possible modifications of rare top quark production cross sections induced by BSM effects at high energies beyond the reach of current experiments can be parameterized in the SM effective field theory (SMEFT).
In the context of SMEFT, \tttt and \ttt production provide particular sensitivity to four-heavy-quark (\ie, involving only \PQt and \PQb quarks) and top-quark--Higgs-boson (\tH) operators~\cite{Degrande:2010kt, Zhang:2017mls, Englert:2019zmt, Banelli:2020iau, Darme:2021gtt, Aoude:2022deh, Aleshko:2023rkv, Aleshko:2025jua, DiNoi:2025uhu}.
In contrast, simplified models can be used to model resonant contributions to rare top quark production from heavy BSM particles decaying to top quarks without relying on specific BSM theories.
Such topphilic heavy resonances can be categorized by their main characteristics: scalar, pseudoscalar, or vector nature, and either color singlets or octets~\cite{Kim:2016plm, Darme:2018dvz, Darme:2021gtt, Darme:2025leu}.
Modifications of the interaction between top quarks and the Higgs boson can also be probed by measuring the strength of the top quark Yukawa coupling and its \CP structure~\cite{Cao:2016wib, Cao:2019ygh}.

The ATLAS~\cite{ATLAS:Detector-2008, ATLAS:2023dns} and CMS~\cite{CMS:Detector-2008, CMS:PRF-21-001} Collaborations have searched for the \tttt production process using proton-proton (\pp) collision data recorded at the CERN LHC during 2015--2018~\cite{CMS:SUS-16-035, CMS:TOP-17-009, ATLAS:2018alq, ATLAS:2018kxv, CMS:TOP-17-019, CMS:TOP-18-003, ATLAS:2020hpj, ATLAS:2021kqb, CMS:TOP-21-005, Blekman:2022jag, ATLAS:2023ajo, CMS:TOP-22-013}.
The decay channels of the \tttt system can be characterized by the number of electrons and/or muons (referred to as ``leptons'' in the following) in the final state.
Dedicated analysis strategies have been employed for the fully hadronic, single-lepton, opposite-sign and same-sign dilepton (\twol), three-lepton (\threel), and four-lepton (\fourl) channels.
The latest searches in the same-sign \twol, \threel, and \fourl channels by both collaborations observed \tttt production for the first time with a statistical significance of more than five standard deviations~\cite{ATLAS:2023ajo, CMS:TOP-22-013}.
The \tttt cross section was measured to be higher than, but statistically consistent with, the SM prediction.
A strong anticorrelation between the \tttt cross section and the combined \ttt cross section was found when both were allowed to vary freely in the fit to data.

Previous constraints on four-heavy-quark operators in the SMEFT framework have been derived by the ATLAS and CMS Collaborations using \tttt measurements~\cite{ATLAS:2018cye, ATLAS:2018alq, ATLAS:2018kxv, CMS:TOP-17-019, ATLAS:2023ajo}.
The \tH operators have been constrained in \ttH measurements in combination with other processes~\cite{CMS:TOP-21-003, ATLAS:2024lyh}.
A CMS measurement targeting \tttt and \ttH production together with other top quark production processes~\cite{CMS:TOP-22-006} has derived simultaneous constraints on four-heavy-quark and \tH operators, as has its recent combination with other CMS measurements from different sectors of the SM~\cite{CMS:SMP-24-003}.
Using global SMEFT fits to sets of many different experimental measurements, constraints were put on four-heavy-quark~\cite{Hartland:2019bjb, Ethier:2021bye, Celada:2024mcf, deBlas:2025xhe} and \tH~\cite{Ellis:2020unq, Ethier:2021bye, Miralles:2021dyw, Celada:2024mcf, deBlas:2025xhe} operators.

Direct searches for heavy resonances decaying to top quark pairs ($\PX\to\ttbar$) have been performed in the context of various theoretical models and different signatures.
Targeting the \tttt signature via $\pp\to\ttbar\PX\to\tttt$, searches have been performed for scalar and pseudoscalar resonances in the context of two-Higgs-doublet models~\cite{ATLAS:2018alq, CMS:TOP-18-003, ATLAS:2022rws}, as well as for vector resonances~\cite{ATLAS:2023taw}.
The results in Refs.~\cite{CMS:TOP-18-003, ATLAS:2023taw} also consider the resonant contribution to the \ttt channel as part of the signal.
Related searches have been performed for the $\pp\to\PX\to\ttbar$~\cite{CMS:B2G-16-015, ATLAS:2018rvc, CMS:B2G-17-017, ATLAS:2019npw, CMS:HIG-17-027, ATLAS:2020lks, ATLAS:2024vxm, CMS:TOP-24-007, CMS:HIG-22-013, ATLAS:2025kmo, ATLAS:2026dbe, CMS:B2G-25-009} and $\pp\to\PX\PZ\to\ttbar\PZ$~\cite{ATLAS:2023zkt, CMS:B2G-23-006} signatures.

The top quark Yukawa coupling has been measured by the ATLAS and CMS Collaborations in top-quark-associated Higgs boson production~\cite{CMS:HIG-19-013, ATLAS:2020ior, CMS:HIG-19-008, CMS:HIG-21-006, ATLAS:2023cbt, CMS:HIG-19-011}, in combinations of various Higgs boson production measurements~\cite{CMS:HIG-19-009, CMS:HIG-22-001, ATLAS:2022vkf, ATLAS:2022tnm}, and from \tttt production measurements~\cite{CMS:TOP-17-009, CMS:TOP-18-003, ATLAS:2023ajo}.
The collaborations also extracted the top quark Yukawa coupling from differential distributions of \ttbar production~\cite{CMS:TOP-17-004, CMS:TOP-19-008, ATLAS:2025ciy}.
Additionally, ATLAS extracted the top quark Yukawa coupling together with the Higgs boson width \GammaH from a combination of Higgs boson and \tttt production measurements~\cite{ATLAS:2024mhs}.
Combinations of Higgs boson production measurements find the top quark Yukawa coupling compatible with the SM expectation at a precision of about 6\%~\cite{CMS:HIG-21-018}.
Extractions from \tttt production are less precise with an upper limit on the absolute value of the top quark Yukawa coupling of 1.7 times the SM expectation reported in Ref.~\cite{CMS:TOP-18-003}, but reduce the dependence on assumptions about \GammaH.
From dedicated analyses targeting the \CP structure of the top quark Yukawa coupling in top-quark-associated Higgs boson production, the purely \CP-odd case is excluded with a significance of 3.9 standard deviations~\cite{ATLAS:2020ior, CMS:HIG-19-011}.

In this paper, we present an interpretation of the \tttt production measurement from Ref.~\cite{CMS:TOP-22-013} in three BSM scenarios: SMEFT framework, narrow topphilic heavy resonances, and modified top quark Yukawa coupling.
In the SMEFT interpretation, we evaluate constraints on six operators that modify four-heavy-quark and \tH interactions, including these six operators together for the first time in an analysis optimized for \tttt production.
We consider six simplified models of topphilic heavy resonances, representative for a wide range of BSM theories that would enhance \tttt and \ttt production, and evaluate exclusion limits on the coupling strength as a function of the heavy resonance mass.
The top quark Yukawa coupling is extracted using a parameterization with \CP-even and \CP-odd terms.
Both \tttt and \ttt production are treated as signal processes in the three scenarios, taking into account not only that the experimental signature of the processes is very similar, but also that they are modified in similar ways by the considered BSM scenarios.
With respect to the experimental analysis of Ref.~\cite{CMS:TOP-22-013}, some improvements in modeling and experimental methods have been implemented.
In the most signal-enriched regions, two observables are employed, optimizing separately for sensitivity to the different BSM scenarios.

This paper is organized as follows.
The CMS detector and event reconstruction are described in Section~\ref{sec:cms}, and the experimental analysis is summarized in Section~\ref{sec:analysis}.
The interpretation in the SMEFT framework is presented in Section~\ref{sec:smeft}.
Exclusion limits on narrow topphilic heavy resonances are evaluated in Section~\ref{sec:topphilic}, and the extraction of the top quark Yukawa coupling is discussed in Section~\ref{sec:yukawa}.
The results of all three interpretations are summarized in Section~\ref{sec:summary}.
Tabulated results are provided in the HEPData record for this analysis~\cite{hepdata}.

\section{The CMS detector and event reconstruction}
\label{sec:cms}

The CMS apparatus~\cite{CMS:Detector-2008, CMS:PRF-21-001} is a multipurpose, nearly hermetic detector, designed to trigger on~\cite{CMS:TRG-17-001, CMS:TRG-12-001, CMS:TRG-19-001} and identify electrons, muons, photons, and (charged and neutral) hadrons~\cite{CMS:EGM-17-001, CMS:MUO-16-001, CMS:TRK-11-001}.
A global ``particle-flow'' (PF) algorithm~\cite{CMS:PRF-14-001} aims to reconstruct all individual particles in an event, combining information provided by the all-silicon inner tracker and by the crystal electromagnetic and brass-scintillator hadron calorimeters, operating inside a 3.8\unit{T} superconducting solenoid, with data from the gas-ionization muon detectors embedded in the flux-return yoke outside the solenoid.
The reconstructed particles (PF candidates) are used to build \PGt leptons, jets, and missing transverse momentum~\cite{CMS:TAU-16-003, CMS:JME-13-004, CMS:JME-17-001, CMS:JME-18-001}.

The reconstruction methods and selection criteria applied to jets, jets originating from \PQb quarks, electrons, and muons are identical to those described in Ref.~\cite{CMS:TOP-22-013}, and are summarized briefly in the following.
Jets are clustered from the PF candidates using the anti-\kt algorithm~\cite{Cacciari:2008gp, Cacciari:2011ma} with a distance parameter of 0.4, discarding charged PF candidates identified to be originating from additional \pp interactions within the same or nearby bunch crossing (pileup) and applying an offset correction for pileup contributions from neutral PF candidates~\cite{CMS:JME-18-001}.
Jet energy corrections derived from simulation and simulation-to-data comparisons are applied such that the jet energy scale in data and simulation matches, and a smearing procedure is applied to match the jet energy resolution (JER) in simulation to the one in data~\cite{CMS:JME-13-004, CMS:DP-2021-033}.
Compared to Ref.~\cite{CMS:TOP-22-013}, an updated JER for the 2017 data set is used in this analysis.
Reconstructed jets are considered in the analysis if they have transverse momentum $\pt>25\GeV$, absolute pseudorapidity $\abseta<2.4$, and are separated by $\DR=\sqrt{\smash[b]{(\Deta)^2+(\Dphi)^2}}>0.4$ from any identified lepton.
Here, \Deta and \Dphi are the differences in \sigeta and azimuthal angle between the jet and lepton directions.

The \DeepJet algorithm~\cite{CMS:BTV-16-002, Bols:2020bkb, CMS:DP-2023-005} is applied to identify \PQb quark jets, using three working points (WPs).
The ``loose'' WP, which is used for the event selection, has a selection efficiency for \PQb quark jets of 90\%.
The corresponding misidentification rate for \PQc quark jets is 49\%, and the misidentification rates for light-quark and gluon jets is 18\%.
Different from Ref.~\cite{CMS:TOP-22-013}, jets that fail a pileup rejection criterion~\cite{CMS:JME-18-001} are not considered for \PQb tagging, \ie, they fail all WPs and their \DeepJet score is set to the minimal possible value.
In the following, ``\PQb jet'' is used to refer to a jet passing the loose WP.

Electrons are reconstructed in the range $\abseta<2.5$, excluding the barrel--endcap transition region $1.44<\abseta<1.57$~\cite{CMS:EGM-17-001, CMS:DP-2020-021}, and muons are reconstructed in the range $\abseta<2.4$~\cite{CMS:MUO-16-001}.
For electrons, the charge is estimated with three different methods and required to be the same for all three methods, thereby reducing the background from charge mismeasurements significantly~\cite{CMS:EGM-13-001}.
Charge mismeasurements of muons, in contrast, are negligible~\cite{CMS:CFT-09-014, CMS:MUO-17-001}.
We require electrons and muons to have $\pt>10\GeV$, be compatible with the primary interaction vertex, and fulfill a set of loose identification (ID) and isolation criteria~\cite{CMS:EGM-17-001, CMS:MUO-16-001}.
Additional ID criteria are formulated to distinguish between ``prompt'' leptons from the direct decays of top quarks or massive bosons and ``nonprompt'' leptons from background sources such as genuine leptons produced in hadron decays and photon conversions or jet constituents misidentified as leptons.
Specifically, we employ the ID discriminant based on boosted decision trees (BDTs) described in Ref.~\cite{CMS:TOP-22-013}, applying the methods developed for previous measurements and searches in multilepton final states~\cite{CMS:HIG-17-018, CMS:TOP-18-008, CMS:HIG-19-008, CMS:SUS-19-012, CMS:SMP-20-012, CMS:TOP-20-010} and studied in detail for the case of muons in Ref.~\cite{CMS:MUO-22-001}.
A ``tight'' ID WP is defined by requiring that the BDT score for the lepton passes a threshold value, and a ``loose'' ID WP by requiring that the lepton either passes the tight ID or a set of thresholds on two input variables of the BDT.

\section{Experimental analysis}
\label{sec:analysis}

The analyzed data sample was recorded in 2016--2018 and corresponds to an integrated luminosity of 138\fbinv~\cite{CMS:LUM-17-003, CMS:LUM-17-004, CMS:LUM-18-002}.
Events are collected with a combination of triggers that require the presence of one, two, or three leptons.
The simulated event samples for the SM signal and background processes, event selection and categorization, the BDTs employed to distinguish between SM \tttt signal and background events, background estimation, systematic uncertainties, and fit setup are mostly unchanged with respect to Ref.~\cite{CMS:TOP-22-013}.
In the following, we provide a concise summary of the main aspects of the experimental analysis and describe in detail all differences from the previously published measurement.

\subsection{Simulated event samples}

Simulated event samples of the signal and background processes are generated with Monte Carlo event generators, using the NNPDF3.1NNLO parton distribution functions (PDFs)~\cite{NNPDF:2017mvq} in the matrix-element (ME) calculation and including a parton shower (PS) simulation performed with \PYTHIA8.230~\cite{Sjostrand:2014zea} using the CP5 tune~\cite{CMS:GEN-17-001}.
Simulated pileup collisions are overlaid to match the observed pileup distribution in data, and the samples are processed with a full CMS detector simulation based on \GEANTfour~\cite{GEANT4:2002zbu}.
In all cases, separate samples are generated for the different data-taking years to account for differences in the detector conditions.
These include separate samples for the 2016 periods before and after a modification of the APV25 readout chip settings that affected the track hit reconstruction efficiency~\cite{CMS:DP-2020-045}.

Events of SM \tttt production are generated at next-to-leading order (NLO) in quantum chromodynamics (QCD) with \MGvATNLO2.6.5~\cite{Alwall:2014hca}, with the top quark decays simulated with \MADSPIN~\cite{Artoisenet:2012st}.
In the ME calculation, the renormalization and factorization scales (\muR and \muF) are dynamically chosen as half the scalar sum of the transverse masses of all final-state particles.
This event generation setup for \tttt production is identical to the one used in Ref.~\cite{CMS:TOP-22-013}.
The \tttt production cross section in the SM has been calculated at NLO in QCD and electroweak (EW) theory~\cite{Bevilacqua:2012em, Alwall:2014hca, Maltoni:2015ena, Frederix:2017wme, Jezo:2021smh, Dimitrakopoulos:2024qib}.
Including soft gluon resummation corrections at next-to-leading logarithmic (NLL) accuracy and constant nonlogarithmic contributions that do not vanish at the threshold (denoted together as \NLLpr), the inclusive cross section is predicted to be $13.4\,^{+1.0}_{-1.8}\unit{fb}$~\cite{vanBeekveld:2022hty}, where the quoted uncertainty is from \muR and \muF variations and the PDFs.
A recent calculation using the invariant-mass scheme instead of the total-mass scheme results in a smaller prediction of $10.4\,^{+2.5}_{-2.2}\unit{fb}$ when using the same central scale choice and without including PDF uncertainties~\cite{vanBeekveld:2025ghw}.
For consistency with Ref.~\cite{CMS:TOP-22-013}, we have not changed the SM prediction used in the analysis.

Different simulated event samples for \ttt production from those in Ref.~\cite{CMS:TOP-22-013} are used: events are generated at NLO in QCD instead of leading order (LO) with \MGvATNLO separately for the four processes $\ttbar\PQt\PWm$, $\ttbar\PAQt\PWp$, $\ttbar\PQt\PAQq$, and $\ttbar\PAQt\PQq$.
The five-flavor scheme is employed, where \PQb quarks are considered as sea quarks of the proton.
To avoid overlap with the \tttt production samples when including real emissions at NLO, the diagram removal scheme ``DR1''~\cite{Frixione:2008yi, Demartin:2016axk} is applied, in which the squared resonant term from the squared amplitude is removed together with the interference term between the resonant and nonresonant contributions, as implemented in Ref.~\cite{Durieux:2023ttt}.
No uncertainty in the choice of the overlap removal prescription is considered since no alternative implementation was available\footnote{After the submission of this paper, a new prescription has been proposed in Ref.~\cite{Durieux:2026kus} and will be considered for future analyses.}.
The same dynamical scale choice is used in the ME calculation as for the simulated \tttt event samples.
The SM \ttt production cross sections have been calculated at LO in QCD and EW theory in Refs.~\cite{Barger:2010uw, Chen:2014ewl, Malekhosseini:2018fgp, Boos:2021yat}, and the latest calculation in Ref.~\cite{Durieux:2023ttt} extends these calculations with NLO QCD corrections.
At NLO in QCD and with EW corrections at LO, the \tttW (\tttq) production cross section is predicted to be $1.32\,^{+0.21}_{-0.20}$ ($0.70\,^{+0.08}_{-0.07}$)\unit{fb}~\cite{Durieux:2023ttt}.
The quoted uncertainty accounts for \muR and \muF variations and the PDFs.
These SM predictions were also employed in Ref.~\cite{CMS:TOP-22-013}.

We use the identical simulated event samples for all SM background processes as in Ref.~\cite{CMS:TOP-22-013} with one exception.
For \ttW production, events are simulated with \MGvATNLO at NLO as $\ttbar\Pell\PGn$ with invariant mass $m(\Pell\PGn)>30\GeV$, including subleading electroweak contributions as described in Ref.~\cite{CMS:TOP-21-011}.

\subsection{Event selection and categorization}

\begin{figure}[!htp]
\centering
\includegraphics[width=\textwidth]{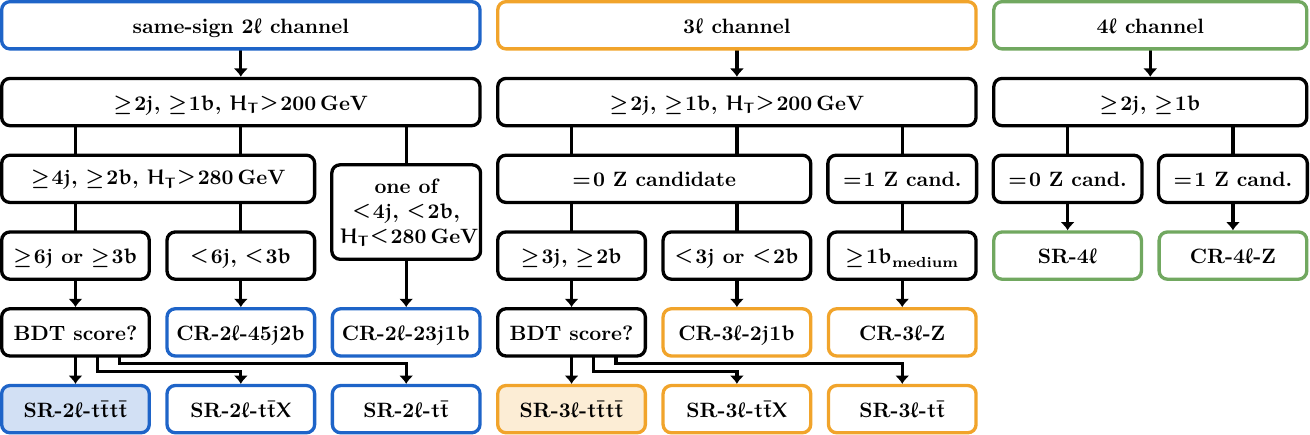}
\caption{Schematic representation of the event selection and categorization.}
\label{fig:eventselection}
\end{figure}

Events are required to have exactly two same-sign, three, or four tight leptons, no additional loose leptons, and at least two jets, of which at least one passes the loose \PQb tagging WP.
The two highest \pt leptons are required to have $\pt>25$ and 20\GeV, which ensures a high trigger efficiency.
To reduce background from leptonic decays of low-mass resonances, events in the \twol (\threel and \fourl) channels are removed if the dilepton invariant mass \mll of any lepton pair is below 20 (12)\GeV.
We use several exclusive signal and control regions (SRs and CRs), as sketched in Fig.~\ref{fig:eventselection} and summarized in the following, with full definitions provided in Ref.~\cite{CMS:TOP-22-013}.
This SR and CR categorization is based on the number of jets and \PQb jets (\Njets and \Nbjets) and the scalar sum of the \pt of all selected jets (\HT), as well as on the number of unique \PZ boson candidates, defined as opposite-sign same-flavor lepton pairs with $\abs{\mll-91.2\GeV}<15\GeV$.
The categorization was optimized in Ref.~\cite{CMS:TOP-22-013} for sensitivity to the SM \tttt signal in the SRs and the various relevant background processes in different CRs.

In the same-sign \twol channel, the \SRtwo is defined by requiring $\Njets\geq4$, $\Nbjets\geq2$, $\HT>280\GeV$, and additionally $\Njets\geq6$ and/or $\Nbjets\geq3$.
The \CRtwoHigh and \CRtwoLow are defined with lower \Njets, \Nbjets, and \HT requirements.
In the \threel channel, the \SRthree is defined by requiring $\Njets\geq3$, $\Nbjets\geq2$, $\HT>200\GeV$, and that there is no \PZ boson candidate.
The \CRthreeLow is defined by lower requirements on \Njets and \Nbjets, and the \CRthreeZ by requiring that there is one \PZ boson candidate.
In the \fourl channel, the only requirement for \SRfour in addition to the baseline selection of $\Njets\geq2$ and $\Nbjets\geq1$ is that there is no \PZ boson candidate, whereas events with exactly one \PZ boson candidate form the \CRfourZ.

Two multiclassification BDTs trained with the \textsc{tmva} program~\cite{Voss:2007jxm} are employed, separately for the \twol and the combined \threel{+}\fourl channels, to separate between three classes of events:\ the SM \tttt signal, associated top quark pair production with a heavy boson (\ttZ, \ttW, and \ttH, labeled together as \ttX), and \ttbar production (representative for nonprompt-lepton and charge-misidentified backgrounds).
Input variables include \PQb tagging information of the selected jets, angular separations between leptons and \PQb jets, and invariant masses of two- and three-jet systems consistent with the decay of a hadronically decaying \PW boson or top quark.
More details on the input variables and the BDT training are provided in Ref.~\cite{CMS:TOP-22-013}.
The BDT scores can be interpreted as measures of how likely an event is to originate from the corresponding classes.
Events selected in \SRtwo and \SRthree are further split according to which BDT class has the highest score.
The corresponding regions are labeled by appending the class name to \SRtwo or \SRthree, as shown in Fig.~\ref{fig:eventselection}.
Because of the low expected number of events in the \SRfour, we do not split it further into classes, thus deviating from the approach in Ref.~\cite{CMS:TOP-22-013}.

The main sensitivity to the \tttt and \ttt signal processes is expected from the \SRtwoTTTT and \SRthreeTTTT.
The expected number of \tttt and \ttt signal events in the SM scenario in the \SRfour, in contrast, is less than one, and thus the \SRfour distributions do not add relevant sensitivity in comparison to \SRtwoTTTT and \SRthreeTTTT.
Thus, the focus of the studies presented here is on modifications of the expected signal distributions in the \SRtwoTTTT and \SRthreeTTTT caused by the different BSM scenarios.

\subsection{Background estimation}

Background contributions with at least one nonprompt lepton are estimated with a ``tight-to-loose'' ratio method~\cite{CMS:SUS-15-008, CMS:HIG-19-008, CMS:TOP-21-011} from orthogonal data samples.
The probabilities for a loose lepton to also pass the tight ID are measured in a data sample enriched in events composed uniquely of jets produced through the strong interaction.
These probabilities are then applied as scale factors to data events in a sideband of the SRs and CRs, defined by requiring that one or more of the selected leptons fail the tight ID but pass the loose ID, to estimate the contribution in the SRs and CRs.
Closure tests performed with simulated \ttbar and \Zjets production samples find good agreement at the level of 30\%~\cite{CMS:TOP-22-013}, which is applied as normalization uncertainty.
This background contribution is labeled as ``NP'' on figures.

Contributions with only prompt leptons but where one electron has a mismeasured charge are a relevant background only in the same-sign \twol channel.
The probability for an electron to have its charge mismeasured is estimated in simulation.
To predict the charge-mismeasured background contribution in the SRs and CRs, this probability is applied as a scale factor to data events in a sideband of the SRs and CRs, defined by requiring opposite-sign instead of same-sign leptons.
Comparisons of data and simulation in a \Zjets event selection are used to derive additional correction factors.
After applying the corrections, agreement at the level of 15\% is found, which is applied as normalization uncertainty.
This background contribution is labeled as ``CMID'' on figures.

All other background contributions are estimated from simulated event samples normalized according to the measured integrated luminosity and state-of-the-art cross section predictions.
The main background contributions with prompt leptons are \ttZ production in the \threel and \fourl channels with smaller contributions in the \twol channel, \ttW production in the \twol and \threel channels, and \ttH production in all channels.
Contributions with top quarks other than \ttZ, \ttW, and \ttH are grouped as ``other \PQt'', contributions from the production of two or three heavy vector bosons as ``\VVV'', and contributions with photon conversions as ``\Xgamma''; and these three groups are shown together as ``other'' on figures.
The largest contribution in the \VVV category are from \WZ and \ZZ diboson production, and we apply scale factors as functions of \Njets derived from two dedicated data validation regions to improve the data-to-simulation agreement.

\subsection{Systematic uncertainties}

Systematic uncertainties affect the acceptance and reconstruction efficiency in simulation, the background event yields, and the distributions of the fitted observables.
We consider experimental uncertainties in the integrated luminosity, the efficiency of the electron and muon selection, the trigger efficiency, jet energy scale, JER, missing transverse momentum, and \PQb tagging efficiency.
For the background with nonprompt leptons, statistical uncertainties in the measured tight-to-loose ratio and in the sideband yields are considered, as well as normalization uncertainties that correspond to 30\% in total.
The background with charge-misidentified electrons is affected by the statistical uncertainty in the sideband yields and a normalization uncertainty of 15\%.
Normalization uncertainties for the backgrounds with prompt leptons of 7.7\% for \ttH, 6.2\% for \WZ, 20\% for other \PQt, 6\% for \VVV, and 5\% for \Xgamma are included, as in Ref.~\cite{CMS:TOP-22-013}.
Additional systematic uncertainties are considered for the \Njets-dependent correction factors applied to the \WZ and \ZZ background prediction.
The \ttZ and \ttW background normalizations are not constrained a priori, but included as free parameters in the fit.

We consider various modeling uncertainties for the signal and background predictions.
Different from Ref.~\cite{CMS:TOP-22-013}, the choice of \muR and \muF in the ME calculation is considered with two separate systematic uncertainties corresponding to the individual variations of \muR and \muF by a factor of 2 up and down, separately for each process.
The remaining modeling uncertainties are identical to those applied in Ref.~\cite{CMS:TOP-22-013}, these include: the PDF uncertainty with NNPDF replicas~\cite{NNPDF:2017mvq} following the PDF4LHC prescription~\cite{Butterworth:2015oua} correlated between all processes, the choice of \muR for initial-state (separately for each process) and final-state radiation (correlated between processes involving top quarks) in the PS simulation, and the additional uncertainty for additional \PQb quark radiation in \ttX production based on the results of Refs.~\cite{CMS:TOP-18-002, CMS:TOP-22-009}.
For the per-process modeling uncertainties, \tttt, \tttW, and \tttq are treated as three separate processes.
Concerning the additional uncertainty in the \Njets distribution in \ttW production based on the comparison with Ref.~\cite{Frederix:2021agh}, we apply the same procedure as in Ref.~\cite{CMS:TOP-22-013}.
Given the improved \ttW simulation employed in this analysis, the maximal variation is reduced from a factor of 1.55 to 1.32 for events with $\Njets\geq7$ in the \twol channel.

Depending on the fit scenario, additional normalization uncertainties are introduced for the SM cross section of \tttt and \ttt production, corresponding to the uncertainty in the state-of-the-art theory predictions.
The resulting normalization uncertainties are ${}^{+8\%}_{-13\%}$ for \tttt~\cite{vanBeekveld:2022hty}, ${}^{+16\%}_{-15\%}$ for \tttW~\cite{Durieux:2023ttt}, and ${}^{+12\%}_{-10\%}$ for \tttq~\cite{Durieux:2023ttt}, and are implemented as multiplicative variations of the normalization of the SM distributions.
In the SM fits, only normalization uncertainties for those processes whose cross sections are not parameters of interest of the fit are included.
In the BSM fits, all three normalization uncertainties are always included, in addition to the modifications introduced by the BSM scenarios.

\subsection{Fit setup}

We construct likelihood functions $L(r,\theta)$ to quantify the agreement between the predicted yields and observed data in the SRs and CRs for each SM and BSM scenario.
The set $r$ denotes the parameters of interest, \eg, the \tttt production cross section in the SM scenario, or the model parameters specific to the considered BSM scenario.
The set $\theta$ of the nuisance parameters represents the systematic uncertainties.
Following the method described in Ref.~\cite{CMS:NOTE-2011-005}, each $L$ is constructed as the product of the Poisson probabilities to obtain the observed yields given the estimated background contribution and the scenario-specific signal contribution, with additional factors for the nuisance parameters.
Each bin in the fit distributions of each SR and CR is included separately per data-taking year in $L$.
One nuisance parameter is added for each bin to account for the statistical uncertainties in the predicted yields, jointly for all processes~\cite{Barlow:1993dm, Conway:2011in}.
The negative log-likelihood difference $q(r)=-2\ln L(r,\widehat{\theta}_r)/L(\widehat{r},\widehat{\theta})$ (also denoted as $-2\,\Delta\ln L$ on figures), constructed from the fitted set $\widehat{\theta}_r$ that maximizes the considered $L$ for a fixed set $r$, and using the fitted sets $\widehat{r}$ and $\widehat{\theta}$ that simultaneously maximize the considered $L$, is employed as test statistic.
We employ the CMS statistical analysis tool \textsc{combine}~\cite{CMS:CAT-23-001}, which is based on the \textsc{RooFit}~\cite{Verkerke:2003ir} and \textsc{RooStats}~\cite{Moneta:2010pm} frameworks, to implement $L$ and $q$ for all scenarios, to perform the fits, and to evaluate the fit results.

To evaluate 68 and 95\% confidence level (\CL) intervals for the parameters of interest, we can evaluate $q$ either via an asymptotic approximation~\cite{CMS:NOTE-2011-005, Cowan:2010js} or with a Monte Carlo method (``toys'').
The asymptotic approximation is based on the assumption, known as Wilks' theorem~\cite{Wilks:1938dza}, that $q$ follows a $\chi^2$ distribution with degrees of freedom equal to the number of profiled parameters of interest.
This assumption holds for the measurement of the \tttt production cross section in the SM scenario, where the parameter of interest is a signal strength that scales the \tttt production yields linearly.
It has been shown, however, that a nonlinear parameterization of yields, such as the quadratic dependence on the operator strength often obtained in the SMEFT framework, can result in a violation of Wilks' theorem, meaning that the coverage of asymptotically approximated \CL intervals is not guaranteed to match 68 and 95\%~\cite{Bernlochner:2022oiw}.
Since all three BSM scenarios considered here introduce nonlinear yield modifications of \tttt and \ttt production, we have studied the coverage in all cases using toys.
We then present results as 68 and 95\% \CL intervals in cases where we have explicitly evaluated the intervals with toys or have confirmed the correct coverage with toys, or by specifying the $q$ threshold at which the interval is evaluated in cases where the large number of toys required to assess coverage would entail an impractically high computational demand.

The fit distributions are the BDT score \ttX in \SRtwoTTX, \SRthreeTTX, and \SRfour; the BDT score \ttbar in \SRtwoTT, \SRthreeTT, \CRtwoHigh, and \CRtwoLow; the \Njets distribution in \CRthreeZ and \CRfourZ; and the event yields with positive or negative sum of lepton charges in \CRthreeLow.
Of these regions, the \SRtwoTTX and \SRthreeTTX are dominated by \ttZ, \ttW, and \ttH production, the \CRthreeZ and \CRfourZ by \ttZ and \VVV production, and the \CRtwoHigh, \CRtwoLow, and \CRthreeLow have significant contributions from nonprompt-lepton backgrounds.
The inclusion of these regions in the fit thus allow for constraints of background normalization nuisance parameters from data.
The \SRtwoTTTT and \SRthreeTTTT have the largest contributions from the \ttttPLUSttt signal process, and the fit distributions are either the BDT score \tttt (as in Ref.~\cite{CMS:TOP-22-013}) or \HT, optimized for the considered scenario.
In all cases, the \SRtwoTTTT and \SRtwoTTX are further split according to the lepton flavors.

\begin{figure}[!t]
\centering
\includegraphics[width=0.45\textwidth]{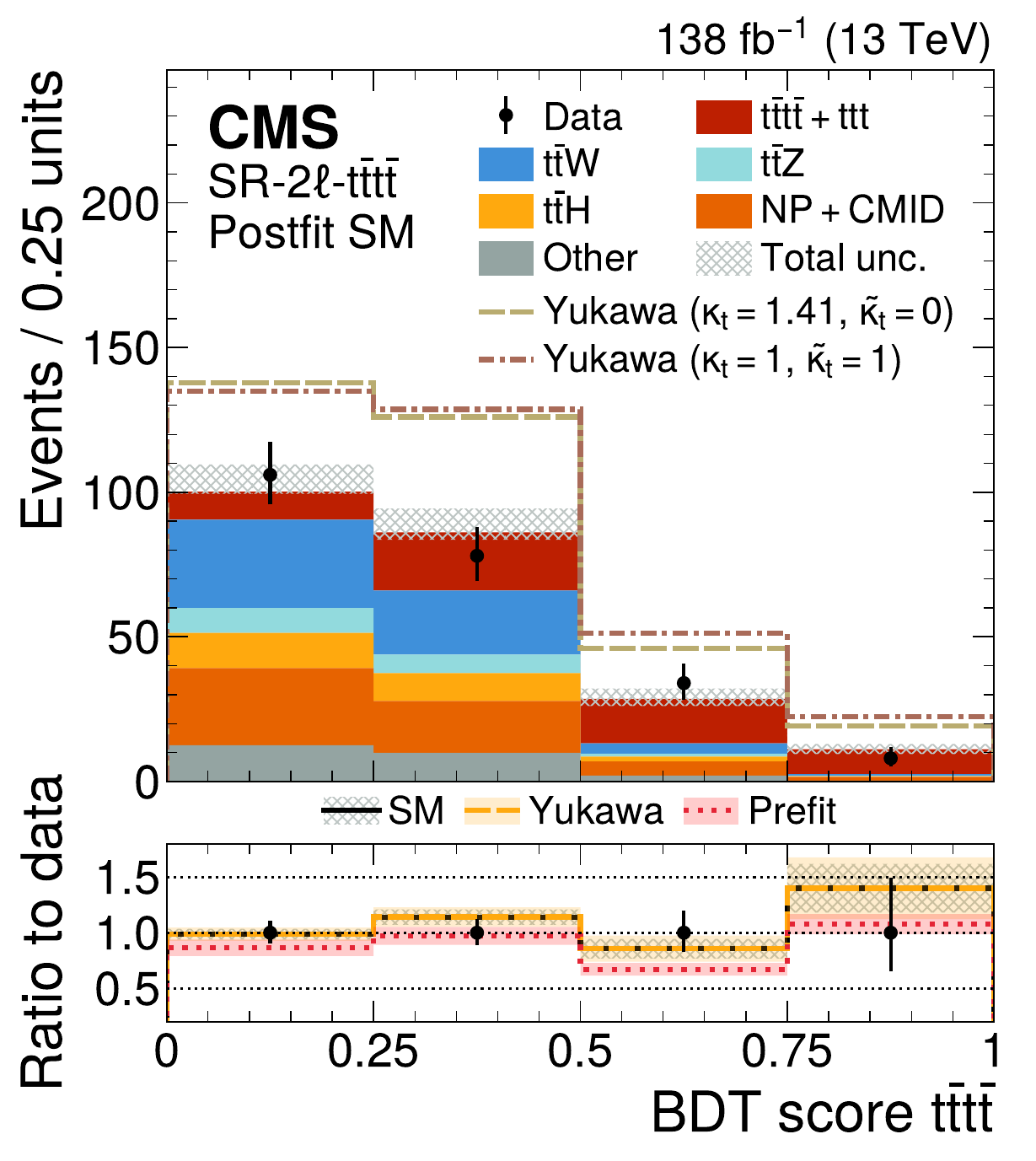}%
\hspace*{0.05\textwidth}
\includegraphics[width=0.45\textwidth]{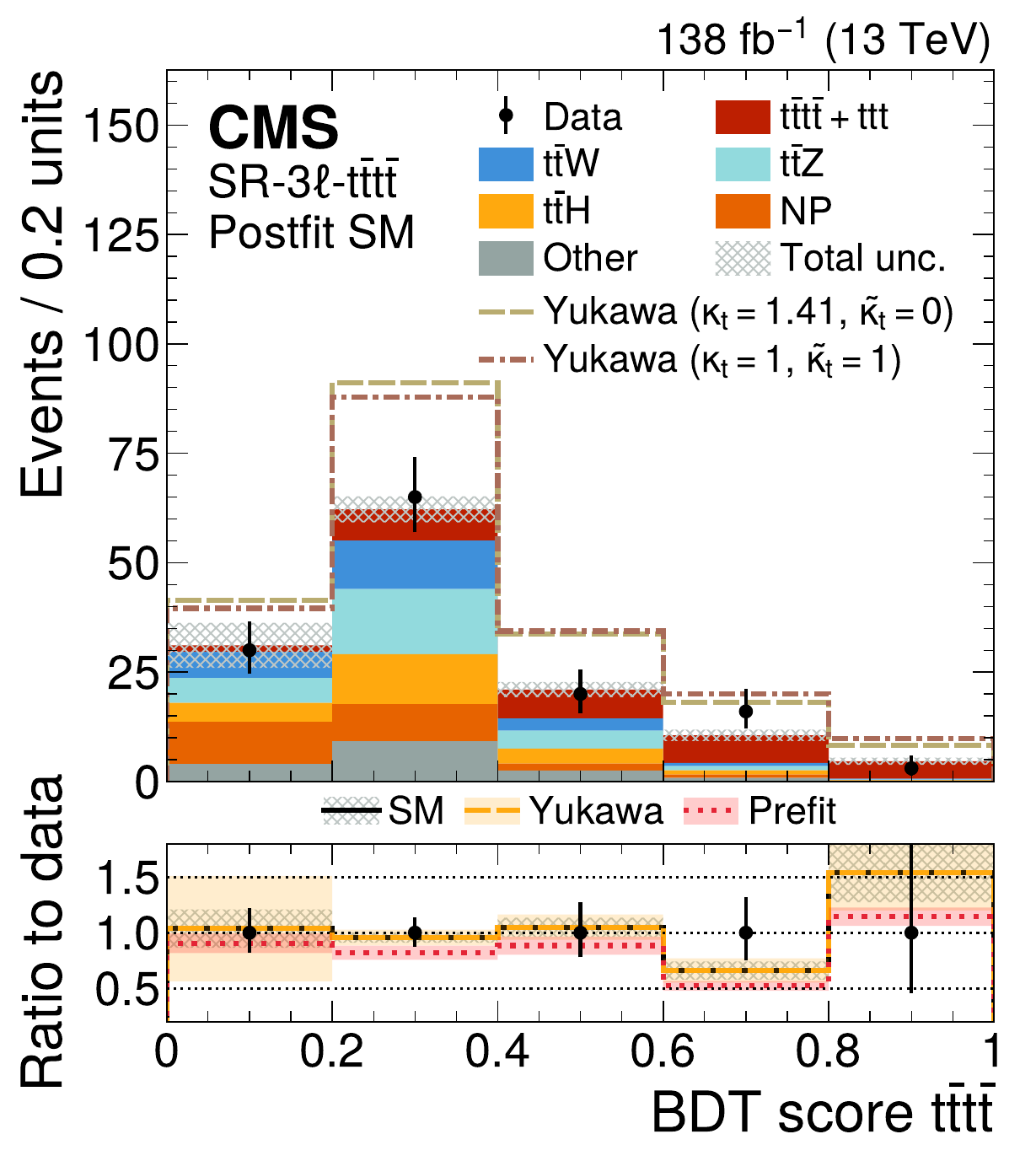} \\
\includegraphics[width=0.45\textwidth]{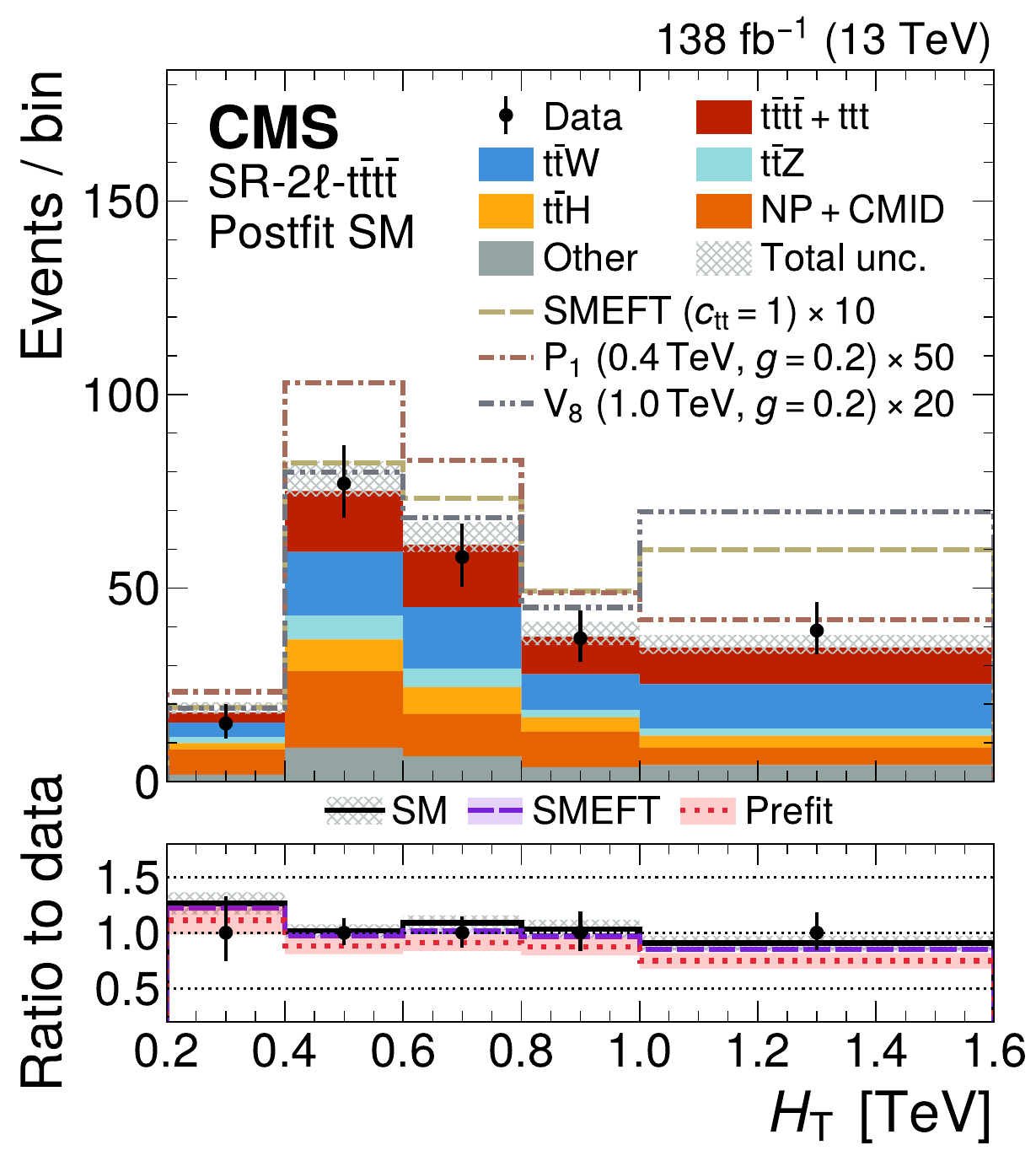}%
\hspace*{0.05\textwidth}
\includegraphics[width=0.45\textwidth]{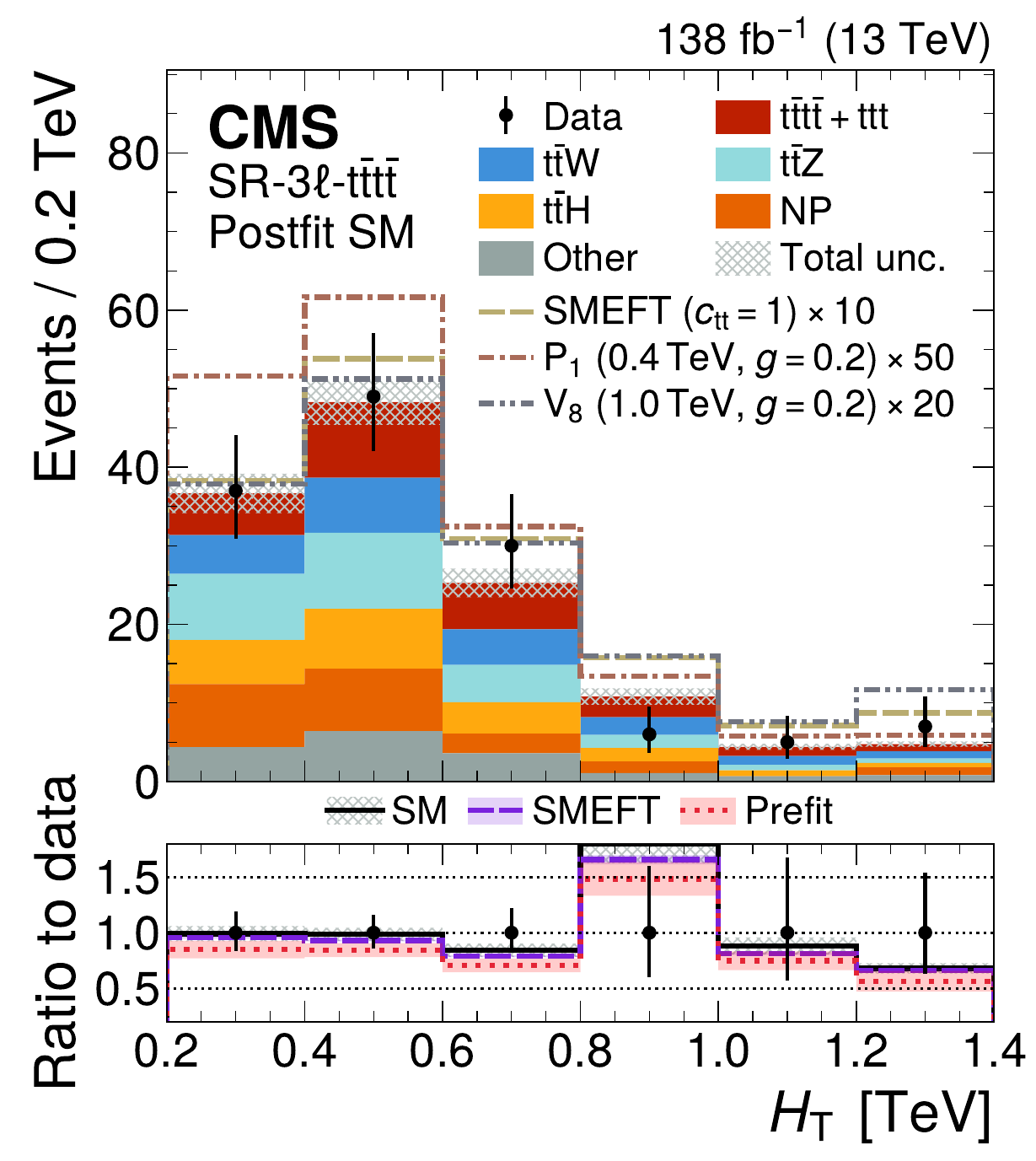}
\caption{%
    Comparison of the number of observed (points) and predicted (colored histograms) events in the BDT score \tttt (upper row) or \HT (lower row) distribution, shown for the \SRtwoTTTT (left, merged across all lepton flavor categories) and the \SRthreeTTTT (right).
    The last bins of the \HT distributions include the overflow contribution.
    The vertical bars on the points represent the statistical uncertainties in the data, and the hatched bands the total uncertainty in the predictions.
    The SM yields are shown with their best fit normalizations from the simultaneous fit to the data (``postfit'') for the SM fit.
    The dashed/dotted lines show the enhancement of \ttttPLUSttt production in different new-physics scenarios.
    The lower panels show the ratio of the total prediction to data for three postfit scenarios---SM, Yukawa coupling extraction, and SMEFT---and also using the SM yields before any fit to the data (``prefit'').
}
\label{fig:fitdistributions}
\end{figure}

The two choices of input distributions used in the \SRtwoTTTT and \SRthreeTTTT are shown in Fig.~\ref{fig:fitdistributions}.
The tails of the BDT score \tttt distributions have higher purity in predicted SM \tttt and \ttt production events.
In contrast, the tails of the \HT distributions provide better sensitivity to enhanced \tttt and \ttt production at higher energy scales, as modeled with four-heavy-quark SMEFT operators or topphilic heavy resonances.
For optimal sensitivity, we thus use the BDT score \tttt distributions as fit variables for the SM scenario and the extraction of the Yukawa coupling, but the \HT distributions for the SMEFT interpretation and the evaluation of exclusion limits on topphilic heavy resonances.

The \ttt contribution to the total \ttttPLUSttt yield is 8\% both in the \SRtwoTTTT and in the \SRthreeTTTT.
In the last bins of the BDT score \tttt distributions, the \ttt contribution falls to about 5\%.
In the last bins of the \HT distributions, in contrast, the \ttt contribution remains at about 7\%.

\subsection{Fit results in the SM scenario}

To cross-check the previously published result in Ref.~\cite{CMS:TOP-22-013} given the changes to the experimental analysis discussed in this section, we perform fits without considering a BSM model.
First, we consider the \tttt production cross section as single parameter of interest, constrain the \ttt production cross section to its SM prediction, and use the BDT score \tttt distributions as the fit distributions in \SRtwoTTTT and \SRthreeTTTT.
The observed (expected) significance of the \tttt signal is 6.2 (4.9) standard deviations, and the measured cross section of \tttt production is
\begin{equation}
    \sigma(\tttt)=18.9\,^{+3.7}_{-3.5}\stat\,^{+2.5}_{-2.1}\syst\unit{fb}=18.9\,^{+4.5}_{-4.0}\unit{fb},
\end{equation}
consistent with the results reported in Ref.~\cite{CMS:TOP-22-013}.
The obtained normalization parameters for \ttZ and \ttW production agree with those reported in Ref.~\cite{CMS:TOP-22-013}.
The leading systematic uncertainties in $\sigma(\tttt)$ are related to the \ttZ normalization, \PQb tagging, final-state radiation modeling, and jet energy calibration.
Compared to Ref.~\cite{CMS:TOP-22-013}, the impact of the uncertainty in the modeling of additional jet and \PQb jet radiation in \ttW production is reduced, which follows from the improved \ttW simulation employed here.

The measured \tttt production cross section is 1.3 standard deviations higher than the SM prediction of $13.4\,^{+1.0}_{-1.8}\unit{fb}$ from Ref.~\cite{vanBeekveld:2022hty}.
The interpretations performed here characterize this discrepancy in terms of the different considered BSM scenarios, with a clear expectation to find best fit values about one standard deviation away from the SM configuration in each scenario.
Using instead the SM prediction of $10.4\,^{+2.5}_{-2.2}\unit{fb}$ from Ref.~\cite{vanBeekveld:2025ghw}, the measured \tttt production cross section is 1.8 standard deviations higher, and the BSM interpretations would find a correspondingly larger deviation from the SM.

\begin{figure}[!htp]
\centering
\includegraphics[width=0.45\textwidth]{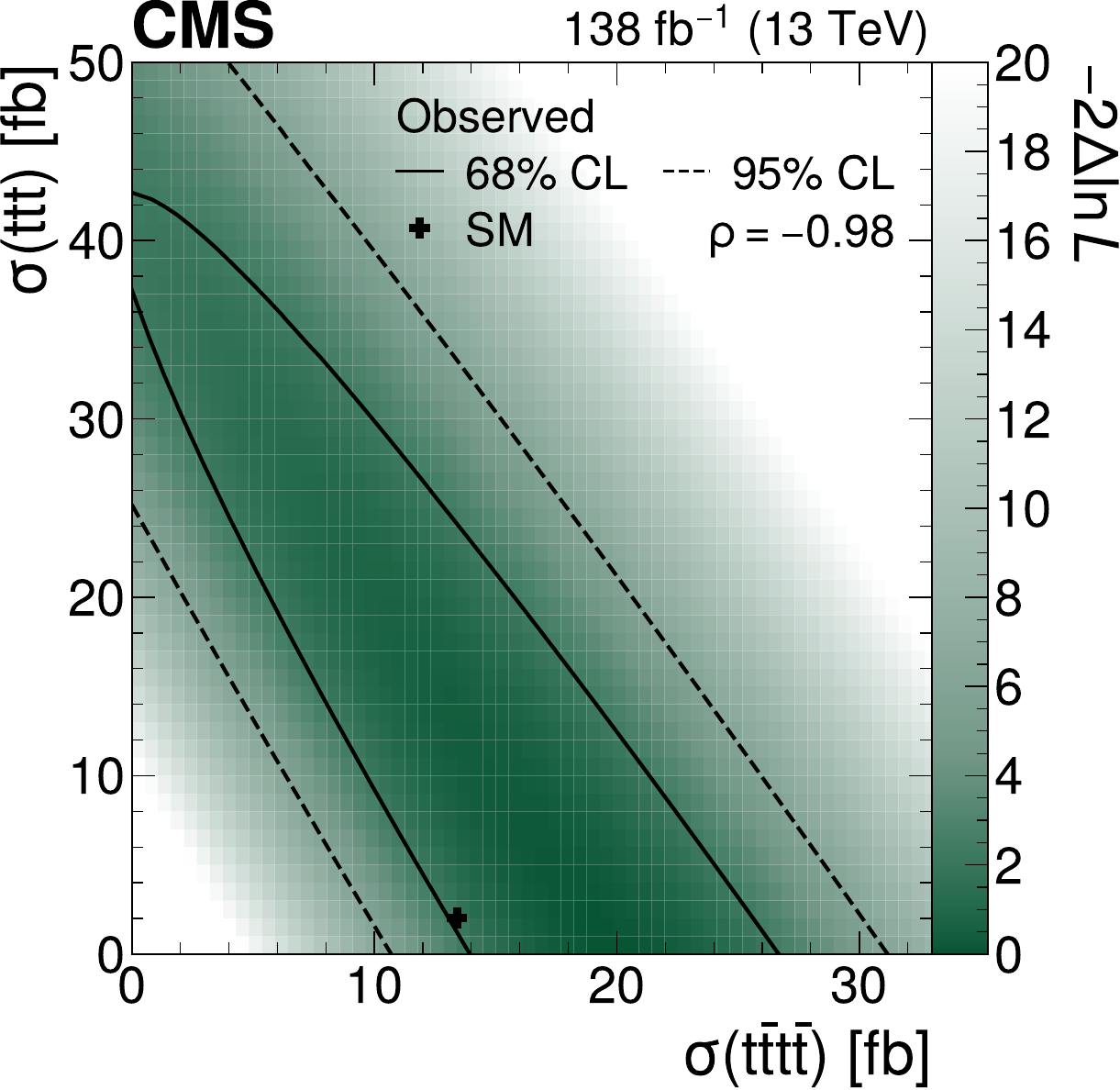}
\caption{%
    Two-dimensional scan of the \tttt and \ttt cross sections.
    The color scale shows the negative log-likelihood difference with respect to the best fit point, and the contour lines show the 68\% (solid) and 95\% (dashed) \CL intervals.
    The SM prediction is indicated with a black cross.
    The correlation $\rho$ between the two measured cross sections is $-0.98$.
}
\label{fig:sm2dfit}
\end{figure}

We performed a second fit where both the \tttt and \ttt production cross sections are parameters of interest.
The fit result is shown in Fig.~\ref{fig:sm2dfit}.
A very strong anticorrelation between the \tttt and \ttt production cross sections is observed, indicating that the fit is not able to distinguish between the contributions from the two processes.
Both the scenario with no \ttt production and only \tttt production or vice versa are compatible with the data within one standard deviation.
Given the similarity of the experimental signatures of both processes, the approach in our BSM interpretations to treat both \tttt and \ttt production as signal processes is well motivated.

\section{Interpretation in the SMEFT framework}
\label{sec:smeft}

\subsection{Theoretical framework}
\label{sec:smeft:theory}

In the SMEFT framework, the SM Lagrangian \LagrangianSM is expanded by terms of higher-dimensional operators \OperatorD, where $d\geq5$ is the mass dimension.
The \OperatorD are built from products of SM fields and respect the SM symmetries, and are suppressed by a BSM energy scale $\Lambda$.
The expansion of \LagrangianSM in powers of $\Lambda^{-1}$ is written as
\begin{equation}
    \Lagrangian_{\text{SMEFT}}=\LagrangianSM+\sum_{d,i}\frac{\WilsonD}{\Lambda^{d-4}}\OperatorD,
\end{equation}
where \WilsonD are the Wilson coefficients (WCs) that control the size of the effect of the respective operator.
After excluding operators of odd dimension that violate baryon or lepton number, the leading BSM contributions arise from dimension-6 operators, with higher-dimensional operators suppressed by factors of $\Lambda^{-4}$ or more~\cite{Degrande:2012wf}.
We will thus consider only dimension-6 operators and refer to the corresponding WCs as $c_i$.
If not specified otherwise, we use a fixed value of $\Lambda=1\TeV$.

The ME of a scattering process in the presence of SMEFT operators can be written as
\begin{equation}\label{eq:matelementsmeft}
    \MatElement=\MatElementSM+\sum_i\frac{c_i}{\Lambda^2}\MatElementEFT+\mathcal{O}\left(\frac{c^2}{\Lambda^4}\right),
\end{equation}
where \MatElementSM is the SM ME and the \MatElementEFT are the MEs corresponding to BSM interactions.
To obtain the cross section of the scattering process, the ME is squared, resulting in
\begin{equation}
    \sigma\propto\abs{\MatElement}^2=\abs{\MatElementSM}^2+2\sum_i\frac{c_i}{\Lambda^2}\Re\big(\MatElementEFT\MatElementSM^\ast\big)+\sum_{i,j}\frac{c_ic_j}{\Lambda^4}\Re\big(\MatElementEFT\MatElementEFT[j]^\ast\big)+\mathcal{O}\left(\frac{c^3}{\Lambda^6}\right).
\end{equation}
Leaving out the higher-order terms, the cross section $\sigma_p^k$ of a process $p$ in a bin $k$ of some distribution can thus be written as
\begin{equation}\label{eq:eftxsec}
    \sigma_{p,\text{SMEFT}}^k=\sigma_{p,\text{SM}}^k+\sigma_{p,\text{int}}^k(\vec{c})+\sigma_{p,\text{EFT}}^k(\vec{c})=\sigma_{p,\text{SM}}^k\bigg(1+\sum_iA_{p,i}^k\frac{c_i}{\Lambda^2}+\sum_{i,j}B_{p,ij}^k\frac{c_ic_j}{\Lambda^4}\bigg),
\end{equation}
where the $\sigma_p^k$ indices denote the SM contribution (``SM''), the contribution purely from di\-men\-sion-6 SMEFT operators (``EFT''), and the contribution from the interference between the SM and EFT MEs (``int'').
The final part of the formula shows a parameterization into a part that is linear in the WCs and part that is quadratic in them, with the strength of the linear and quadratic contributions described by the constants $A_{p,i}^k$ and $B_{p,ij}^k$.

We consider six dimension-6 operators to which \tttt and \ttt production provide significant sensitivity.
Five of these operators are four-heavy-quark operators, defined following the SMEFTsim 3.0 convention~\cite{Brivio:2017btx, Brivio:2020onw} as
\begin{equation}\begin{aligned}
    \Operator_{\PQt\PQt}&=\big(\PAQt\gamma_\mu\PQt\big)\big(\PAQt\gamma^\mu\PQt\big), &
    \Operator_{\PQQ\PQQ}^{(1)}&=\big(\PAQQ\gamma_\mu\PQQ\big)\big(\PAQQ\gamma^\mu\PQQ\big), &
    \Operator_{\PQQ\PQQ}^{(8)}&=\big(\PAQQ T^a\gamma_\mu\PQQ\big)\big(\PAQQ T^a\gamma^\mu\PQQ\big), \\
    \Operator_{\PQQ\PQt}^{(1)}&=\big(\PAQQ\gamma_\mu\PQQ\big)\big(\PAQt\gamma^\mu\PQt\big), &
    \Operator_{\PQQ\PQt}^{(8)}&=\big(\PAQQ T^a\gamma_\mu\PQQ\big)\big(\PAQt T^a\gamma^\mu\PQt\big),
\end{aligned}\end{equation}
where \PQQ denotes the $SU(2)$ doublet of left-handed third-generation quarks, \PQt the $SU(2)$ singlet of the right-handed top quark, and $T^a$ the $SU(3)$ generators.
Following the observation in Ref.~\cite{Zhang:2017mls} that the contribution of these five operators to \tttt production can be fully described with only four degrees of freedom, we consider only the four WCs \ctt, \cQQone, \cQtone, and \cQteight.
In addition, we consider the \tH operator
\begin{equation}
    \Operator_{\PQt\PH}=\big(\PH^\dagger\PH\big)\big(\PAQQ\tilde{\PH}\PQt\big),
\end{equation}
where \PH is the scalar Higgs doublet field, which modifies the Yukawa coupling of the top quark.
To consider both \CP-even and \CP-odd Yukawa coupling modifications, we define a complex WC $\ctH=\ctHRe+\iu\ctHIm$ with two real-valued coefficients \ctHRe, \ctHIm that we will constrain in this analysis.
In total, we thus consider six WCs.

Simulated event samples for \tttt, \tttW, and \tttq production in the SMEFT scenario are produced at LO with \MGvATNLO using the \textsc{topU3l} model provided in SMEFTsim 3.0~\cite{Brivio:2017btx, Brivio:2020onw}, as well as for \ttH production to account for the sensitivity to \ctHRe and \ctHIm.
The ME calculation includes diagrams with no SMEFT vertices (\ie, the SM contribution) and diagrams with one of the six (or two for \ttH) considered SMEFT vertices.
The \MGvATNLO event reweighting technique~\cite{Mattelaer:2016gcx} is employed to include per-event weights in the simulated samples for a large number of configurations of the six WCs.
Using these weights, we can obtain predictions for any detector-level distribution of interest corresponding to a chosen SMEFT configuration, and thus determine the parameterization defined in Eq.~\eqref{eq:eftxsec} relative to the SM prediction included in the same sample.
The parameterization is extracted from the samples simulated at LO, and applied as a function of the observable of interest to the SM samples simulated at NLO.
This approach was developed in Ref.~\cite{CMS:TOP-19-001}, where a more detailed description is provided.
The assumption in this approach of an approximate factorization of higher-order QCD corrections from the SMEFT dependence is not exact, with residual differences expected to not be larger than the scale variations~\cite{Degrande:2020evl}.

\begin{figure}[!ht]
\centering
\includegraphics[width=0.45\textwidth]{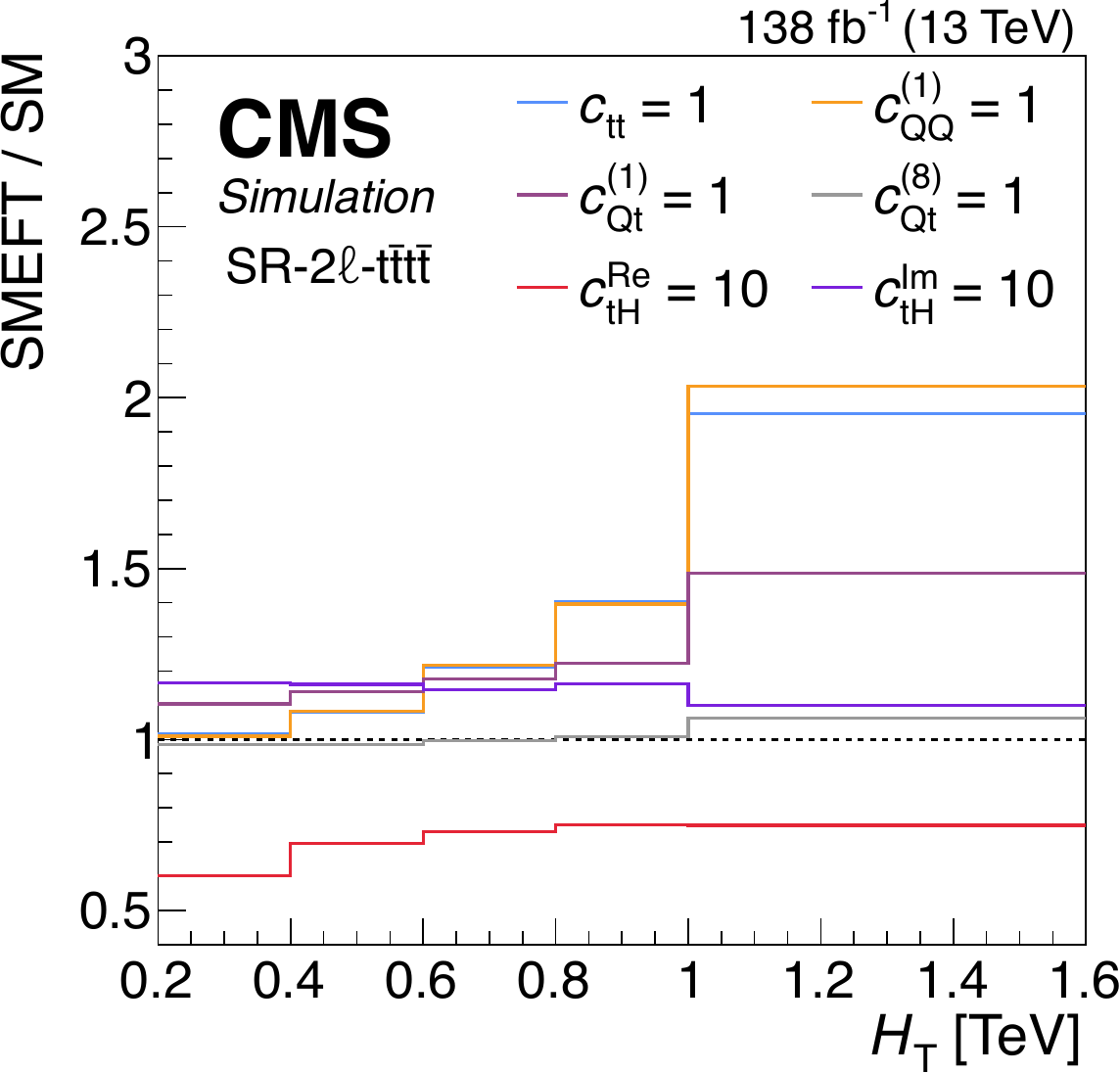}
\caption{%
    Comparison of the \HT distribution in the \SRtwoTTTT for different SMEFT scenarios relative to the SM prediction.
    Each line shows the ratio of the SMEFT prediction for \tttt, \ttt, and \ttH production combined with exactly one WC at a nonzero value to the SM prediction for the same processes.
    The last bin includes the overflow contribution.
}
\label{fig:EFTparameterization}
\end{figure}

The impact of the six considered WCs on the \HT distribution in the \SRtwoTTTT is shown in Fig.~\ref{fig:EFTparameterization}.
Nonzero values of the four-heavy-quark operators result in a larger signal contribution when compared to the SM scenario, which is especially enhanced at high values of \HT.
This indicates that these operators parameterize possible BSM effects at higher energies.
In contrast, the two \ctH WCs have a stronger effect at lower values of \HT, consistent with the expectation that EW contributions are more relevant at lower energies~\cite{Frederix:2017wme}.
A similar behavior is found in \SRthreeTTTT.
Compared to the BDT score distributions, using the \HT distributions improves the sensitivity to the four-heavy-quark operators by 12--20\%.

In the highest \HT bin, the median invariant mass of the four top quarks at parton level is 2\TeV in the \SRtwoTTTT and \SRthreeTTTT, with a central 68\% interval of 1.5--2.6\TeV.
Because $\Lambda=1\TeV$ fixes only the WC normalization and squared dimension-6 terms are retained without the complete set of $\Lambda^{-4}$ contributions, the limits should be interpreted within this SMEFT truncation.

\subsection{Results}

To set exclusion limits on the WCs in the SMEFT scenario, the \HT distribution is used for the fit in the \SRtwoTTTT and \SRthreeTTTT.
The effect of the WCs on \tttt, \ttt, and \ttH production is considered.
Fits are performed with all six WCs included as free parameters of the fit.
The result of the six-dimensional fit is consistent with the SM expectation with a negative log-likelihood difference of $q\approx6$, corresponding to 0.8 standard deviations under the assumption that Wilks' theorem holds.
This matches the observation of a slight excess in the SM fit.
We also evaluate one- and two-dimensional profile-likelihood ratios on single WCs or pairs of WCs, respectively, either by fixing the other WCs to zero or by profiling over the other WCs.

\begin{figure}[!ph]
\centering
\includegraphics[width=0.45\textwidth]{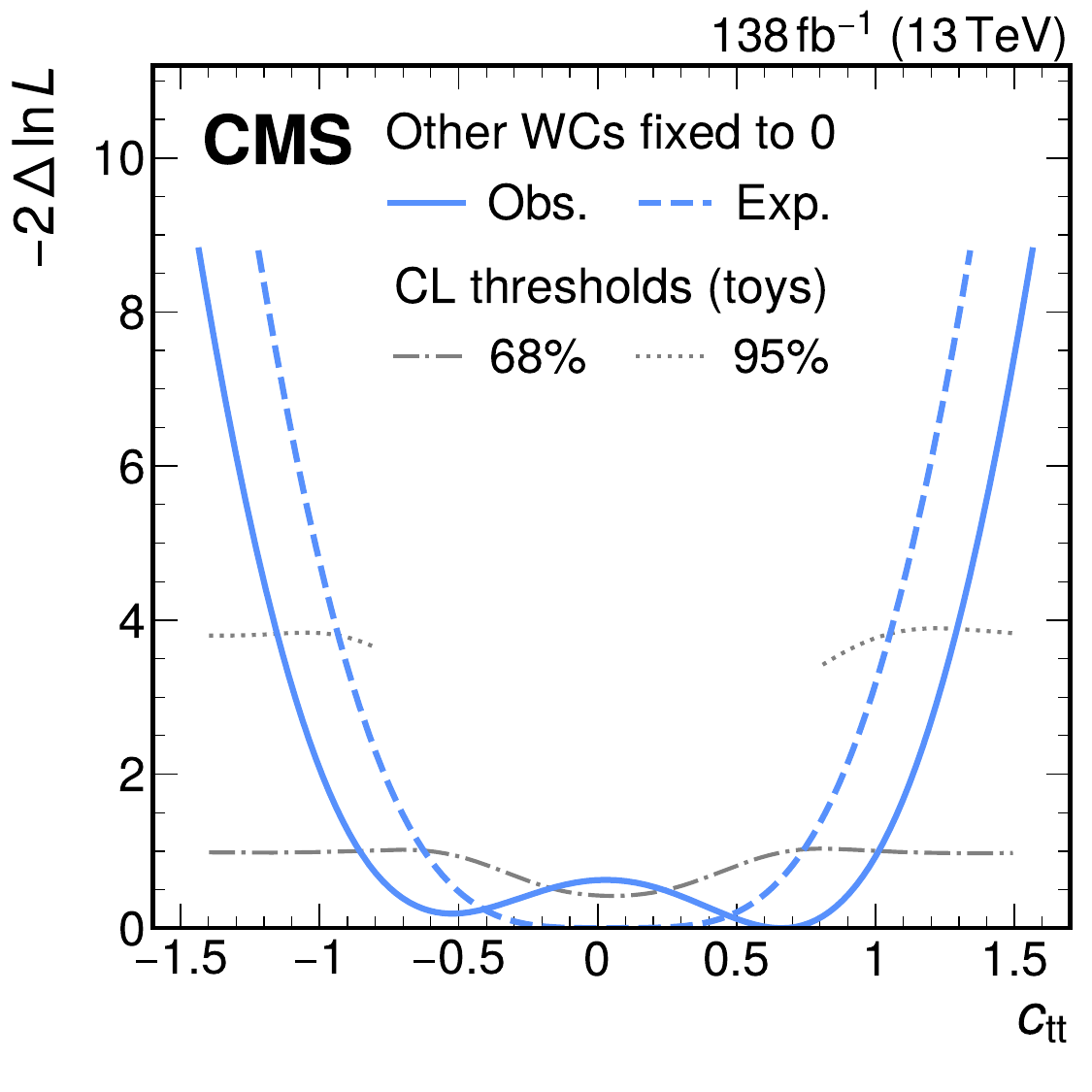}
\hspace*{0.05\textwidth}%
\includegraphics[width=0.45\textwidth]{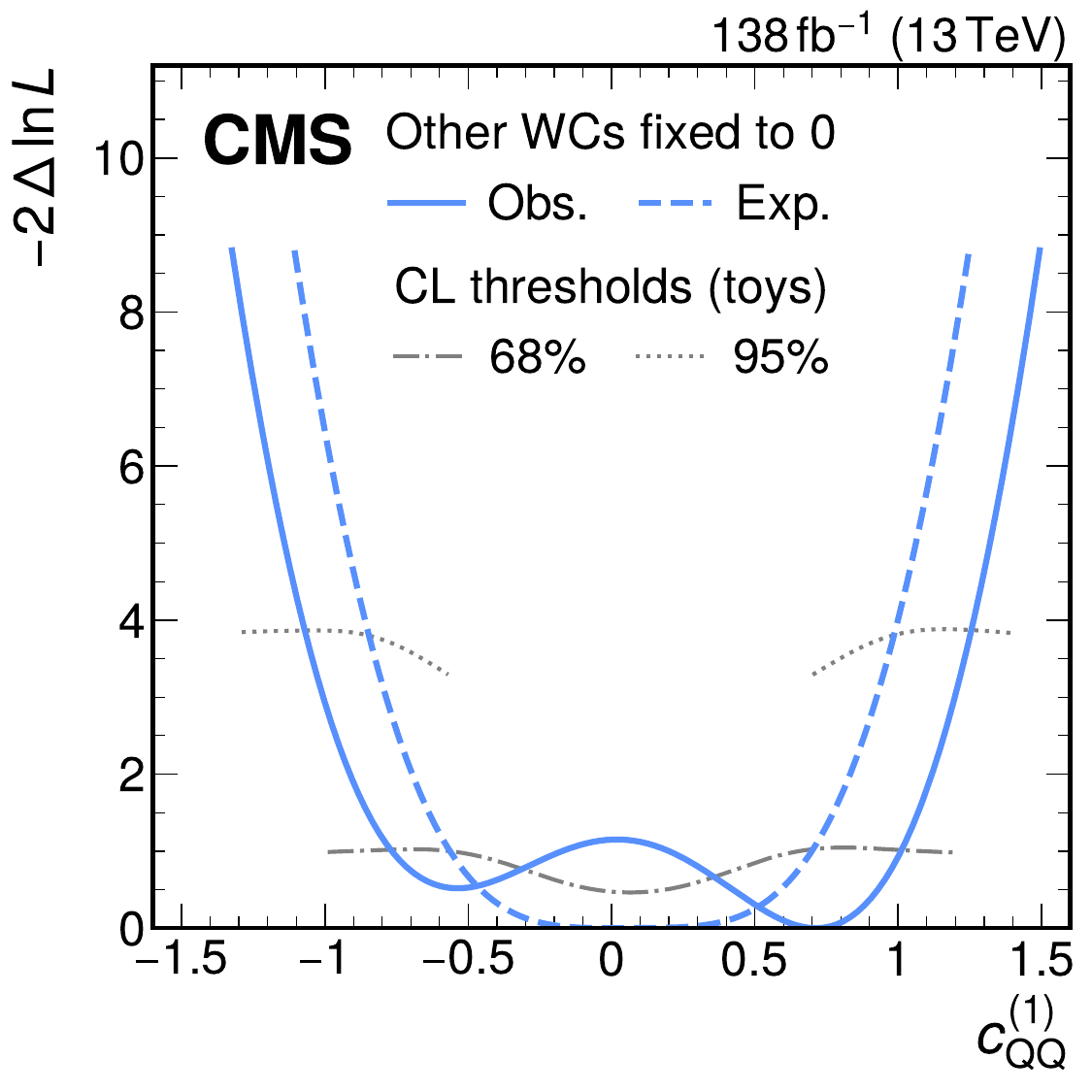} \\
\includegraphics[width=0.45\textwidth]{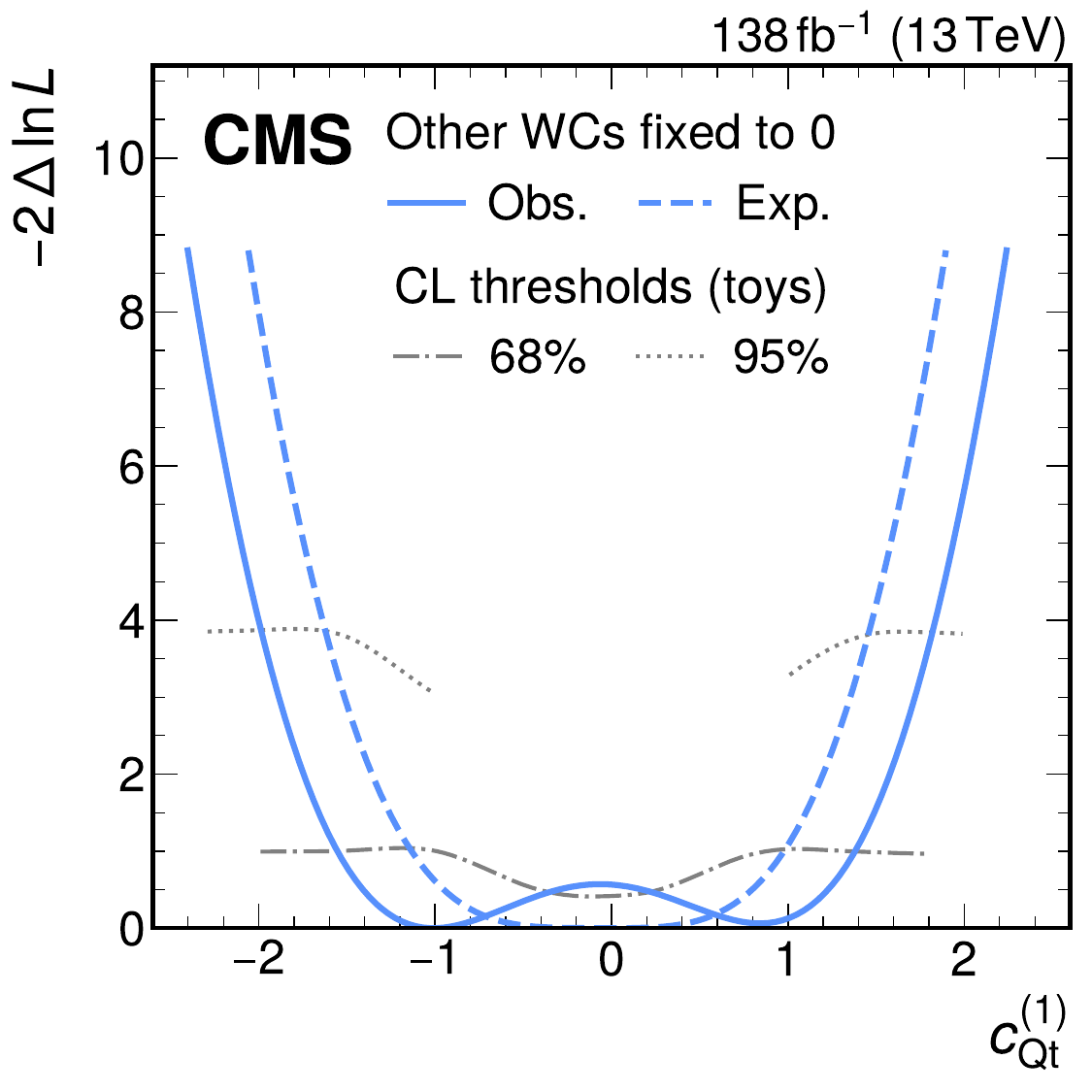}%
\hspace*{0.05\textwidth}%
\includegraphics[width=0.45\textwidth]{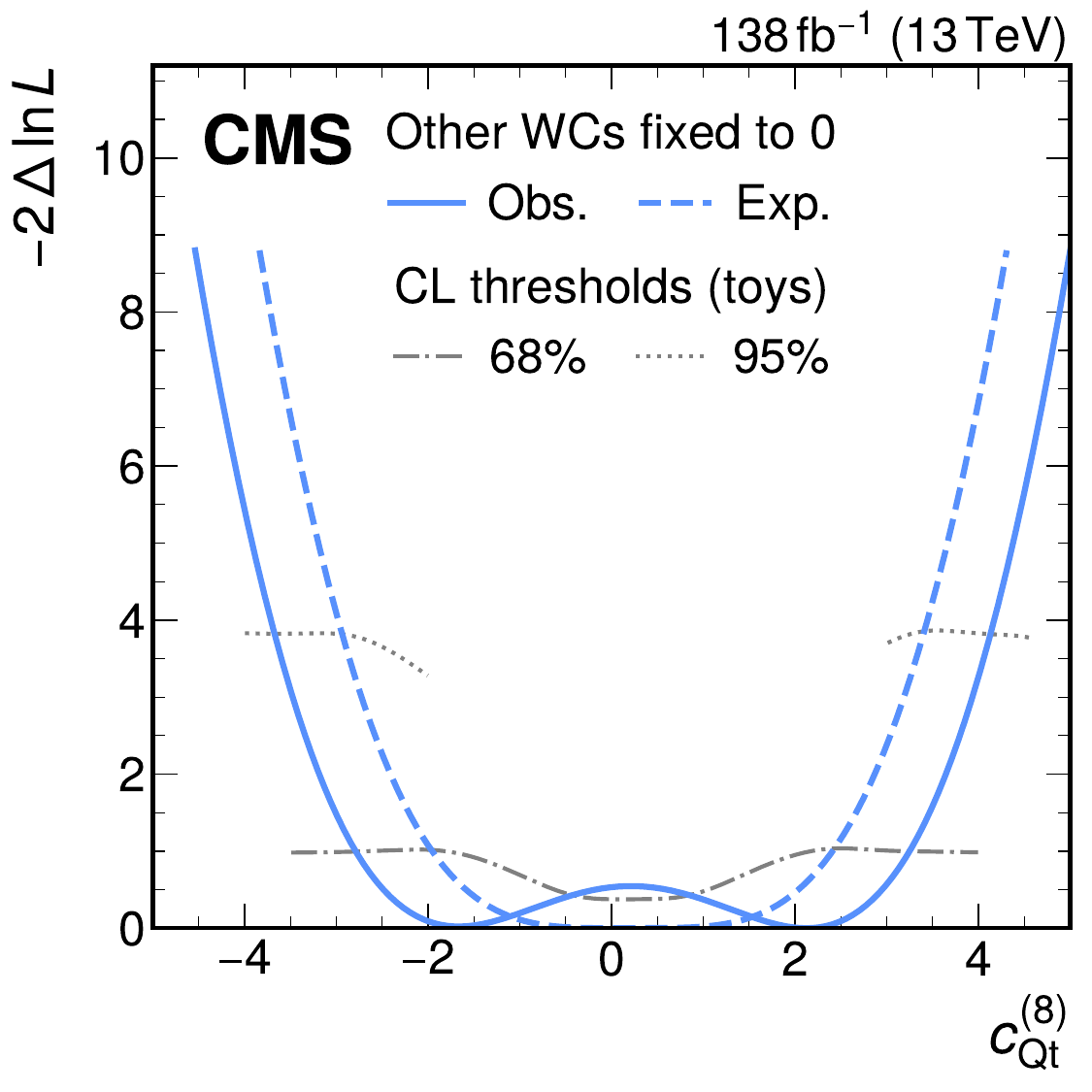} \\
\includegraphics[width=0.45\textwidth]{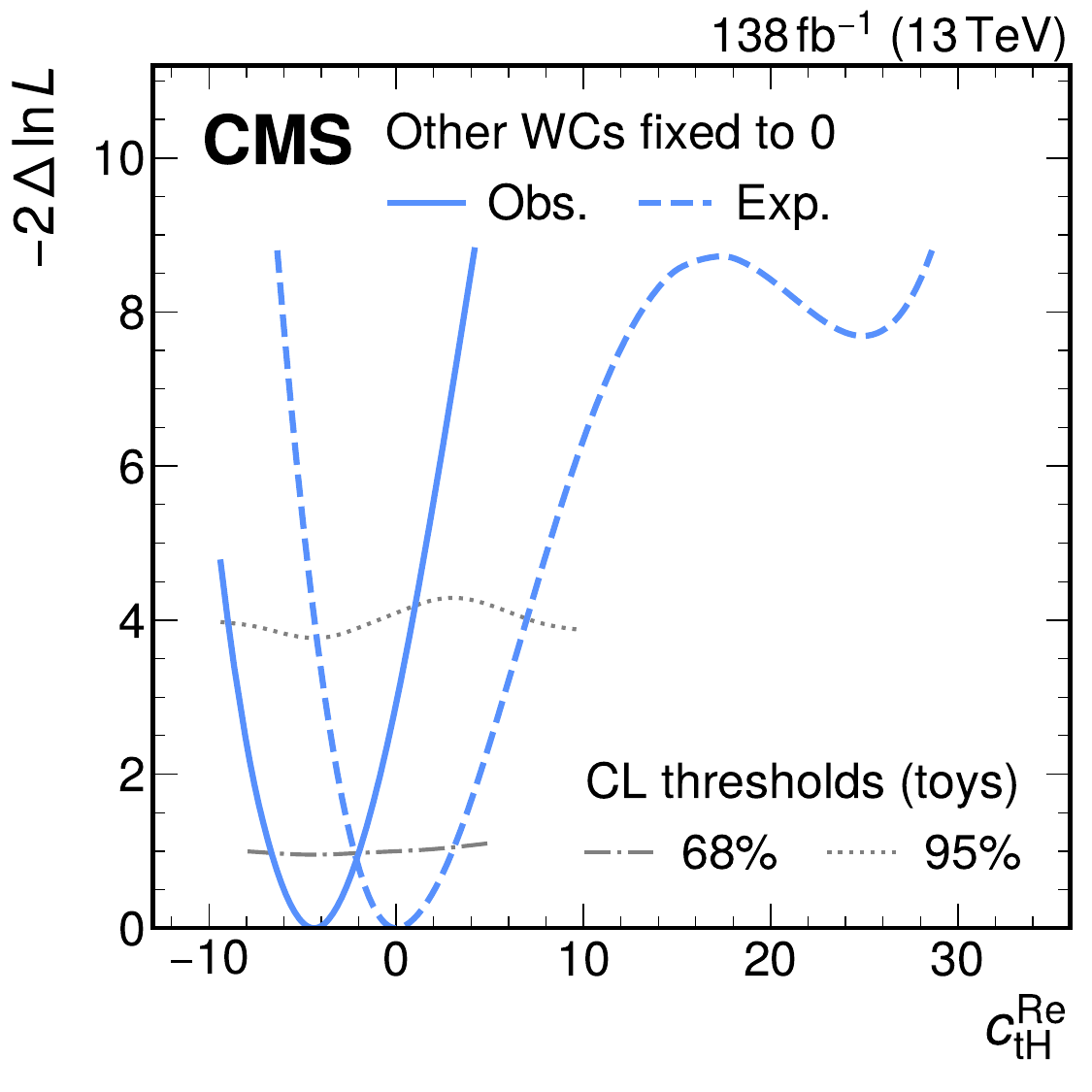}%
\hspace*{0.05\textwidth}%
\includegraphics[width=0.45\textwidth]{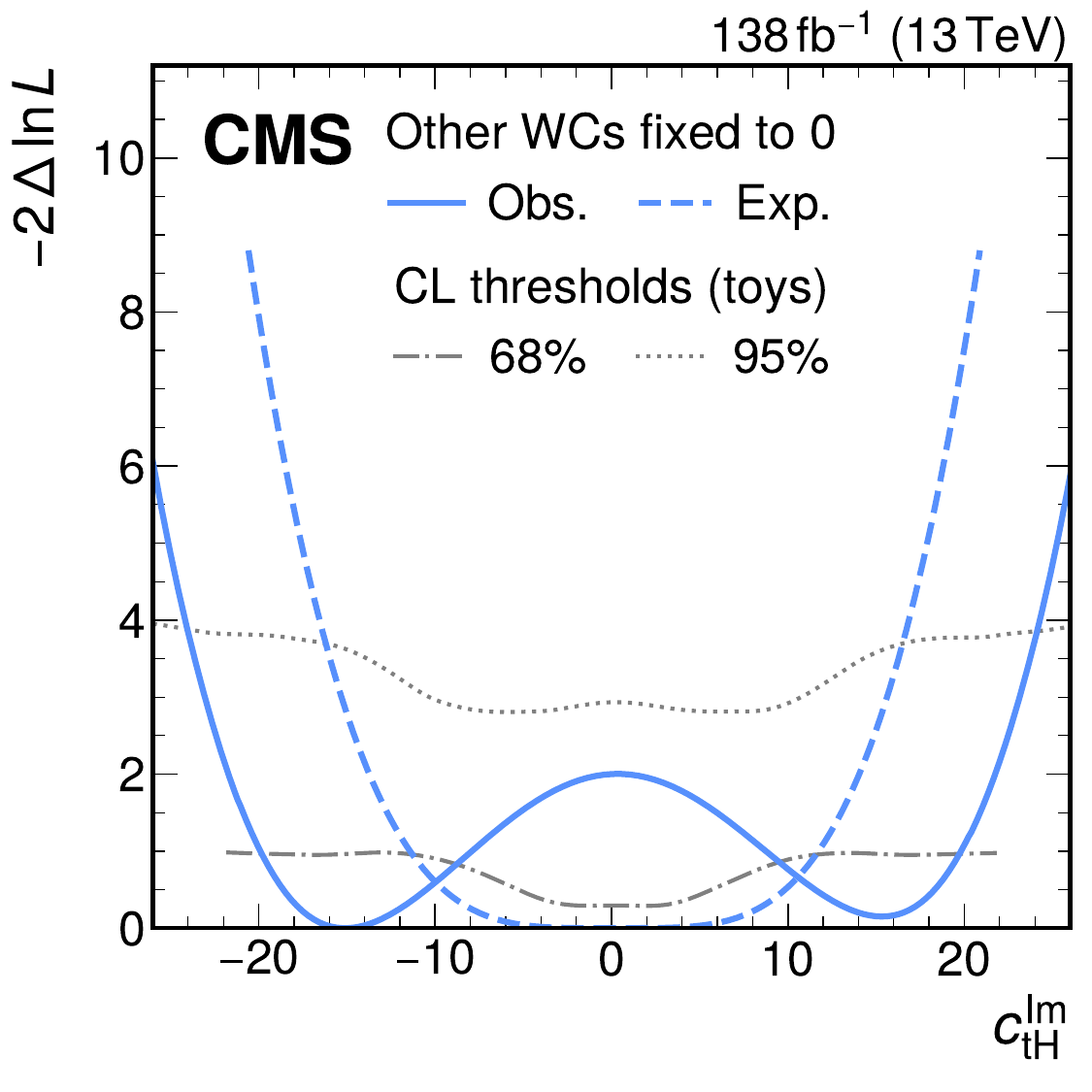}
\caption{%
    Negative log-likelihood difference from the best fit value for the one-dimensional scans of the WCs \ctt (upper left), \cQQone (upper right), \cQtone (middle left), \cQteight (middle right), \ctHRe (lower left), and \ctHIm (lower right), where the other WCs are fixed to zero.
    Shown are the expected (blue dashed line) and observed (blue solid line) results, as well as the threshold values for 68\% (gray dash-dotted line) and 95\% (gray dotted lines) \CL intervals as evaluated with toys.
}
\label{fig:eft1dfrozen}
\end{figure}

\begin{figure}[!ph]
\centering
\includegraphics[width=0.45\textwidth]{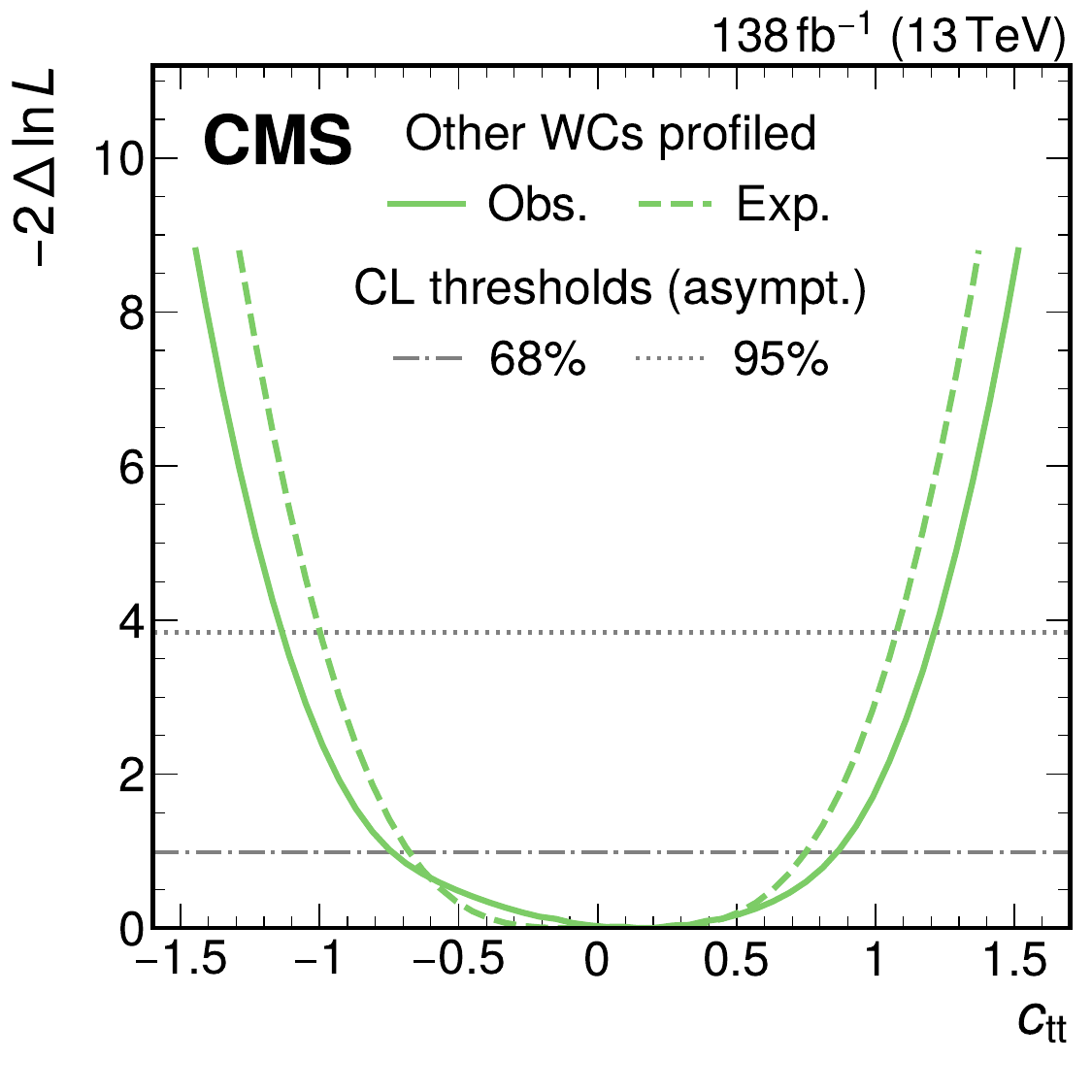}
\hspace*{0.05\textwidth}%
\includegraphics[width=0.45\textwidth]{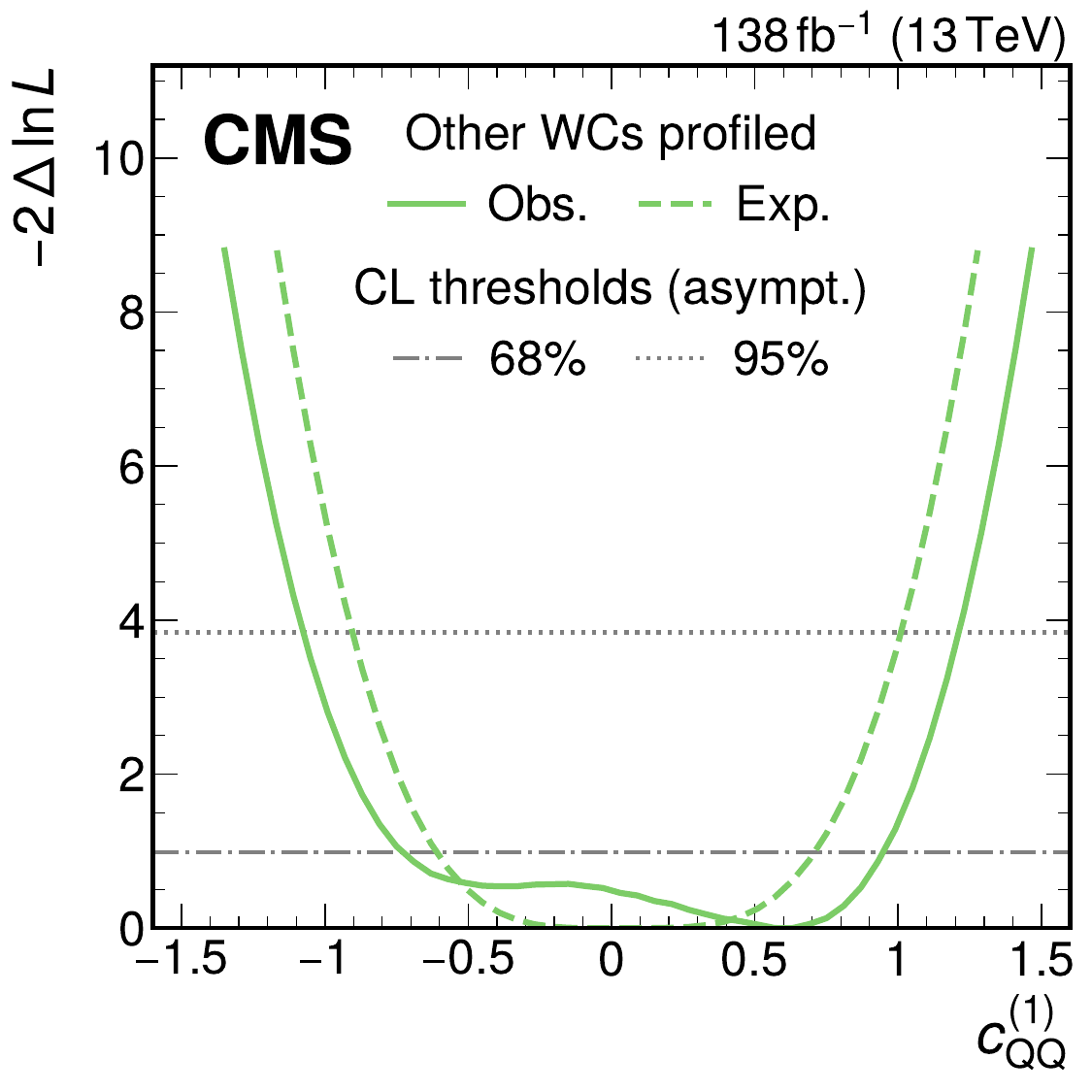} \\
\includegraphics[width=0.45\textwidth]{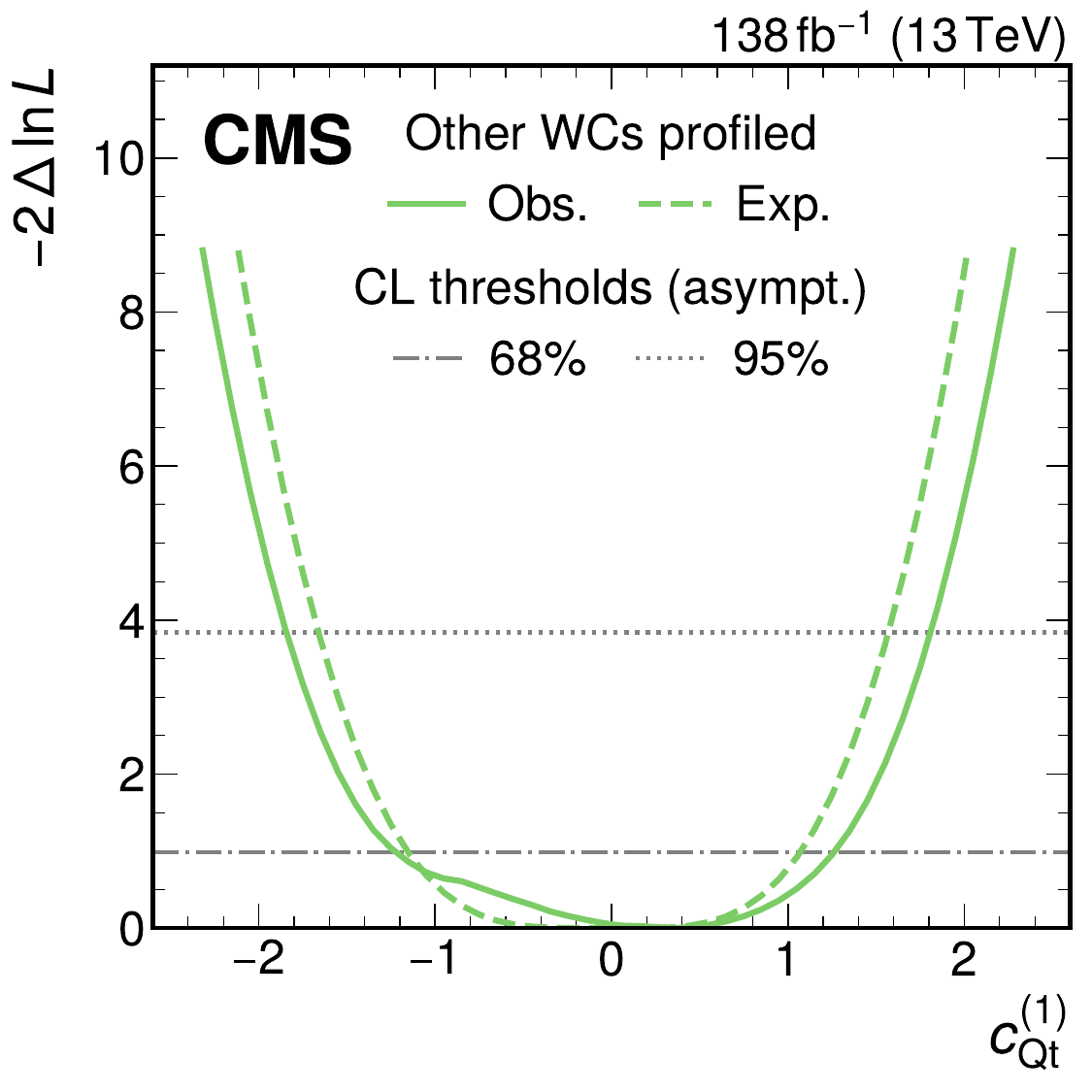}%
\hspace*{0.05\textwidth}%
\includegraphics[width=0.45\textwidth]{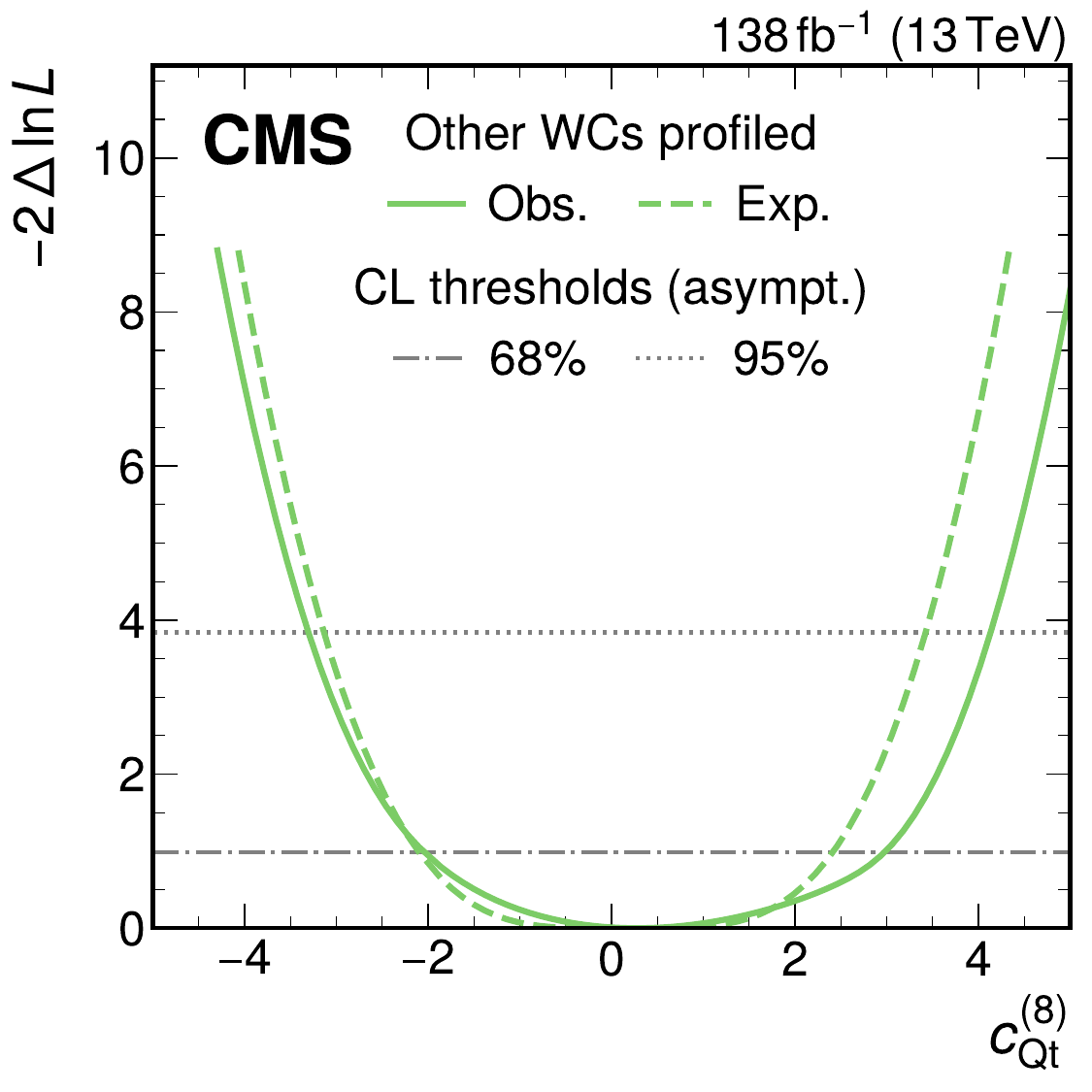} \\
\includegraphics[width=0.45\textwidth]{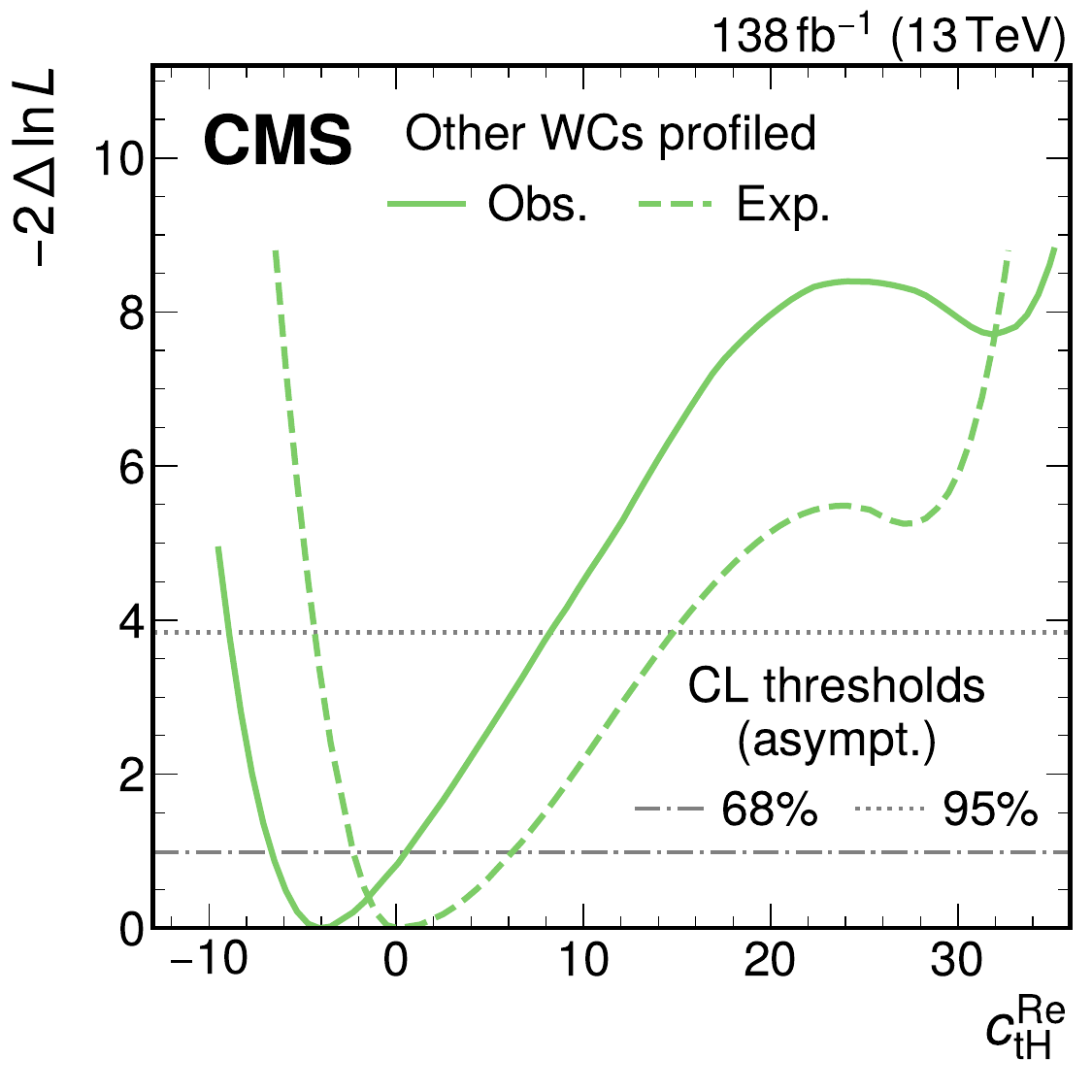}%
\hspace*{0.05\textwidth}%
\includegraphics[width=0.45\textwidth]{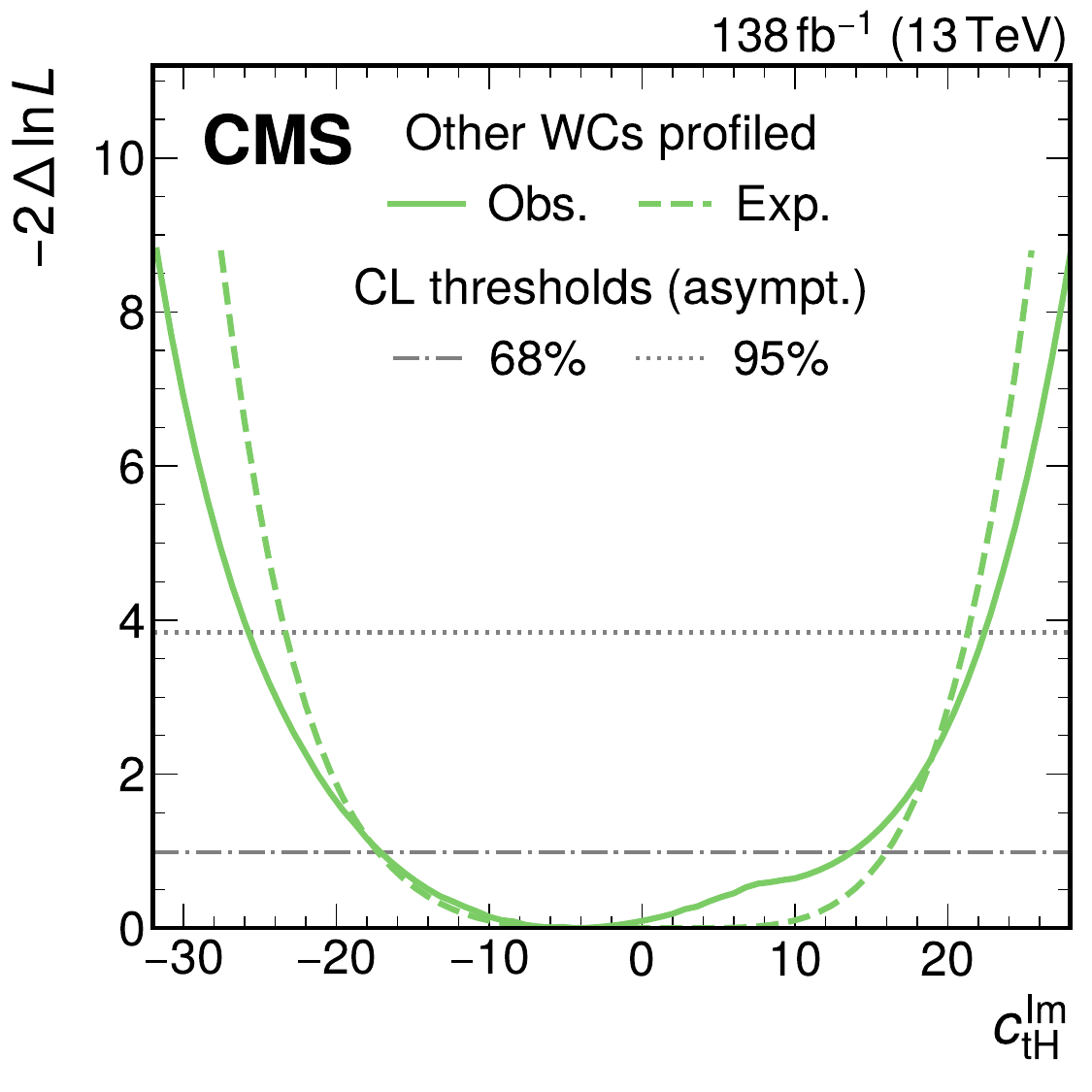}
\caption{%
    Negative log-likelihood difference from the best fit value for the one-dimensional scans of the WCs \ctt (upper left), \cQQone (upper right), \cQtone (middle left), \cQteight (middle right), \ctHRe (lower left), and \ctHIm (lower right), where the other WCs are profiled.
    Shown are the expected (green dashed line) and observed (green solid line) results, as well as the threshold values that would apply for 68\% (gray dash-dotted line) and 95\% (gray dotted line) \CL intervals if the asymptotic approximation was valid.
}
\label{fig:eft1dprofiled}
\end{figure}

The results of the one-dimensional scans for each WC with fixing the other WCs to zero are shown in Fig.~\ref{fig:eft1dfrozen}.
For this case, we have evaluated 68 and 95\% \CL exclusion limits using toys and display the threshold values on $q$ as a function of the parameter of interest in the relevant ranges with gray dash-dotted and dotted lines.
Consistent with the general observations in Ref.~\cite{Bernlochner:2022oiw}, the deviation from the asymptotic approximation is most notable around WC values of zero but becomes small for large deviations from the SM hypothesis.
The results when profiling the other WCs are shown in Fig.~\ref{fig:eft1dprofiled}, where a full toy evaluation would have been computationally too demanding, and we have thus only compared with the threshold values expected from the asymptotic approximation without checking the correct coverage.

\begin{figure}[!ph]
\centering
\includegraphics[width=0.8\textwidth]{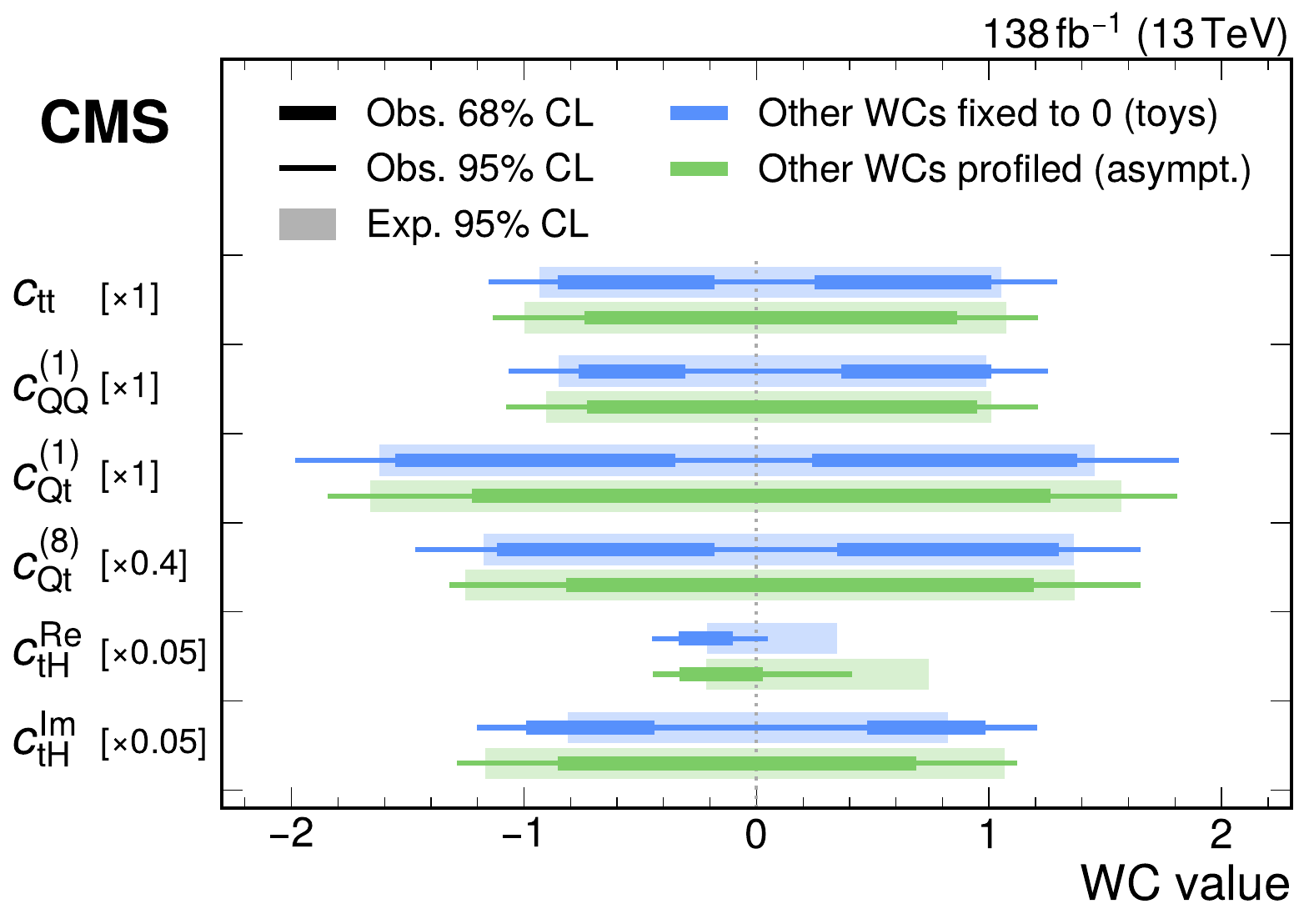}
\caption{%
    Constraints on the individual WCs, obtained by either fixing the other WCs to zero (blue) or profiling them (green).
    The lines and shaded areas indicate the observed and expected \CL intervals, respectively.
    For the case where the other WCs are fixed to zero, the 68 and 95\% \CL intervals are evaluated with toys.
    For the case where the other WCs are profiled, the intervals are instead evaluated by applying the asymptotic approximation.
    The constraints are scaled to ensure that all six WCs can be visualized on the same axis range.
}
\label{fig:eftsummary}
\end{figure}

In all cases, the data are statistically compatible with the SM expectation of zero for all WCs.
For each operator, the result when fixing the other WCs to zero slightly prefers nonzero values, whereas the profiled results have the best fit point closer to the SM expectation of zero.
This is consistent with the observation of a small excess of data over the expected signal prediction in the SM fit, which can be explained similarly well by either a strong modification of a single operator or a simultaneous small modification of all of them.
For \ctHRe, there are additional local minima in the range $25<\ctHRe<35$ in all scans, where the SMEFT prediction approaches the SM scenario but with a top quark Yukawa coupling with inverted sign.
When comparing to an alternative fit where the \ttt production processes are fixed to the SM, the width of the 95\% \CL exclusion limit is 1\% narrower for \ctt, 10\% narrower for \cQQone, 6\% narrower for \cQtone, 5\% narrower for \cQteight, very similar for \ctHRe, and 3\% narrower for \ctHIm, indicating the added sensitivity from the inclusion of \ttt production as signal processes.
We summarize the obtained one-dimensional \CL intervals (using toys for the case where the other WCs are fixed to zero and the asymptotic approximation for the profiled results) in Fig.~\ref{fig:eftsummary}.

\begin{figure}[!ph]
\centering
\includegraphics[width=0.45\textwidth]{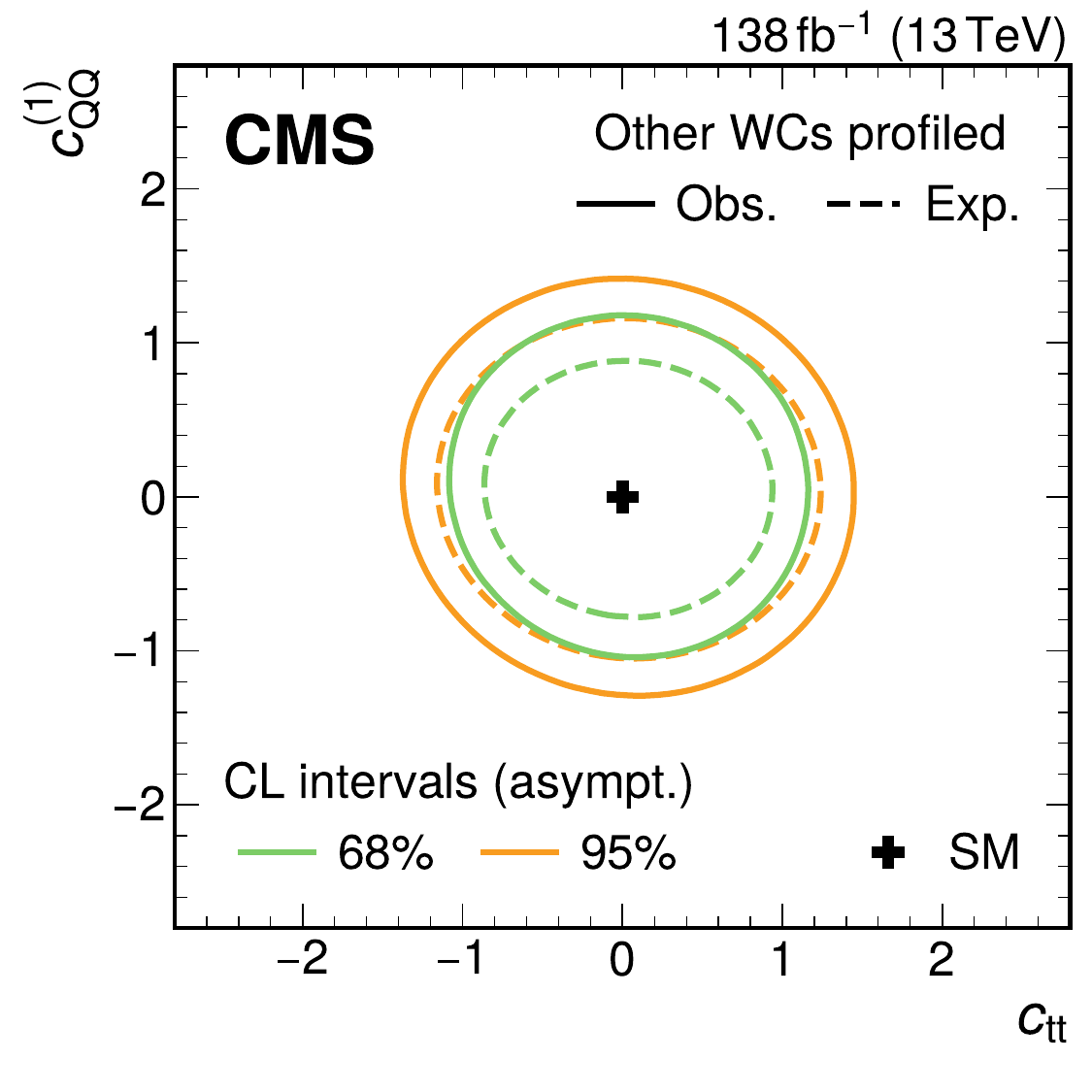}%
\hspace*{0.05\textwidth}%
\includegraphics[width=0.45\textwidth]{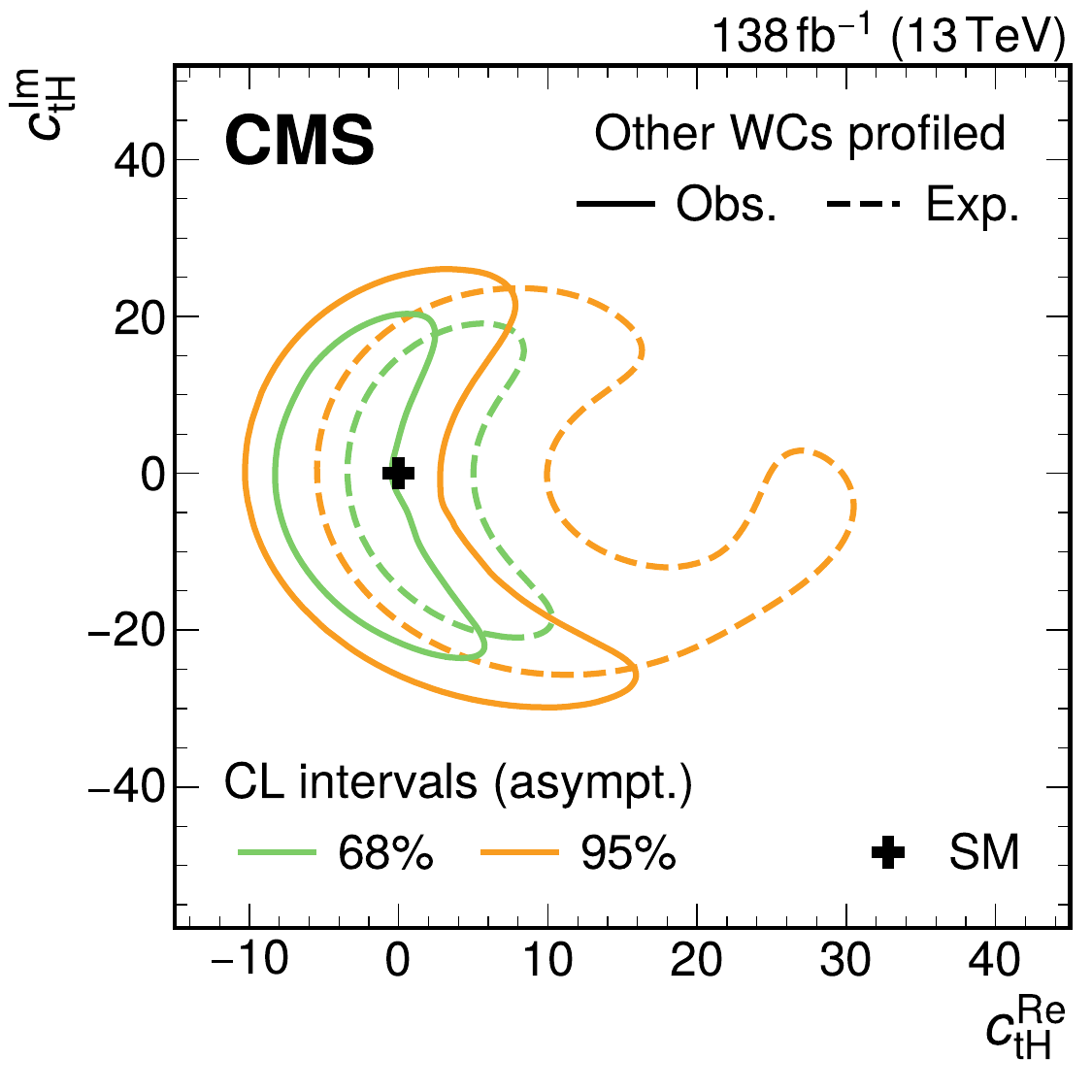}
\caption{%
    Expected (dashed lines) and observed (solid lines) exclusion contours for the two-dimensional scans of the WCs \ctt and \cQQone (left) and \ctHRe and \ctHIm (right), with the other WCs profiled in both cases.
    Shown are the \CL intervals where the test statistic falls below 2.3 (green lines) and 6.2 (orange lines), \ie, corresponding to the 68 and 95\% \CL intervals if the asymptotic approximation was valid.
}
\label{fig:eft2d}
\end{figure}

Two examples of two-dimensional scans are shown in Fig.~\ref{fig:eft2d}, where the other WCs are profiled and the \CL intervals are obtained using the asymptotic approximation.
For the four-heavy-quark operators, as well as for one four-heavy-quark operator and one \ctH operator, no strong correlations in the \CL intervals are found.
Only in the case of the two-dimensional scan of \ctHRe and \ctHIm, shown in Fig.~\ref{fig:eft2d} (right), there is a small correlation found, which is induced by SMEFT effects in the \ttt production processes.

To relate the sensitivity of our results to previous experimental constraints on the considered WCs, we compare the width of the 95\% \CL intervals between our analysis and relevant previous measurements.
The CMS \ttX multilepton EFT measurement in Ref.~\cite{CMS:TOP-22-006} is based on the same data and includes \tttt production in addition to \ttH and other associated top quark production processes as signal.
Our observed (expected) exclusion intervals on the four-heavy-quark operators are 12--25 (20--34)\% tighter than the limits (both profiled or only one WC considered at a time) quoted in Ref.~\cite{CMS:TOP-22-006}, reflecting that the \ttX measurement is not optimized for a high purity in \tttt events.
In contrast, the observed and expected \ctHRe intervals in Ref.~\cite{CMS:TOP-22-006} are up to 50\% tighter than ours, corresponding to the larger \ttH contributions to the event selection in the \ttX measurement.
Recently, a combination of different CMS SMEFT measurements was performed~\cite{CMS:SMP-24-003}, including the analysis from Ref.~\cite{CMS:TOP-22-006} and further dedicated Higgs boson cross section measurements.
Compared to the limits quoted in the combination that include both linear and quadratic terms (similar to our approach) and only consider one WC at a time, our observed (expected) exclusion intervals on the four-heavy-quark operators are 11--21 (6--17)\% tighter, whereas the \ctHRe intervals are up to three times tighter there.
The ATLAS \tttt cross section measurement~\cite{ATLAS:2023ajo}, which uses a single graph-neural-network discriminant for the combined same-sign \twol, \threel, and \fourl SR, and is not further optimized for EFT sensitivity, includes 95\% \CL intervals for the heavy-quark operators where only one WC is varied at a time.
Compared to those results, our observed (expected) exclusion intervals are 64--89 (31--55)\% tighter.
Our results provide the first constraints on \ctHIm from a \tttt production analysis within a SMEFT framework that includes both linear and quadratic SMEFT contributions.

\section{Search for narrow topphilic heavy resonances}
\label{sec:topphilic}

\subsection{Theoretical framework}
\label{sec:topphilic:theory}

\begin{figure}[!ht]
\centering
\includegraphics[width=0.26\textwidth]{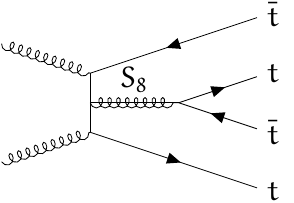}%
\hfill%
\includegraphics[width=0.26\textwidth]{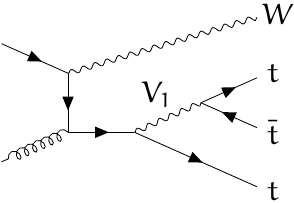}%
\hfill%
\includegraphics[width=0.26\textwidth]{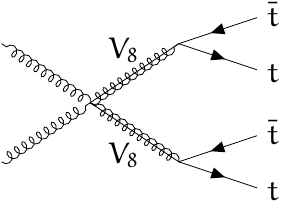}
\caption{%
    Example LO Feynman diagrams for \tttt production with resonant $\Seight\to\ttbar$ decay (left), \tttW production with resonant $\Vone\to\ttbar$ decay (center), and doubly resonant \tttt production as \Veight pair production with subsequent $\Veight\to\ttbar$ decays (right).
}
\label{fig:feynmanbsm}
\end{figure}

Six simplified models for topphilic heavy resonances are introduced in Ref.~\cite{Darme:2021gtt}: a color-singlet scalar boson \Sone, a color-singlet pseudoscalar boson \Pone, a color-singlet vector boson \Vone, a color-octet scalar boson \Seight, a color-octet pseudoscalar boson \Peight, and a color-octet vector boson \Veight.
Each simplified model has three free parameters: the mass \mX and width \GammaX of the new boson, and the strength of the coupling to the top quarks.
The interaction terms with top quarks are written as:
\begin{equation}\begin{aligned}
    \Lagrangian(\Sone)&=\ySone\PAQt\Sone\PQt, &
    \Lagrangian(\Pone)&=\iu\yPone\PAQt\gamma_5\Pone\PQt, &
    \Lagrangian(\Vone)&=\PAQt\gamma_\mu\big(\gVoneL P_{\mathrm{L}}+\gVoneR P_{\mathrm{R}}\big)\Vone^\mu\PQt, \\
    \Lagrangian(\Seight)&=\ySeight\PAQt T^a\Seight^a\PQt, &
    \Lagrangian(\Peight)&=\iu\yPeight\PAQt\gamma_5T^a\Peight^a\PQt, &
    \Lagrangian(\Veight)&=\PAQt\gamma_\mu\big(\gVeightL P_{\mathrm{L}}+\gVeightR P_{\mathrm{R}}\big)T^a\Veight^{a,\mu}\PQt,
\end{aligned}\end{equation}
where (different from the convention in Section~\ref{sec:smeft:theory}) \PQt and \PAQt are the top quark Dirac spinor and its adjoint, respectively.
The spin-0 bosons have Yukawa couplings \ySone, \yPone, \ySeight, and \yPeight, and we consider only real values.
For the spin-1 bosons, $P_{\mathrm{L}/\mathrm{R}}$ are the left- and right-handed projectors.
We consider only cases with equal left- and right-handed couplings and define $\gVone=\gVoneL=\gVoneR$ and $\gVeight=\gVeightL=\gVeightR$.
The masses of the new bosons are denoted as \mSone, \mPone, \mVone, \mSeight, \mPeight, and \mVeight, and we consider only masses larger than twice the top quark mass such that the new bosons decay to \ttbar with a branching fraction of 100\%.
In addition to interactions with top quarks, the color-octet bosons also obtain interaction diagrams with gluons resulting in the possibility of doubly resonant \tttt production.
Example Feynman diagrams for \tttt and \ttt production via these bosons are shown in Fig.~\ref{fig:feynmanbsm}.

We generate simulated event samples of \tttt, \tttW, and \tttq production in the topphilic heavy resonances scenario with \MGvATNLO at LO using the model provided by the authors of Ref.~\cite{Darme:2021gtt}.
In the ME calculation, we only include terms that are linear or quadratic in diagrams involving the new boson, \ie, excluding the pure SM contribution.
We can then add these samples to the SM samples for \tttt, \tttW, and \tttq production to obtain the prediction including all SM, interference, and pure BSM contributions.
Separate samples are generated for all the simplified models, for different fixed \mX and \GammaX values.
The \MGvATNLO reweighting technique~\cite{Mattelaer:2016gcx} is employed to evaluate per-event weights that allow for reweighting to arbitrary coupling values.
Our study focuses on narrow resonances with a total decay width smaller than the experimental resolution, for which we generate samples with $\GammaX=10\GeV$ fixed and $0.4<\mX<1.6\TeV$ in steps of 0.2\TeV.
To evaluate the choice of fixed \GammaX, we additionally generate samples with $40<\GammaX<500\GeV$ for $\mX=1\TeV$ fixed.

\begin{figure}[!ht]
\centering
\includegraphics[width=0.45\textwidth]{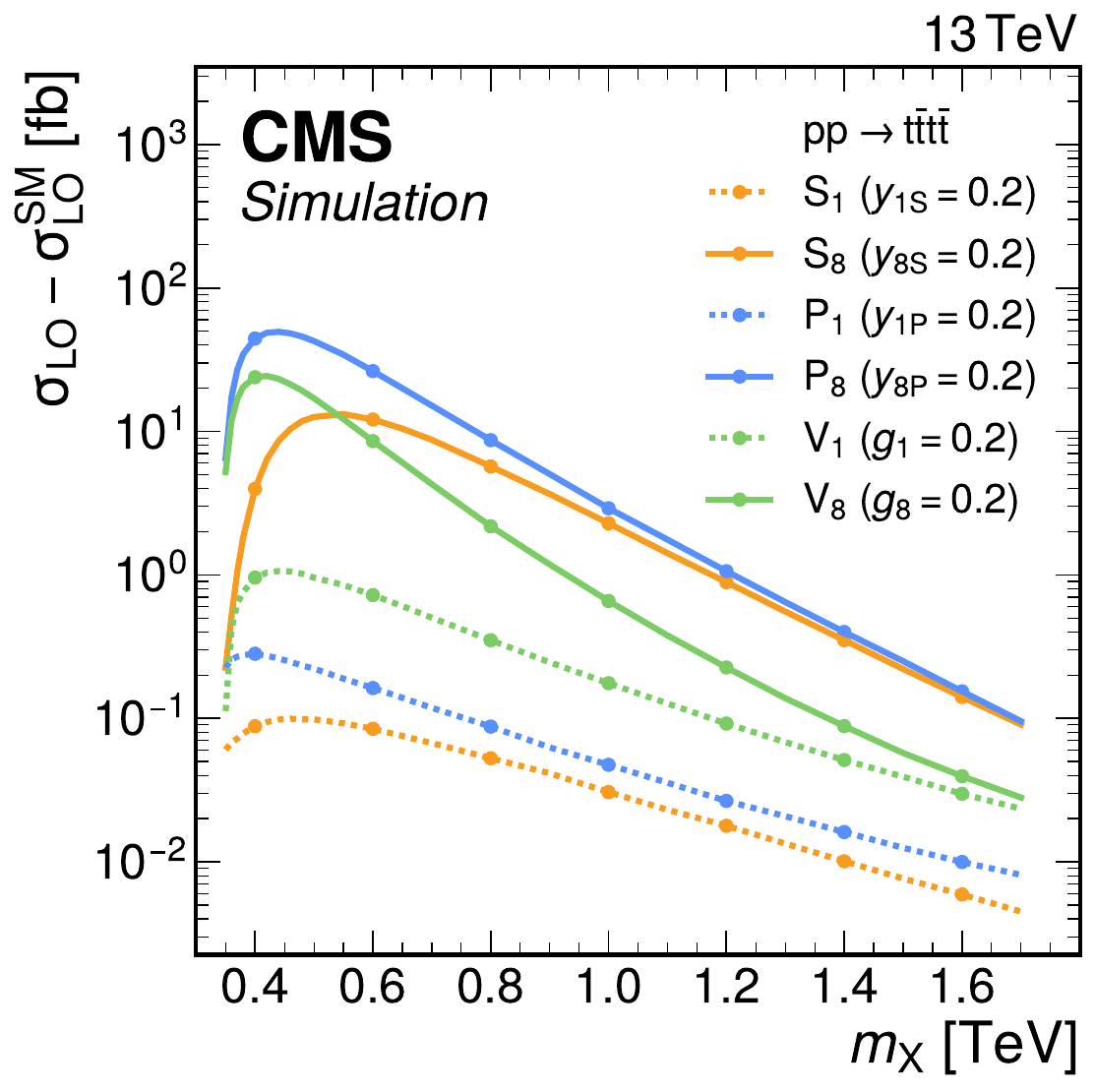}%
\hspace*{0.05\textwidth}%
\includegraphics[width=0.45\textwidth]{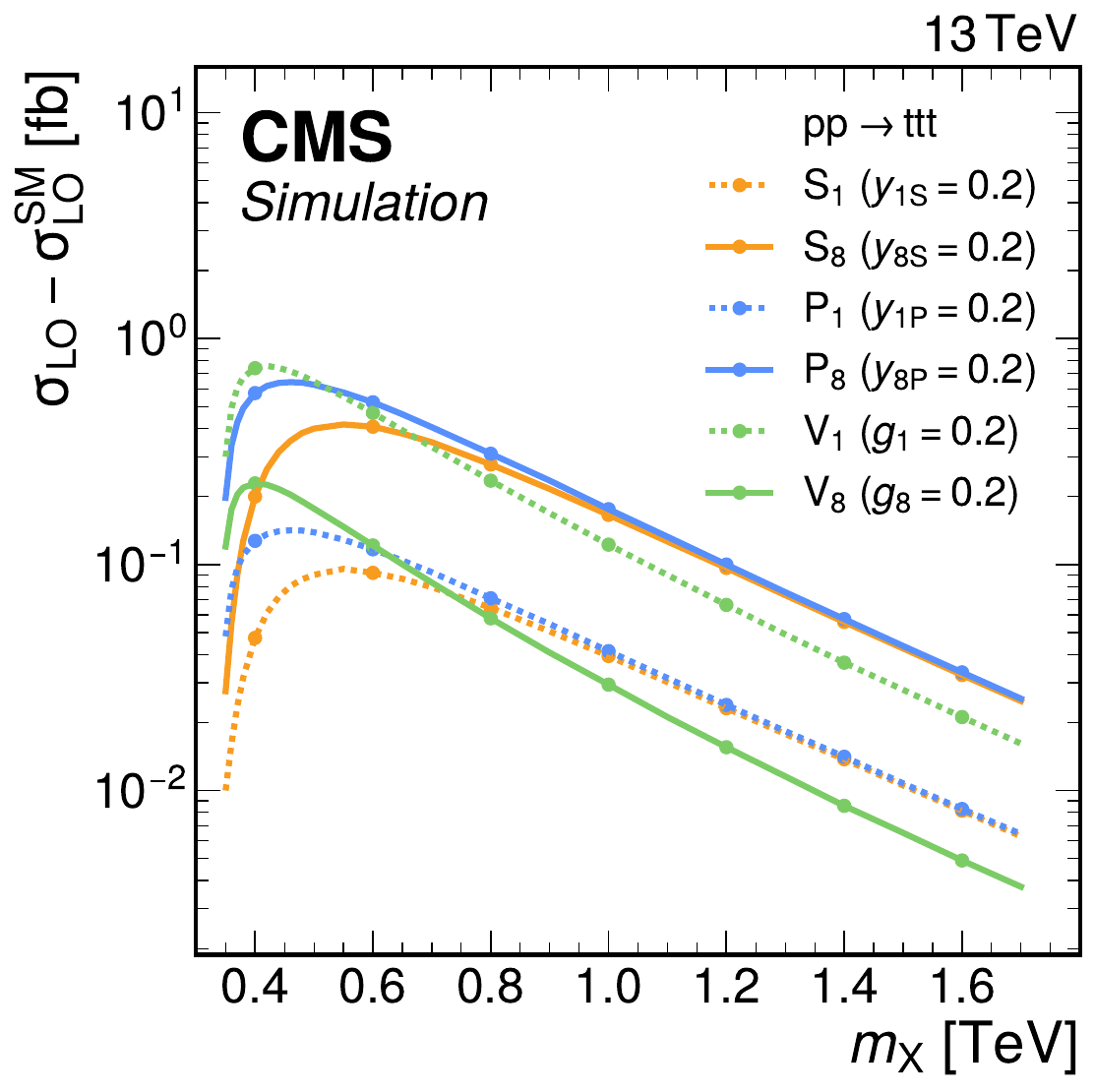}%
\hfill%
\caption{%
    Enhancement of the \tttt (left) and \ttt (right) production cross section in the different scenarios for a narrow topphilic heavy resonance with $\GammaX=10\GeV$ as a function of \mX, evaluated at LO as the difference between the cross section calculated with all SM, BSM, and interference contributions and the SM-only cross section.
    The coupling strength is fixed to a value of 0.2 in all scenarios.
    The points indicate the mass values at which we evaluate exclusion limits.
}
\label{fig:xsecbsm}
\end{figure}

The cross sections obtained for samples with $\GammaX=10\GeV$ for representative coupling strength values are shown in Fig.~\ref{fig:xsecbsm}.
For the singlet scenarios (dotted lines), both \tttt and \ttt production receive significant enhancement, with the \ttt contribution relative to the overall \ttttPLUSttt enhancement in the range 30--60\% for the probed \mX values.
For the octet scenarios (dashed lines), the \tttt enhancement is significantly larger than the \ttt enhancement, with a relative \ttt contribution in the range 1--20\%.
In all scenarios, the relative \ttt contribution increases with higher \mX values.

\subsection{Results}

\begin{figure}[!p]
\centering
\includegraphics[width=0.43\textwidth]{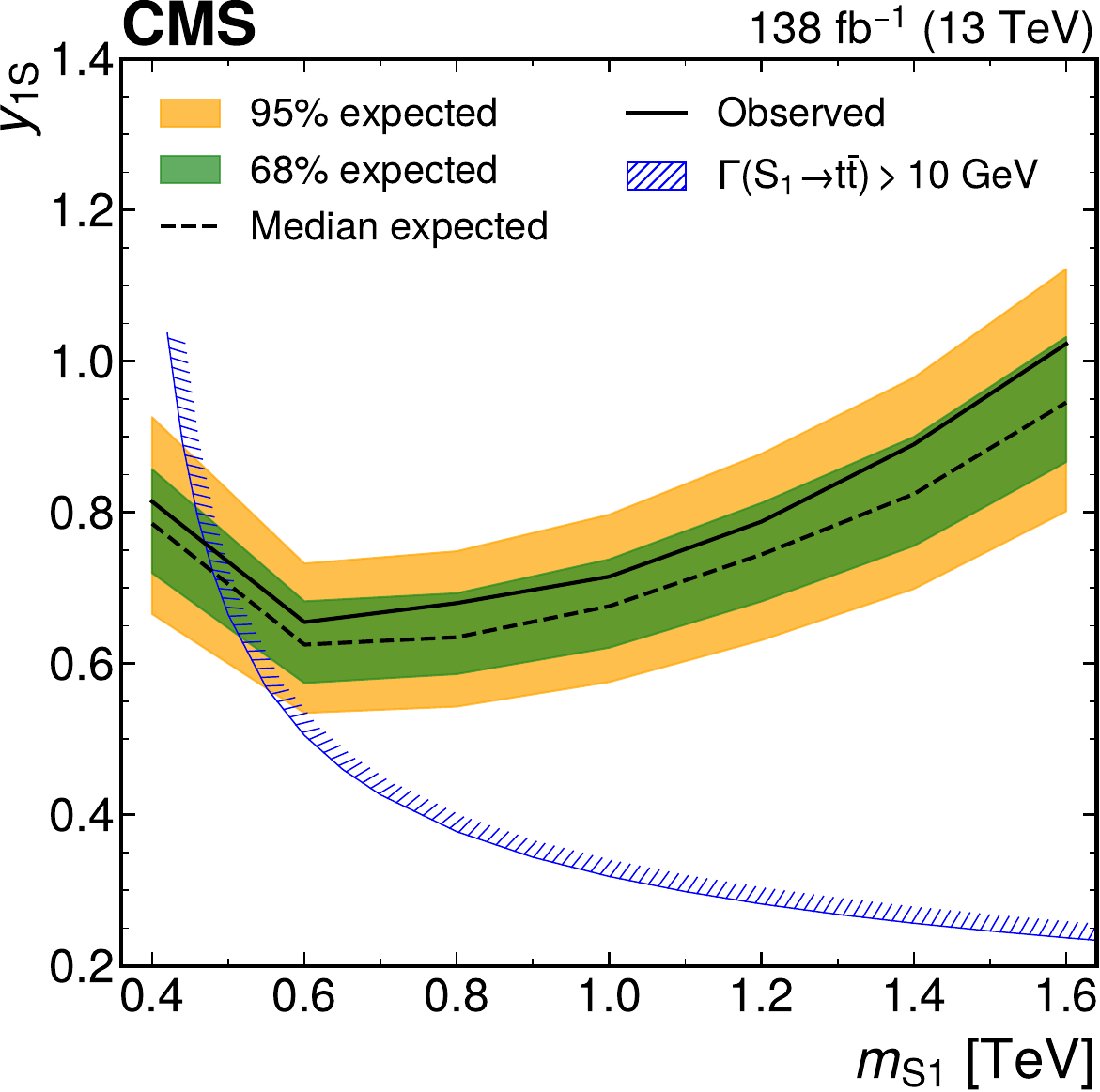}%
\hspace*{0.05\textwidth}%
\includegraphics[width=0.43\textwidth]{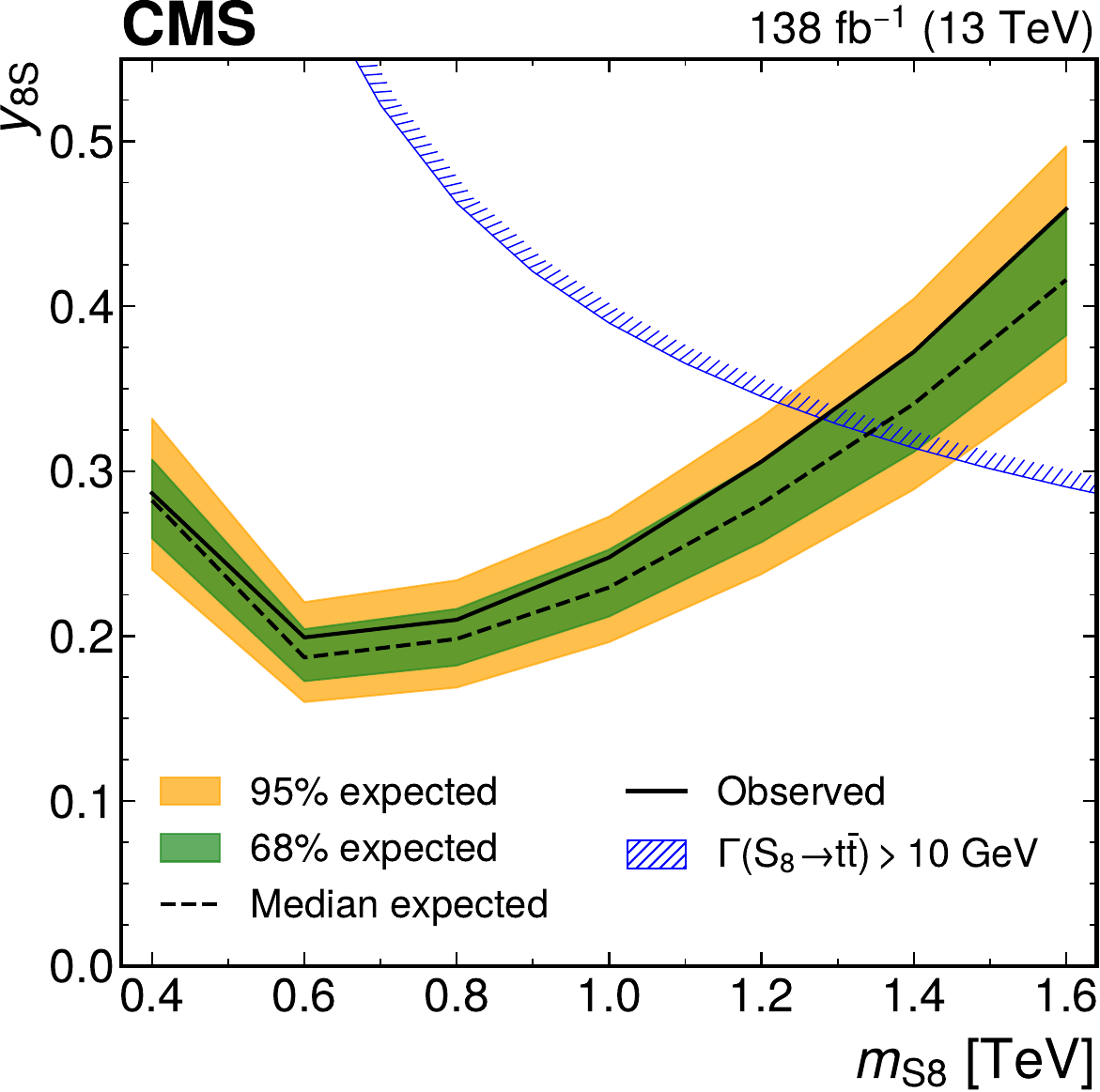} \\[1ex]
\includegraphics[width=0.43\textwidth]{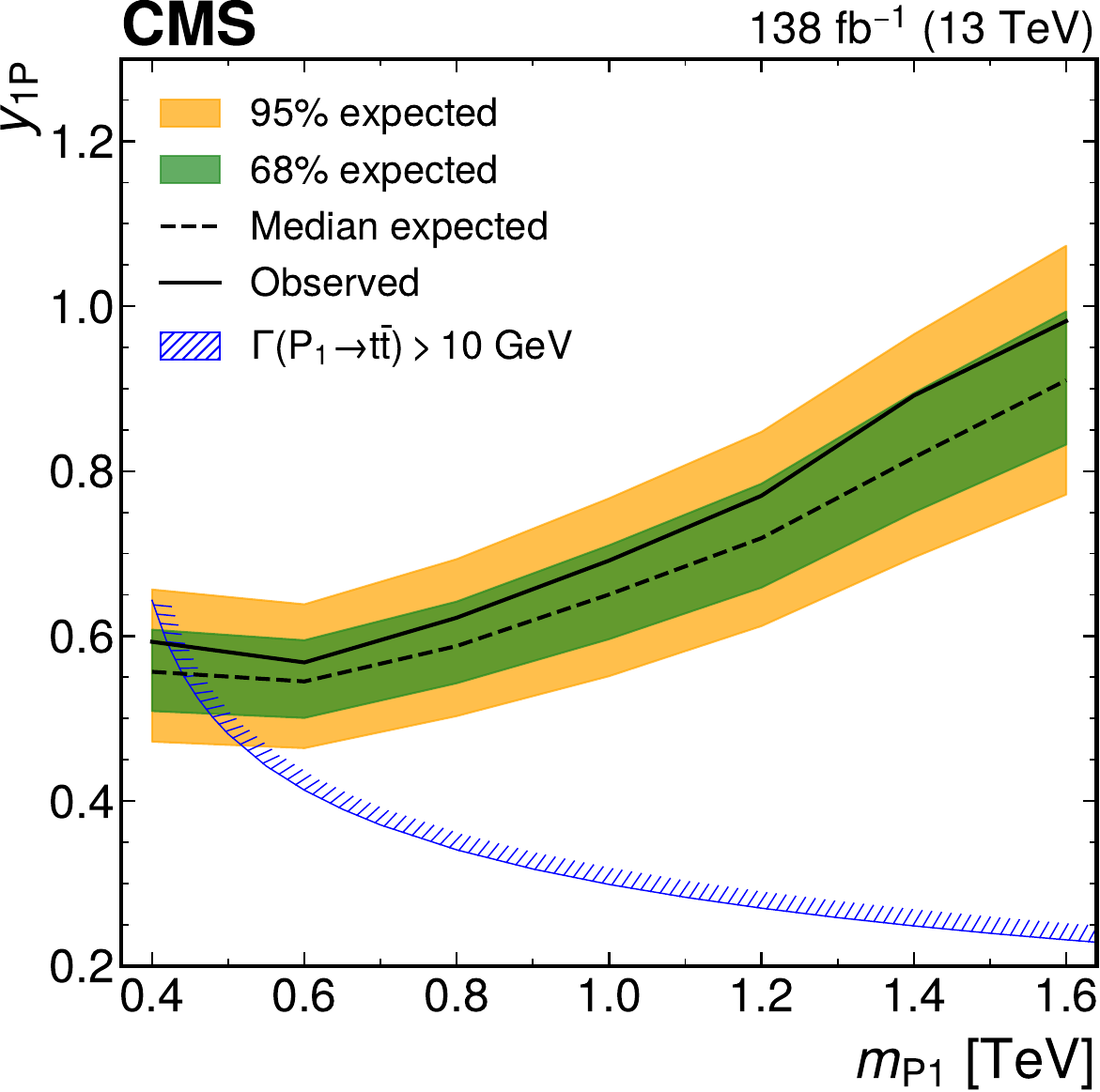}%
\hspace*{0.05\textwidth}%
\includegraphics[width=0.43\textwidth]{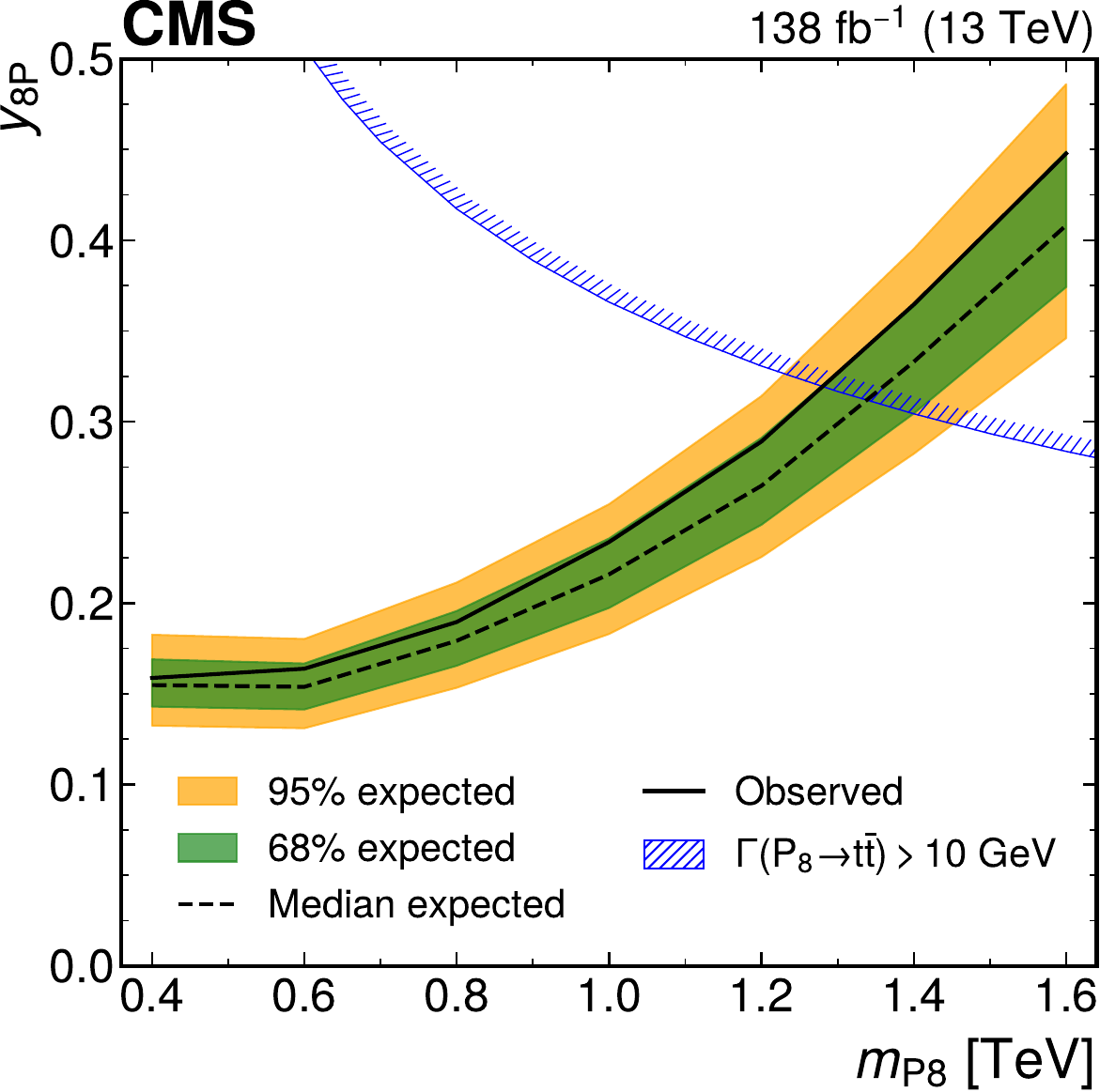} \\[1ex]
\includegraphics[width=0.43\textwidth]{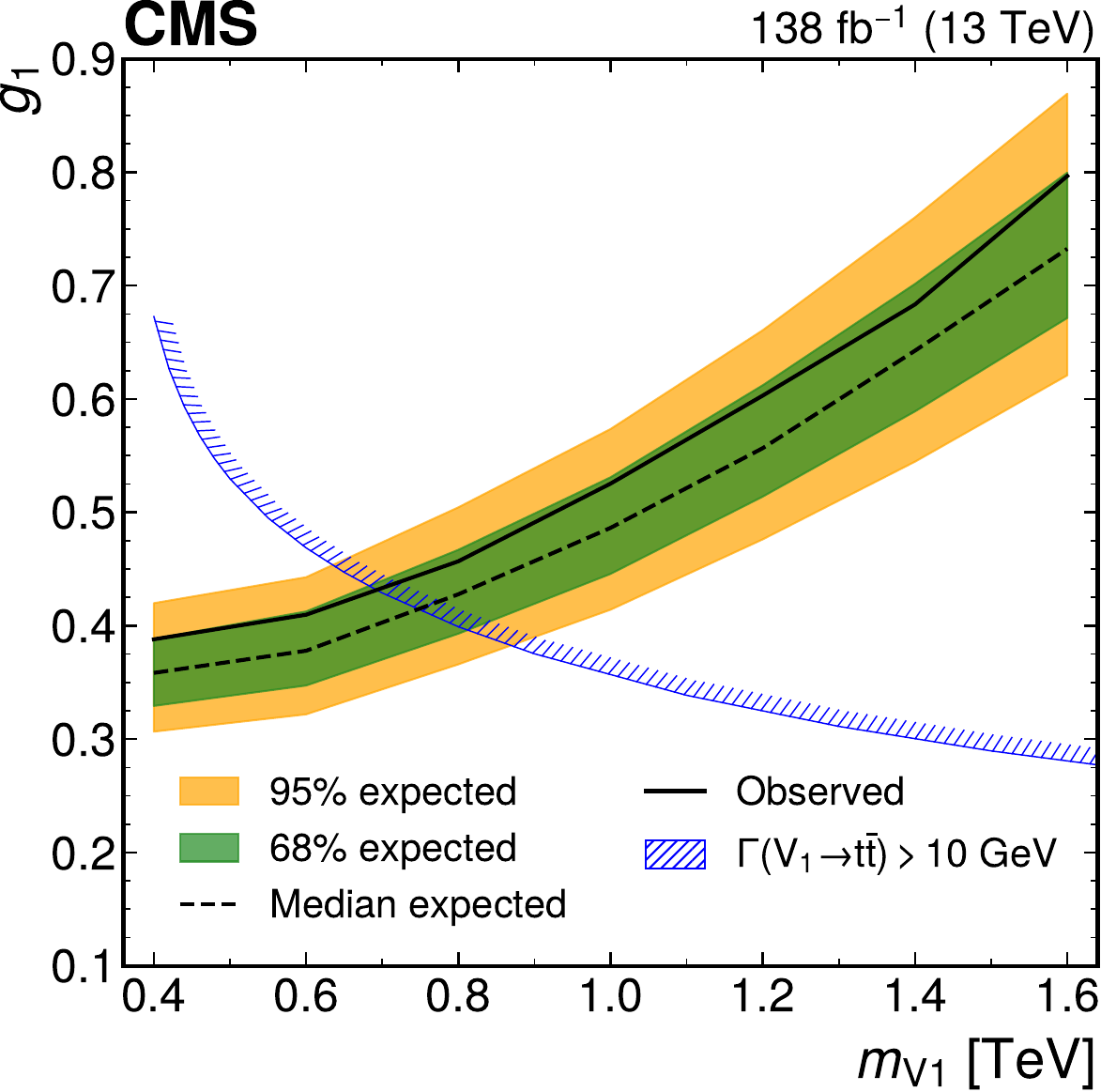}%
\hspace*{0.05\textwidth}%
\includegraphics[width=0.43\textwidth]{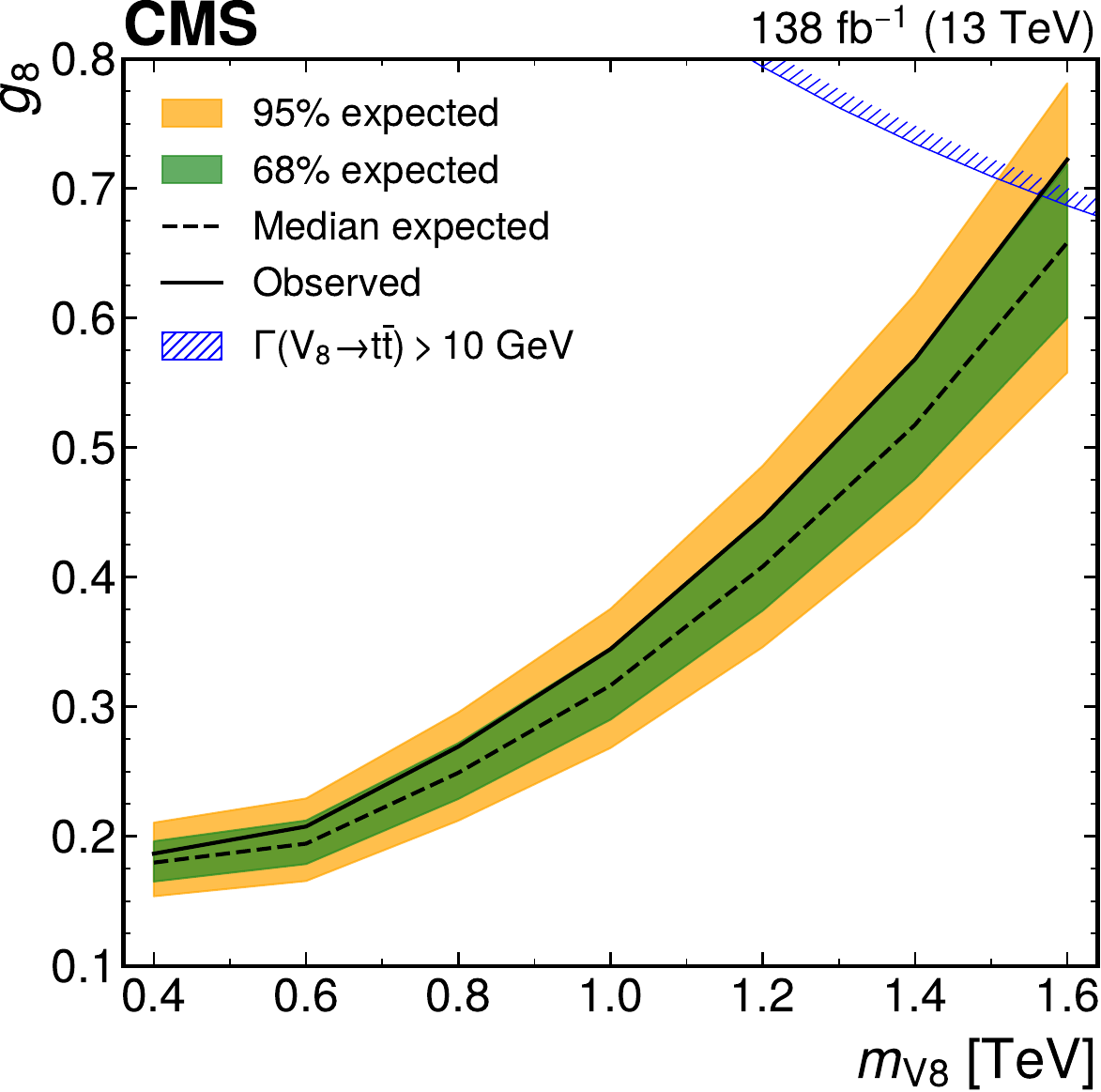}
\caption{%
    The 95\% \CL exclusion limits
    on \ySone as a function of \mSone (upper left),
    on \ySeight as a function of \mSeight (upper right),
    on \yPone as a function of \mPone (center left),
    on \yPeight as a function of \mPeight (center right),
    on \gVone as a function of \mVone (lower left),
    and
    on \gVeight as a function of \mVeight (lower right).
    The area above the solid (dashed) black line indicates the observed (expected) exclusion region.
    The total decay width is fixed to 10\GeV in all scenarios.
    The area above the hatched blue line indicates the nonphysical region of phase space in which the partial width $\Gamma(\PX\to\ttbar)$ becomes larger than 10\GeV.
}
\label{fig:bsm}
\end{figure}

For the evaluation of the simplified models of narrow topphilic heavy resonances, the \HT distribution is used in the fit for the \SRtwoTTTT and \SRthreeTTTT.
Fits are performed separately for the six scenarios, using a grid of mass values.
At each mass value, the normalization of the interference and pure BSM contributions are parameterized separately as a function of the coupling, such that the different scaling of both contributions with the coupling is taken into account.
In the relevant parameter ranges, the interference terms contribute less than 14\% of the combined interference and pure BSM yield.
The coupling is then used as parameter of interest in the fit, and the asymptotic approximation is applied to evaluate at 95\% \CL the largest coupling value at each mass point that is compatible with the data.
The resulting 95\% \CL exclusion limits as a function of the mass are shown in Fig.~\ref{fig:bsm}.
We have validated for several example mass points that the results from the asymptotic approximation agree with the evaluation using toys to better than 0.5\%.

The observed exclusion limits are less tight than the expected ones, consistent with the observation of a slight excess in the SM fit, but are consistent within one standard deviation.
For the scalar resonances, the lowest excluded coupling values are obtained for $\mX=600\GeV$, corresponding to the fact that the \ttttPLUSttt production cross section as shown in Fig.~\ref{fig:xsecbsm} is the largest.
For the pseudoscalar resonances, the lowest excluded coupling values for 400 and 600\GeV are similar, and the limits become less stringent at higher \mX.
For the vector resonances, the lowest excluded coupling values become strictly larger for larger \mX.
Considering that $\GammaX=10\GeV$ is used in the simulated signal samples, the excluded coupling values for the color-singlet models are mostly in a nonphysical region of phase space in which the partial width of the heavy resonance decay to \ttbar is larger than 10\GeV.
Physical scenarios are excluded only for masses below 600 (800)\GeV for the \Sone and \Pone (\Vone) models.
For the color-octet models, in contrast, the exclusion limits are tighter and well in the physical phase space for masses below 1.4 (1.6)\TeV with the \Seight and \Peight (\Veight) models.

The \ttttPLUSttt production cross sections corresponding to the largest excluded coupling values are between 25\unit{fb} at 400\GeV and 4.3\unit{fb} at 1.6\TeV above the SM prediction.
For $\mX=1\TeV$, we have also evaluated the exclusion limits for \GammaX values larger than 10\GeV.
Increasing \GammaX from 10 to 40\GeV, the observed and expected 95\% \CL exclusion limits on the production cross sections stay consistent within 2.5\% for the color-singlet bosons (\Sone, \Pone, and \Vone), but increase by about 25\% for the color-octet bosons (\Seight, \Peight, \Veight).
For even larger \GammaX values, the cross section limits increase roughly linearly, with steeper slopes for the color-octet than for the color-singlet bosons.
When considering the exclusion limits on the coupling values instead, there is a nonlinear dependence on \GammaX in all cases, resulting also from the fact that the predicted cross section depends on \GammaX even for a fixed coupling strength.

The previous \tttt search from the CMS Collaboration reported in Ref.~\cite{CMS:TOP-18-003}, based on the same final state and data set, also evaluated exclusion limits on scalar and pseudoscalar color-singlet bosons in the range $350<\mX<650\GeV$.
Comparing the exclusion limits on the \ttttPLUSttt production cross sections, our results are between 10 and 35\% stricter, following from the various improvements in experimental methods discussed in detail in Ref.~\cite{CMS:TOP-22-013}, as well as from employing an optimized fit observable.

\section{Extraction of the top quark Yukawa coupling}
\label{sec:yukawa}

\subsection{Theoretical framework}

For the extraction of the top quark Yukawa coupling and its \CP structure, we parameterize the \tH interaction as
\begin{equation}
    \Lagrangian_{\ttH}=-\frac{m_{\PQt}}{v}\PAQt\big(\yukeven+\iu\yukodd\gamma_5\big)\PQt\PH,
\end{equation}
where \PQt and \PAQt are the Dirac spinor and its adjoint (as in Section~\ref{sec:topphilic:theory}), $m_{\PQt}$ is the top quark mass, $v$ is the vacuum expectation value of the Higgs field, and \yukeven and \yukodd denote coupling strength modifiers to a purely \CP-even and a purely \CP-odd component, respectively~\cite{Gritsan:2016hjl}.
In the SM, $\yukeven=1$ and $\yukodd=0$.

\begin{figure}[!ht]
\centering
\includegraphics[width=0.26\textwidth]{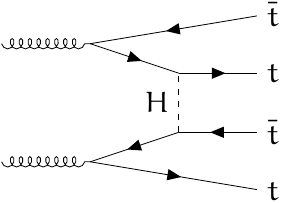}%
\hfill%
\includegraphics[width=0.26\textwidth]{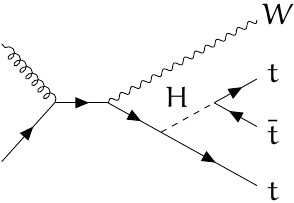}%
\hfill%
\includegraphics[width=0.26\textwidth]{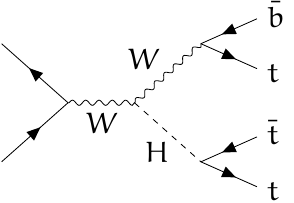}
\caption{%
    Example LO Feynman diagrams for \tttt (left), \tttW (center), and \tttq (right) production that contain the top quark Yukawa coupling.
}
\label{fig:feynmanyukawa}
\end{figure}

\begin{figure}[!ht]
\centering
\includegraphics[width=0.44\textwidth]{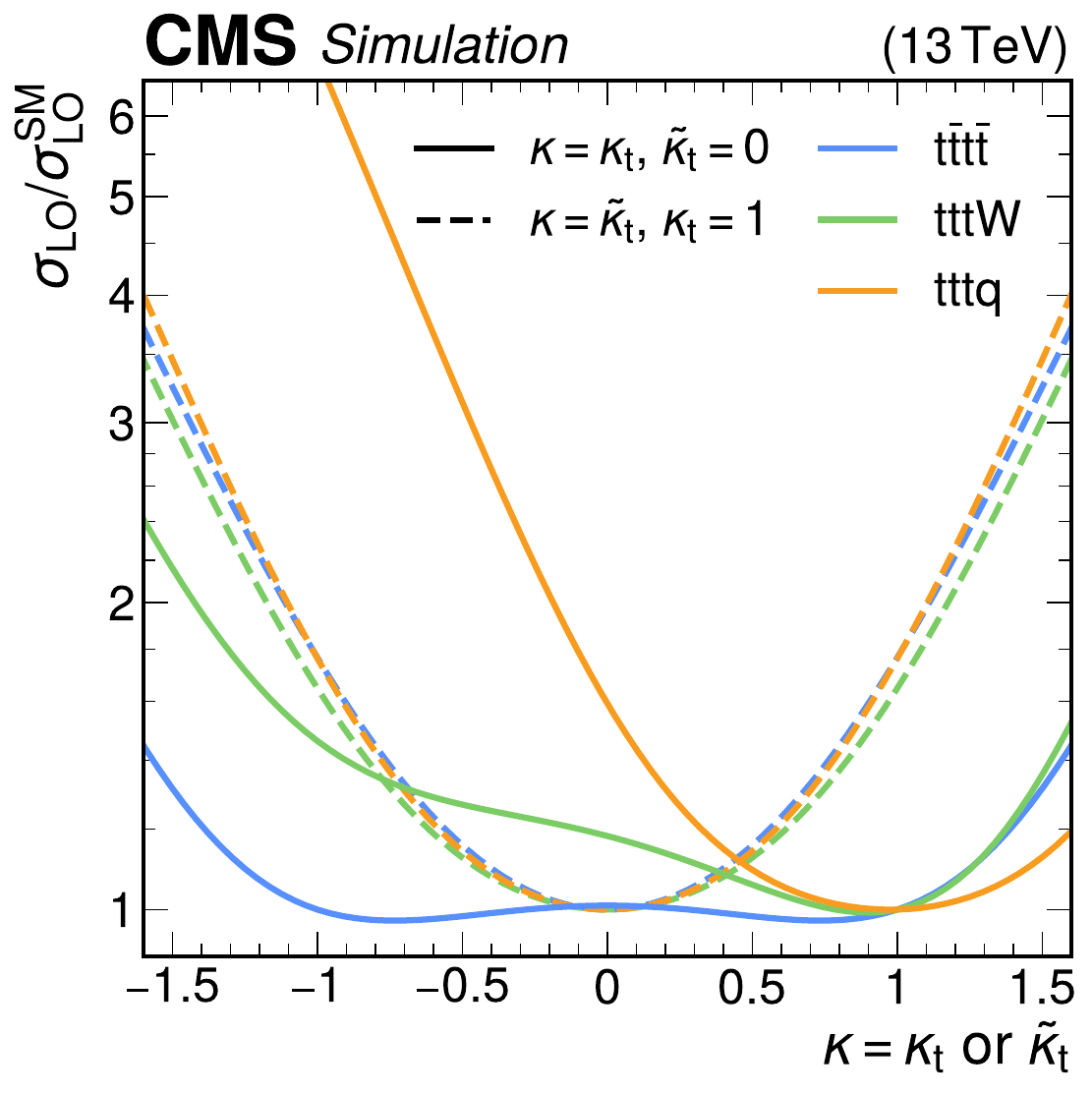}
\caption{%
    Ratio of the \tttt, \tttW, and \tttq cross sections with modified top quark Yukawa couplings to the SM values, evaluated at LO.
    The solid lines show modifications of \yukeven for a fixed value of $\yukodd=0$, and dashed lines modifications of \yukodd for fixed $\yukeven=1$.
}
\label{fig:yukawaparameterization}
\end{figure}

Top quark Yukawa coupling contributions to \tttt and \ttt production arise from diagrams with virtual Higgs boson contributions, with examples depicted in Fig.~\ref{fig:feynmanyukawa}.
We perform LO calculations of the \tttt, \tttW, and \tttq cross sections with \MGvATNLO, using the model from Refs.~\cite{Artoisenet:2013puc, Demartin:2014fia} to evaluate the cross sections with different \yukeven, \yukodd values.
We fit the inclusive cross section dependence on \yukeven, \yukodd with fourth-order polynomial functions, and obtain the following parameterizations:
\begin{equation}\label{eq:yukawa_tttt_ttt}\begin{aligned}
    \sigma_{\text{LO}}(\tttt)={}&\big(
        6.4
        -0.8\yukeven^2
        +1.9\yukodd^2
        +0.74\yukeven^4
        +1.8\yukeven^2\yukodd^2
        +1.2\yukodd^4
    \big)\unit{fb}, \\
    \sigma_{\text{LO}}(\tttW)={}&\big(
        1.14
        -0.2\yukeven
        -0.11\yukeven^2
        +0.07\yukodd^2
        -0.03\yukeven^3
        \\ &
        -0.02\yukeven\yukodd^2
        +0.16\yukeven^4
        +0.39\yukeven^2\yukodd^2
        +0.19\yukodd^4
    \big)\unit{fb}, \\
    \sigma_{\text{LO}}(\tttq)={}&\big(
        0.7
        -0.77\yukeven
        +0.89\yukeven^2
        +0.03\yukodd^2
        -0.514\yukeven^3
        \\ &
        -0.16\yukeven\yukodd^2
        +0.135\yukeven^4
        +0.34\yukeven^2\yukodd^2
        +0.1\yukodd^4
    \big)\unit{fb}.
\end{aligned}\end{equation}
All terms not listed above are found to be zero.
These parameterizations are illustrated in Fig.~\ref{fig:yukawaparameterization}.
While the \tttt cross section depends only quadratically on \yukeven and \yukodd, the \ttt cross sections include also terms with linear dependencies.
This follows from diagrams with only a single \tH interaction vertex, with an example shown in Fig.~\ref{fig:feynmanyukawa} (right).
In particular, the \tttq cross section shows a strong enhancement for negative values of \yukeven.

We further scale the \tttt cross section by a factor of 2.1 to match the NLO(QCD+EW)+\NLLpr prediction from Ref.~\cite{vanBeekveld:2022hty} at the SM point, and we shift the \tttW and \tttq cross sections by 0.36 and 0.26\unit{fb}, respectively, to account for the NLO QCD corrections from Ref.~\cite{Durieux:2023ttt}.
These matching prescriptions are pragmatic approximations, with the multiplicative prescription for \tttt production and the additive prescriptions for \tttW and \tttq production motivated by the different expected Yukawa coupling dependence of the higher-order contributions included in the SM predictions for these processes.

The $\Operator_{\tH}$ operator introduced in Section~\ref{sec:smeft:theory} describes equivalent modifications of the \tH interaction as the Yukawa coupling modifiers introduced here.
The conversion between the WCs and the modifiers is given by~\cite{Brod:2022bww}:
\begin{equation}\label{eq:wctoyukawa}
    \yukeven=1-\frac{v^2}{\Lambda^2}\ctHRe,
    \qquad
    \yukodd=-\frac{v^2}{\Lambda^2}\ctHIm.
\end{equation}
There is, however, a relevant difference in the approach of how the WCs and modifiers are used in the two interpretations: while \yukeven, \yukodd modify all \tH vertices at the same time, the dimension-6 SMEFT approach at first order in WCs as described by Eq.~\eqref{eq:matelementsmeft} only considers Feynman diagrams with up to one \tH vertex replaced by $\Operator_{\tH}$.
As a result, the cross section modification in the SMEFT interpretation is truncated at quadratic order in \ctHRe, \ctHIm, whereas Eq.~\eqref{eq:yukawa_tttt_ttt} includes quartic terms in \yukeven, \yukodd.
For small deviations from the SM case, the two approaches will provide approximately the same results.
For the example of $\yukeven=-1$, \ie, a Yukawa coupling with SM magnitude but opposite sign, the full calculation from Eq.~\eqref{eq:yukawa_tttt_ttt} yields the SM \tttt cross section, but the insertion of Eq.~\eqref{eq:wctoyukawa} and truncation at quadratic order in \ctHRe results in an LO \tttt cross section of 18.19\unit{fb}, almost three times as large.
Thus, the SMEFT limits on \ctHRe, \ctHIm will be significantly different from the \yukeven, \yukodd limits when considering scenarios very different from the SM.

\subsection{Results}

In the fit for the extraction of the Yukawa coupling modifiers \yukeven and \yukodd, the BDT score \tttt distribution is used as the fitted distribution for the \SRtwoTTTT and \SRthreeTTTT.
The cross section dependencies of \tttt and \ttt production are given by Eq.~\eqref{eq:yukawa_tttt_ttt}.
Additionally, the dependence of the \ttH background contribution is parameterized with a scale factor of $\yukodd^2+0.46\yukeven^2$, taken from Ref.~\cite{Cao:2019ygh} by fixing \GammaH to its SM value.
Toys are evaluated for the two-dimensional fit of \yukeven and \yukodd, such that both one- and two-dimensional exclusion limits at 68 and 95\% can be evaluated without relying on the asymptotic approximation.

\begin{figure}[!b]
\centering
\includegraphics[width=0.44\textwidth]{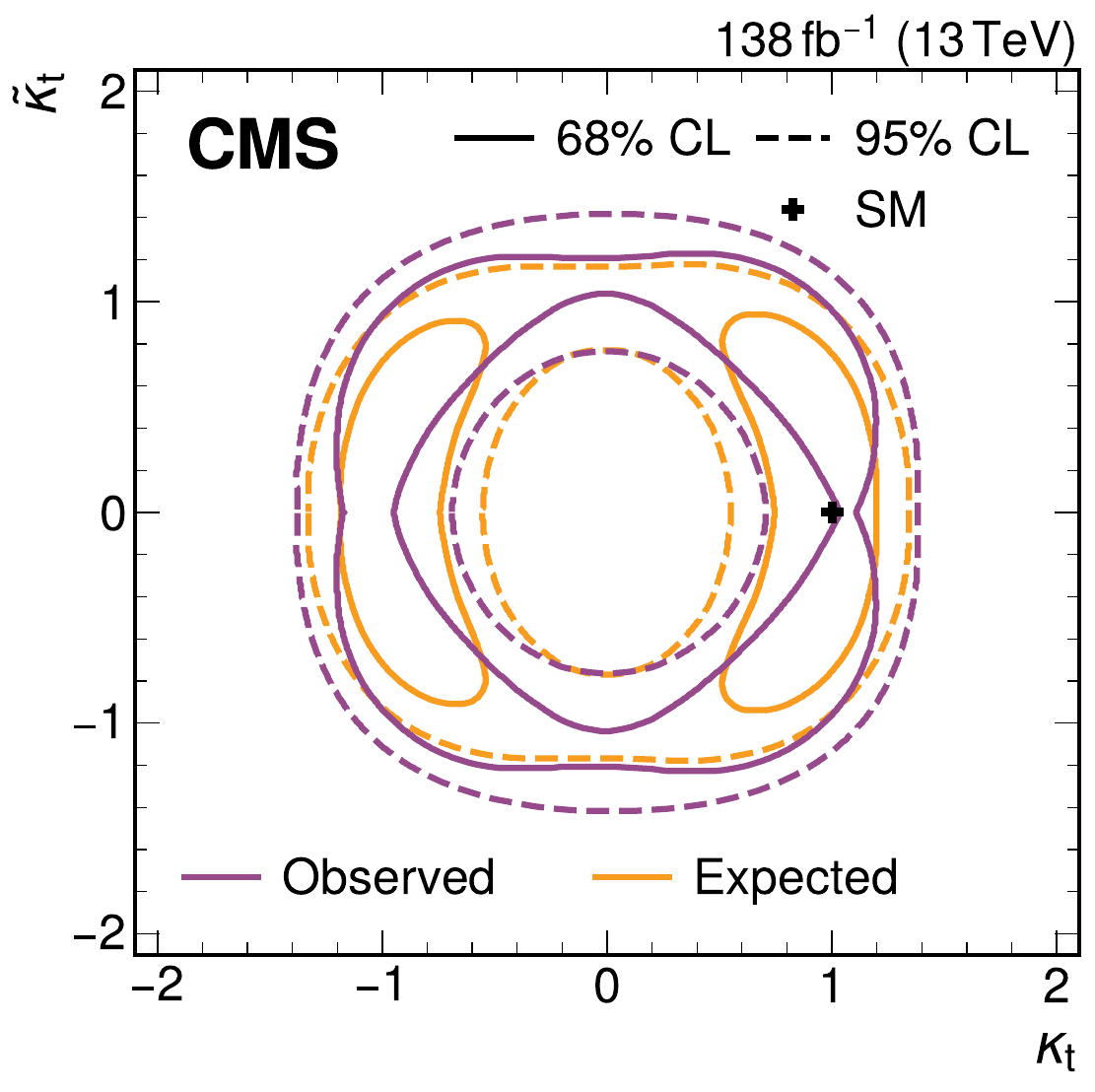}
\caption{%
    Expected (orange lines) and observed (purple lines) exclusion contours on the Yukawa coupling modifiers \yukeven and \yukodd corresponding to the 68\% (solid lines) and 95\% (dashed lines) \CL intervals as evaluated with toys.
    The SM prediction is shown with a black cross.
}
\label{fig:yukawa2d}
\end{figure}

The two-dimensional fit result is shown in Fig.~\ref{fig:yukawa2d}.
Consistent with the slight excess of data over the SM prediction in the SRs, the observed limits allow for an enhanced magnitude of the Yukawa coupling with any combination of signs for \yukeven and \yukodd allowed.
The two compatible best fit points are $\yukeven=+0.808$ and $\yukodd=\pm0.824$, slightly preferred over the configuration with a negative sign of \yukeven.
The agreement with the SM prediction of $(+1,0)$ is at the level of one standard deviation.

\begin{figure}[!thp]
\centering
\includegraphics[width=0.44\textwidth]{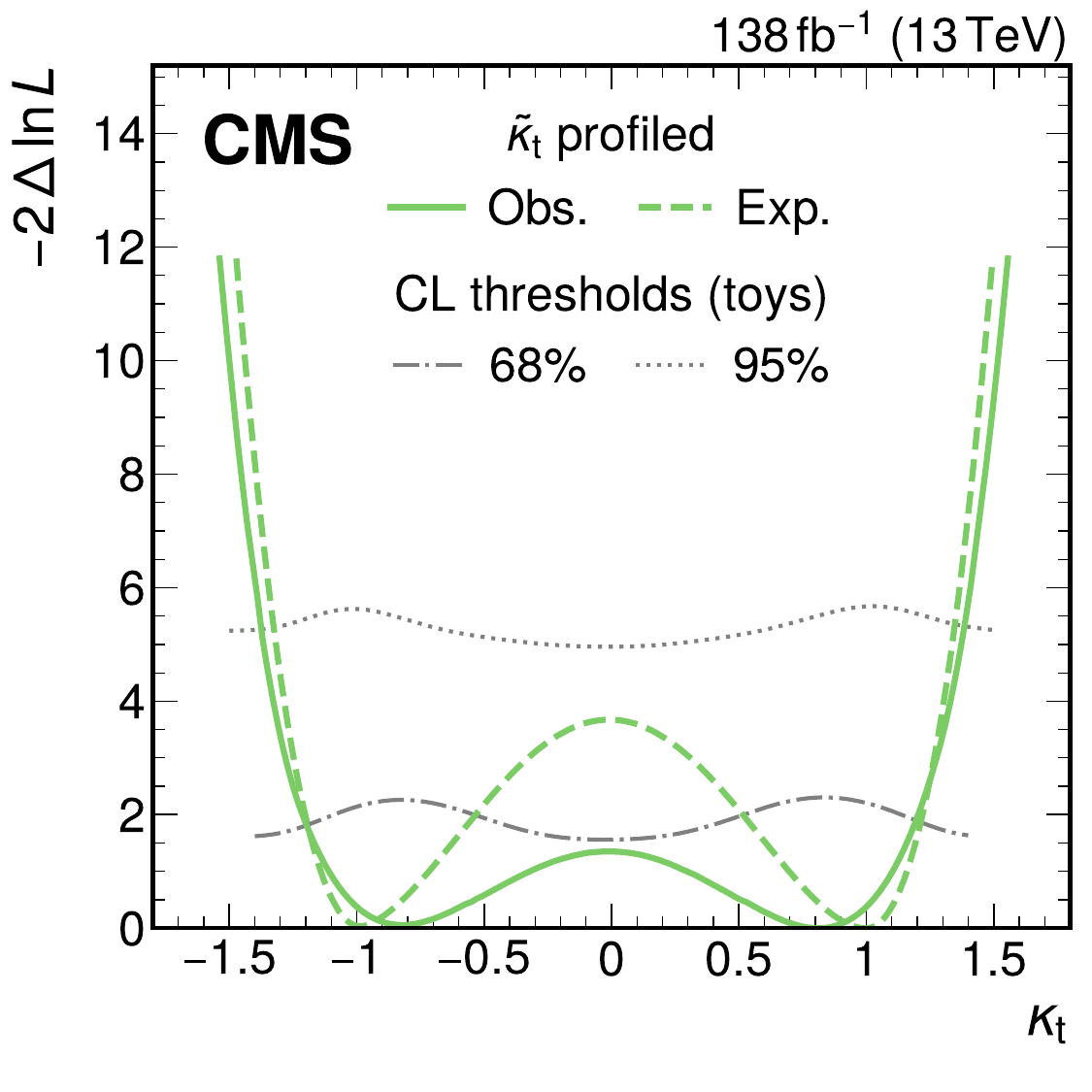}%
\hspace*{0.05\textwidth}%
\includegraphics[width=0.44\textwidth]{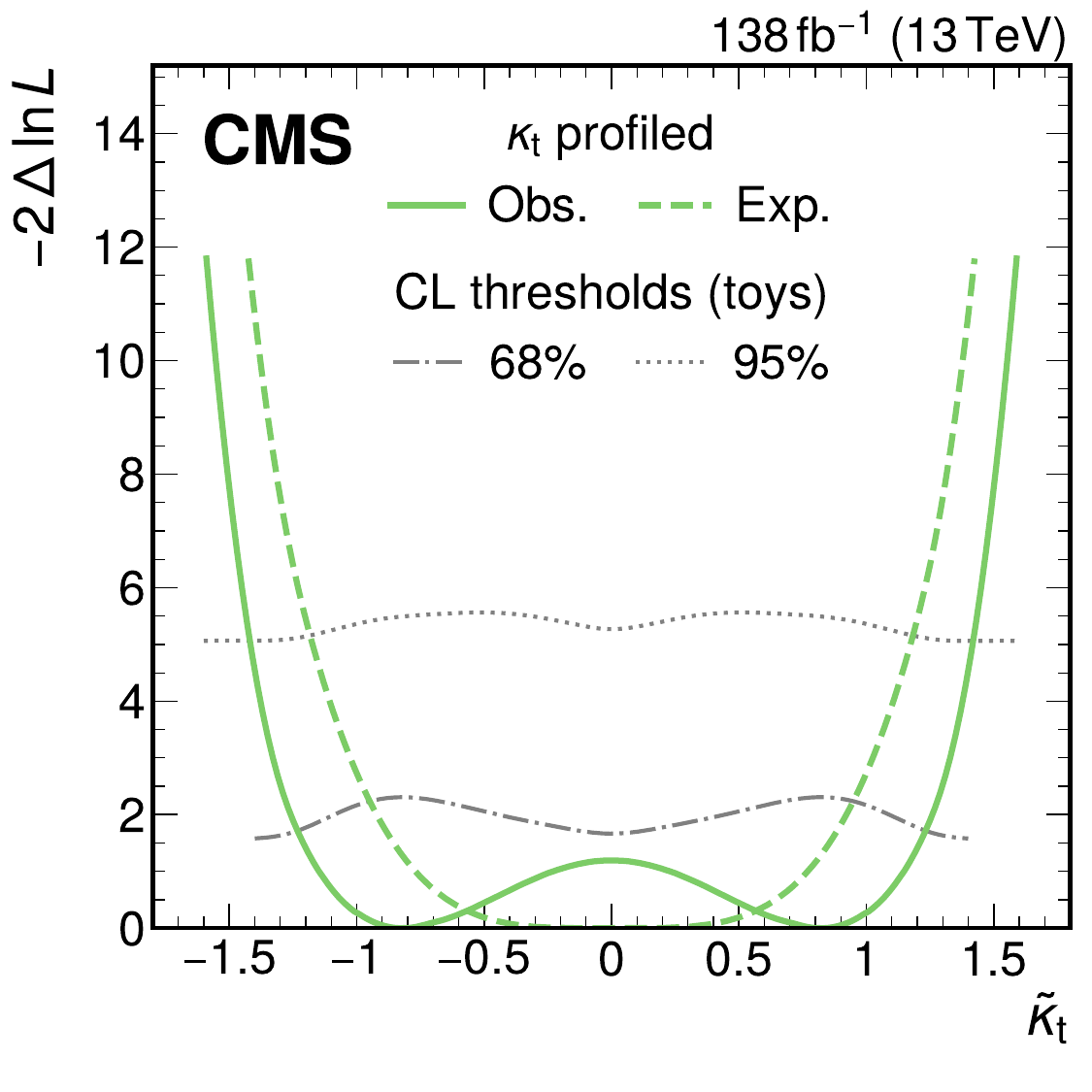} \\
\includegraphics[width=0.44\textwidth]{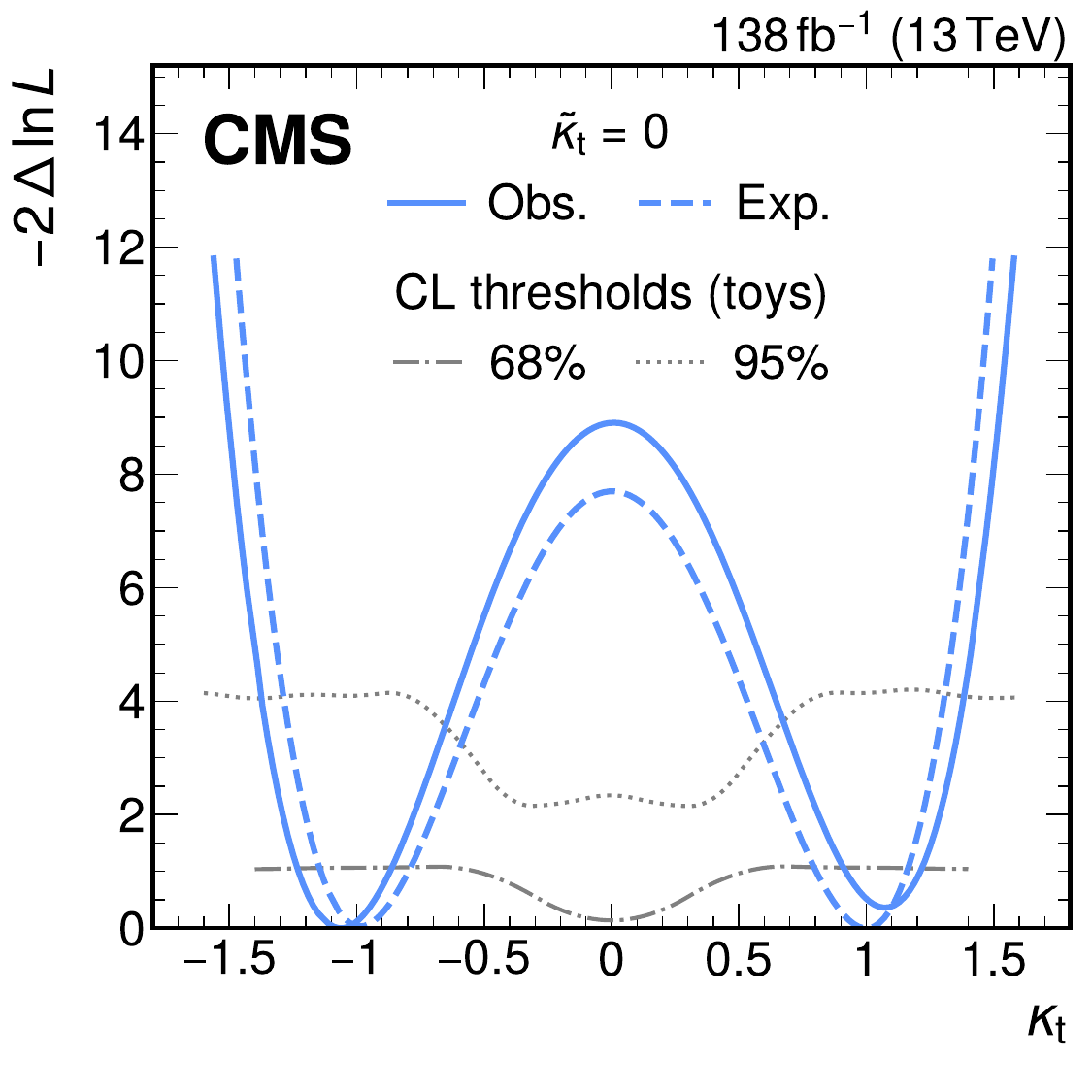}%
\hspace*{0.05\textwidth}%
\includegraphics[width=0.44\textwidth]{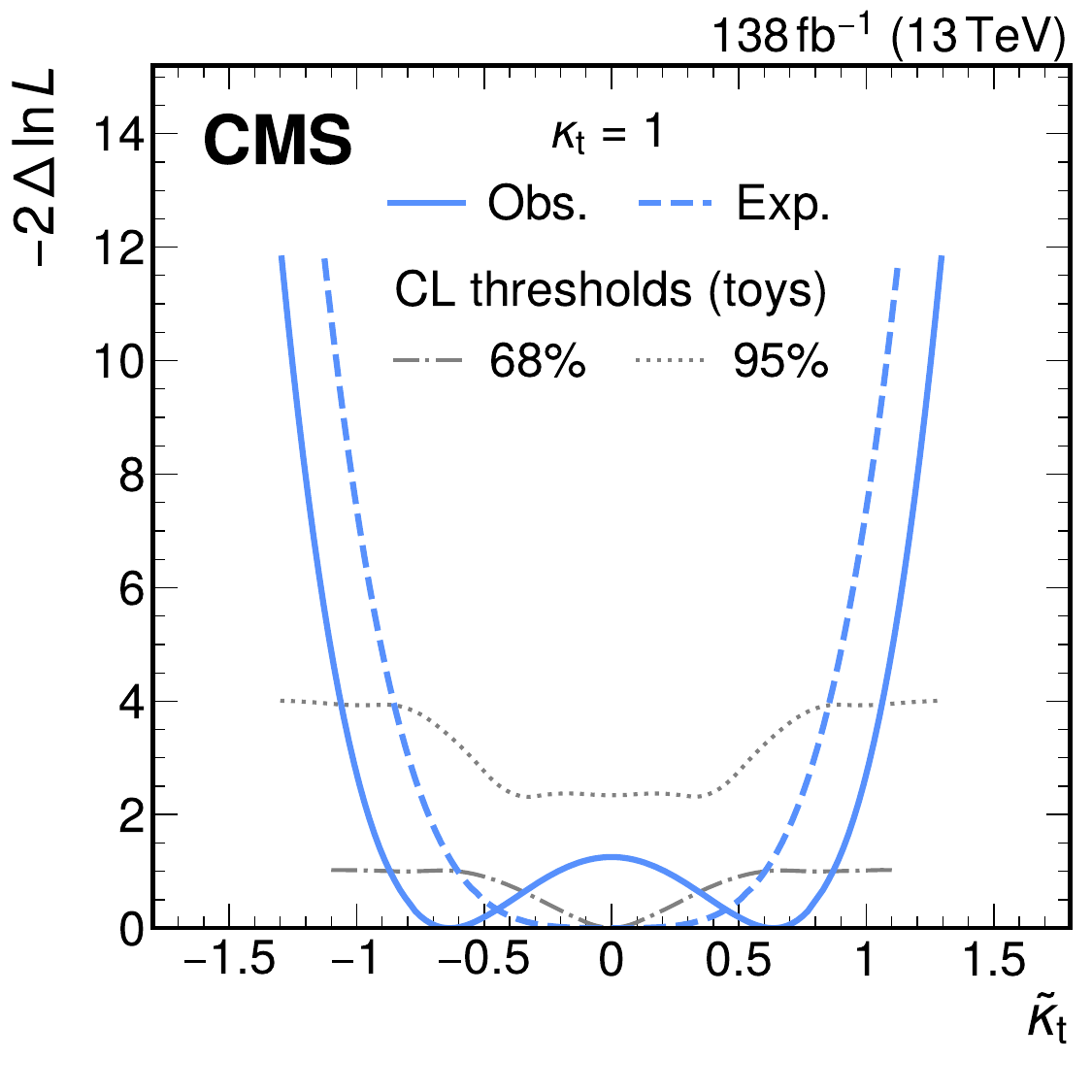}
\caption{%
    Negative log-likelihood difference from the best fit value for the one-dimensional scans of the Yukawa coupling modifiers \yukeven (left) and \yukodd (right), where the other modifier is profiled (upper) or fixed to its SM  prediction (lower).
    Shown are the expected (colored dashed line) and observed (colored solid line) results, as well as the threshold values for 68\% (gray dash-dotted line) and 95\% (gray dotted lines) \CL intervals as evaluated with toys.
}
\label{fig:yukawa1d}
\end{figure}

Two types of one-dimensional projections of the fit results are shown in Fig.~\ref{fig:yukawa1d}, where either the other modifier is profiled or fixed to its SM value.
For \yukeven with $\yukodd=0$ fixed, the observed (expected) 95\% \CL interval is $0.7<\abs{\yukeven}<1.4$ ($0.6<\abs{\yukeven}<1.3$), which is significantly tighter than previous constraints from \tttt production measurements~\cite{CMS:TOP-18-003, ATLAS:2023ajo}.
With fixed $\yukodd=0$, the best fit value for $\abs{\yukeven}$ is larger than 1.
In contrast, profiling over \yukodd results in values smaller than 1 being preferred, corresponding to the observation that the fit prefers nonzero values of \yukodd.
The asymmetry in the \yukeven scans is a result of the dependency of \ttt production on the sign of \yukeven.

The presented limits are derived under the assumption that \GammaH corresponds to its SM value.
Reducing (increasing) \GammaH would result in a larger (smaller) predicted \ttH contribution~\cite{Cao:2019ygh}, and thus change the results of the Yukawa coupling extraction.
We have evaluated this by increasing the \ttH production normalization uncertainty to 30\%, corresponding to experimental uncertainty in \GammaH~\cite{CMS:HIG-17-031}.
The resulting 68 (95)\% \CL ranges allow for up to 8 (13)\% higher values of \yukeven, with no noticeable impact on the allowed ranges for \yukodd.

\section{Summary}
\label{sec:summary}

A search for physics beyond the standard model (beyond the SM, BSM) using four and three top quark (\tttt and \ttt) production events has been reported.
The analyzed proton-proton collision data were recorded at 13\TeV with the CMS detector at the CERN LHC in 2016--2018 and correspond to an integrated luminosity of 138\fbinv.
Following the experimental analysis of Ref.~\cite{CMS:TOP-22-013}, events with two same-sign, three, or four leptons (electrons and/or muons) are selected and categorized in signal and control regions.
The signal regions in the two same-sign and three lepton channels are further split following a machine-learning discriminant trained to distinguish between \tttt production and the main background processes.
Assuming no BSM contributions, a mild excess of events in data in the selection enriched with \tttt and \ttt production events is observed at the level of one standard deviation, consistent with the result from Ref.~\cite{CMS:TOP-22-013}.

To interpret this observation in different models of BSM physics, three interpretations are performed using either the machine-learning discriminant or the scalar sum of the jet transverse momenta, optimized for the considered scenario.
Throughout, \ttt production is treated as signal process alongside \tttt production, accounting for the observation that the existing experimental analysis is not able to distinguish between \tttt and \ttt contributions in the most sensitive signal regions.
Using the SM effective field theory framework, constraints are derived on six Wilson coefficients that modify interactions between four third-generation quarks or between top quarks and the Higgs boson.
This is the first SM effective field theory interpretation that considers these operators simultaneously.
Exclusion limits are set on topphilic heavy resonances of different spin and color states, covering masses between 400\GeV and 1.6\TeV.
The top quark Yukawa coupling is extracted, considering both \CP-even and \CP-odd contributions.
All interpretations provide a good description of the data and are statistically compatible with the SM expectation.

\begin{acknowledgments}
We congratulate our colleagues in the CERN accelerator departments for the excellent performance of the LHC and thank the technical and administrative staffs at CERN and at other CMS institutes for their contributions to the success of the CMS effort. In addition, we gratefully acknowledge the computing centers and personnel of the Worldwide LHC Computing Grid and other centers for delivering so effectively the computing infrastructure essential to our analyses. Finally, we acknowledge the enduring support for the construction and operation of the LHC, the CMS detector, and the supporting computing infrastructure provided by the following funding agencies: SC (Armenia), BMFWF and FWF (Austria); FNRS and FWO (Belgium); CNPq, CAPES, FAPERJ, FAPERGS, and FAPESP (Brazil); MES and BNSF (Bulgaria); CERN; CAS, MoST, and NSFC (China); MINCIENCIAS (Colombia); MSES and CSF (Croatia); RIF (Cyprus); SENESCYT (Ecuador); ERC PRG and PSG, TARISTU24-TK10 and MoER TK202 (Estonia); Academy of Finland, MEC, and HIP (Finland); CEA and CNRS/IN2P3 (France); SRNSF (Georgia); BMFTR, DFG, and HGF (Germany); GSRI (Greece); MATE and NKFIH (Hungary); DAE and DST (India); IPM (Iran); SFI (Ireland); INFN (Italy); MSIT and NRF (Republic of Korea); MES (Latvia); LMTLT (Lithuania); MOE and UM (Malaysia); BUAP, CINVESTAV, CONACYT, LNS, SEP, and UASLP-FAI (Mexico); MOS (Montenegro); MBIE (New Zealand); PAEC (Pakistan); MSHE, NSC, and NAWA (Poland); FCT (Portugal); MESTD (Serbia); MICIU/AEI and PCTI (Spain); MOSTR (Sri Lanka); Swiss Funding Agencies (Switzerland); MST (Taipei); MHESI (Thailand); TUBITAK and TENMAK (T\"{u}rkiye); NASU (Ukraine); STFC (United Kingdom); DOE and NSF (USA).

\begin{sloppypar}
\setlength\emergencystretch{\hsize}
\hyphenation{Rachada-pisek} Individuals have received support from the Marie-Curie program and the European Research Council and Horizon 2020 Grant, contract Nos.\ 675440, 724704, 752730, 758316, 765710, 824093, 101115353, 101002207, 101001205, and COST Action CA16108 (European Union); the Leventis Foundation; the Alfred P.\ Sloan Foundation; the Alexander von Humboldt Foundation; the Science Committee, project no. 22rl-037 (Armenia); the Fonds pour la Formation \`a la Recherche dans l'Industrie et dans l'Agriculture (FRIA) and Fonds voor Wetenschappelijk Onderzoek contract No. 1228724N (Belgium); the Beijing Municipal Science \& Technology Commission, No. Z191100007219010, the Fundamental Research Funds for the Central Universities, the Ministry of Science and Technology of China under Grant No. 2023YFA1605804, the Natural Science Foundation of China under Grant No. 12535004, and USTC Research Funds of the Double First-Class Initiative No.\ YD2030002017 (China); the Ministry of Education, Youth and Sports (MEYS) of the Czech Republic; the Shota Rustaveli National Science Foundation (Georgia); the Deutsche Forschungsgemeinschaft (DFG), among others, under Germany's Excellence Strategy -- EXC 2121 ``Quantum Universe" -- 390833306, and under project number 400140256 - GRK2497; the Hellenic Foundation for Research and Innovation (HFRI), Project Number 2288 (Greece); the Hungarian Academy of Sciences, the New National Excellence Program - \'UNKP, the NKFIH research grants K 131991, K 138136, K 143460, K 143477, K 147557, K 146913, K 146914, K 147048, TKP2021-NKTA-64, and 2025-1.1.5-NEMZ\_KI-2025-00004, and MATE KKP and KKPCs Research Excellence and Flagship Research Groups grants (Hungary); the Council of Science and Industrial Research, India; ICSC -- National Research Center for High Performance Computing, Big Data and Quantum Computing, FAIR -- Future Artificial Intelligence Research, and CUP I53D23001070006 (Mission 4 Component 1), funded by the NextGenerationEU program, the Italian Ministry of University and Research (MUR) under Bando PRIN 2022 -- CUP I53C24002390006, PRIN PRIMULA 2022RBYK7T (Italy); the Latvian Council of Science; the Ministry of Science and Higher Education, project no. 2022/WK/14, and the National Science Center, contracts Opus 2021/41/B/ST2/01369, 2021/43/B/ST2/01552, 2023/49/B/ST2/03273, and the NAWA contract BPN/PPO/2021/1/00011 (Poland); the Funda\c{c}\~ao para a Ci\^encia e a Tecnologia (Portugal); the National Priorities Research Program by Qatar National Research Fund; MICIU/AEI/10.13039/501100011033, ERDF/EU, ``European Union NextGenerationEU/PRTR", projects PID2022-142604OB-C21, PID2022-139519OB-C21, PID2023-147706NB-I00, PID2023-148896NB-I00, PID2023-146983NB-I00, PID2023-147115NB-I00, PID2023-148418NB-C41, PID2023-148418NB-C42, PID2023-148418NB-C43, PID2023-148418NB-C44, PID2024-158190NB-C22, RYC2021-033305-I, RYC2024-048719-I, CNS2023-144781, CNS2024-154769 and Plan de Ciencia, Tecnolog{\'i}a e Innovaci{\'o}n de Asturias, Spain; the Chulalongkorn Academic into Its 2nd Century Project Advancement Project, the National Science, Research and Innovation Fund program IND\_FF\_68\_369\_2300\_097, and the Program Management Unit for Human Resources \& Institutional Development, Research and Innovation, grant B39G680009 (Thailand); the Eric \& Wendy Schmidt Fund for Strategic Innovation through the CERN Next Generation Triggers project under grant agreement number SIF-2023-004; the Kavli Foundation; the Nvidia Corporation; the SuperMicro Corporation; the Welch Foundation, contract C-1845; and the Weston Havens Foundation (USA).
\end{sloppypar}
\end{acknowledgments}\section*{Data availability} Release and preservation of data used by the CMS Collaboration as the basis for publications is guided by the  \href{https://doi.org/10.7483/OPENDATA.CMS.1BNU.8V1W}{CMS data preservation, re-use and open access policy}.

\bibliography{auto_generated}

@ARTICLE{Nilles:1983ge,
	AUTHOR=	"Nilles, H.",
	TITLE=	"Supersymmetry, supergravity and particle physics",
	DOI=	"10.1016/0370-1573(84)90008-5",
	JOURNAL=	"Phys. Rept.",
	VOLUME=	"110",
	PAGES=	"1",
	YEAR=	"1984",
}

@ARTICLE{Farrar:1978xj,
	AUTHOR=	"Farrar, G. and Fayet, P.",
	TITLE=	"Phenomenology of the production, decay, and detection of new hadronic states associated with supersymmetry",
	DOI=	"10.1016/0370-2693(78)90858-4",
	JOURNAL=	"Phys. Lett. B",
	VOLUME=	"76",
	PAGES=	"575",
	YEAR=	"1978",
}

@ARTICLE{Toharia:2005gm,
	AUTHOR=	"Toharia, M. and Wells, J.",
	TITLE=	"Gluino decays with heavier scalar superpartners",
	EPRINT=	"hep-ph/0503175",
	ARCHIVEPREFIX=	"arXiv",
	DOI=	"10.1088/1126-6708/2006/02/015",
	JOURNAL=	"JHEP",
	VOLUME=	"02",
	PAGES=	"015",
	YEAR=	"2006",
}

@ARTICLE{Plehn:2008ae,
	AUTHOR=	"Plehn, T. and Tait, T.",
	TITLE=	"Seeking sgluons",
	EPRINT=	"0810.3919",
	ARCHIVEPREFIX=	"arXiv",
	PRIMARYCLASS=	"hep-ph",
	DOI=	"10.1088/0954-3899/36/7/075001",
	JOURNAL=	"J. Phys. G",
	VOLUME=	"36",
	PAGES=	"075001",
	YEAR=	"2009",
}

@ARTICLE{Calvet:2012rk,
	AUTHOR=	"Calvet, S. and Fuks, B. and Gris, P. and Valery, L.",
	TITLE=	"Searching for sgluons in multitop events at a center-of-mass energy of {8\TeV}",
	EPRINT=	"1212.3360",
	ARCHIVEPREFIX=	"arXiv",
	PRIMARYCLASS=	"hep-ph",
	DOI=	"10.1007/JHEP04(2013)043",
	JOURNAL=	"JHEP",
	VOLUME=	"04",
	PAGES=	"043",
	YEAR=	"2013",
}

@ARTICLE{Beck:2015cga,
	AUTHOR=	"Beck, L. and Blekman, F. and Dobur, D. and Fuks, B. and Keaveney, J. and Mawatari, K.",
	TITLE=	"Probing top-philic sgluons with {LHC} \mbox{Run 1} data",
	EPRINT=	"1501.07580",
	ARCHIVEPREFIX=	"arXiv",
	PRIMARYCLASS=	"hep-ph",
	DOI=	"10.1016/j.physletb.2015.04.043",
	JOURNAL=	"Phys. Lett. B",
	VOLUME=	"746",
	PAGES=	"48",
	YEAR=	"2015",
}

@ARTICLE{Darme:2018dvz,
	AUTHOR=	"Darm{\'e}, L. and Fuks, B. and Goodsell, M.",
	TITLE=	"Cornering sgluons with four-top-quark events",
	EPRINT=	"1805.10835",
	ARCHIVEPREFIX=	"arXiv",
	PRIMARYCLASS=	"hep-ph",
	DOI=	"10.1016/j.physletb.2018.08.001",
	JOURNAL=	"Phys. Lett. B",
	VOLUME=	"784",
	PAGES=	"223",
	YEAR=	"2018",
}

@ARTICLE{Lillie:2007hd,
	AUTHOR=	"Lillie, B. and Shu, J. and Tait, T. M. P.",
	TITLE=	"Top compositeness at the {Tevatron} and {LHC}",
	EPRINT=	"0712.3057",
	ARCHIVEPREFIX=	"arXiv",
	PRIMARYCLASS=	"hep-ph",
	DOI=	"10.1088/1126-6708/2008/04/087",
	JOURNAL=	"JHEP",
	VOLUME=	"04",
	PAGES=	"087",
	YEAR=	"2008",
}

@ARTICLE{Pomarol:2008bh,
	AUTHOR=	"Pomarol, A. and Serra, J.",
	TITLE=	"Top quark compositeness: Feasibility and implications",
	EPRINT=	"0806.3247",
	ARCHIVEPREFIX=	"arXiv",
	PRIMARYCLASS=	"hep-ph",
	DOI=	"10.1103/PhysRevD.78.074026",
	JOURNAL=	"Phys. Rev. D",
	VOLUME=	"78",
	PAGES=	"074026",
	YEAR=	"2008",
}

@ARTICLE{Kumar:2009vs,
	AUTHOR=	"Kumar, K. and Tait, T. M. P. and Vega-Morales, R.",
	TITLE=	"Manifestations of top compositeness at colliders",
	EPRINT=	"0901.3808",
	ARCHIVEPREFIX=	"arXiv",
	PRIMARYCLASS=	"hep-ph",
	DOI=	"10.1088/1126-6708/2009/05/022",
	JOURNAL=	"JHEP",
	VOLUME=	"05",
	PAGES=	"022",
	YEAR=	"2009",
}

@ARTICLE{Cacciapaglia:2015eqa,
	AUTHOR=	"Cacciapaglia, G. and Cai, H. and Deandrea, A. and Flacke, T. and Lee, S. J. and Parolini, A.",
	TITLE=	"Composite scalars at the {LHC}: the {Higgs}, the sextet and the octet",
	EPRINT=	"1507.02283",
	ARCHIVEPREFIX=	"arXiv",
	PRIMARYCLASS=	"hep-ph",
	DOI=	"10.1007/JHEP11(2015)201",
	JOURNAL=	"JHEP",
	VOLUME=	"11",
	PAGES=	"201",
	YEAR=	"2015",
}

@ARTICLE{Cacciapaglia:2011kz,
	AUTHOR=	"Cacciapaglia, G. and Chierici, R. and Deandrea, A. and Panizzi, L. and Perries, S. and Tosi, S.",
	TITLE=	"Four tops on the real projective plane at {LHC}",
	EPRINT=	"1107.4616",
	ARCHIVEPREFIX=	"arXiv",
	PRIMARYCLASS=	"hep-ph",
	DOI=	"10.1007/JHEP10(2011)042",
	JOURNAL=	"JHEP",
	VOLUME=	"10",
	PAGES=	"042",
	YEAR=	"2011",
}

@ARTICLE{BhupalDev:2014bir,
	AUTHOR=	"Bhupal Dev, P. S. and Pilaftsis, A.",
	TITLE=	"Maximally symmetric two {Higgs} doublet model with natural standard model alignment",
	EPRINT=	"1408.3405",
	ARCHIVEPREFIX=	"arXiv",
	PRIMARYCLASS=	"hep-ph",
	DOI=	"10.1007/JHEP12(2014)024",
	JOURNAL=	"JHEP",
	VOLUME=	"12",
	PAGES=	"024",
	YEAR=	"2014",
	NOTE=	"[Erratum: \DOI{10.1007/JHEP11(2015)147}]",
}

@ARTICLE{Dicus:1994bm,
	AUTHOR=	"Dicus, D. and Stange, A. and Willenbrock, S.",
	TITLE=	"Higgs decay to top quarks at hadron colliders",
	EPRINT=	"hep-ph/9404359",
	ARCHIVEPREFIX=	"arXiv",
	DOI=	"10.1016/0370-2693(94)91017-0",
	JOURNAL=	"Phys. Lett. B",
	VOLUME=	"333",
	PAGES=	"126",
	YEAR=	"1994",
}

@ARTICLE{Craig:2015jba,
	AUTHOR=	"Craig, N. and D'Eramo, F. and Draper, P. and Thomas, S. and Zhang, H.",
	TITLE=	"The hunt for the rest of the {Higgs} bosons",
	EPRINT=	"1504.04630",
	ARCHIVEPREFIX=	"arXiv",
	PRIMARYCLASS=	"hep-ph",
	DOI=	"10.1007/JHEP06(2015)137",
	JOURNAL=	"JHEP",
	VOLUME=	"06",
	PAGES=	"137",
	YEAR=	"2015",
}

@ARTICLE{Craig:2016ygr,
	AUTHOR=	"Craig, N. and Hajer, J. and Li, Y.-Y. and Liu, T. and Zhang, H.",
	TITLE=	"Heavy {Higgs} bosons at low \tanb: from the {LHC} to {100\TeV}",
	EPRINT=	"1605.08744",
	ARCHIVEPREFIX=	"arXiv",
	PRIMARYCLASS=	"hep-ph",
	DOI=	"10.1007/JHEP01(2017)018",
	JOURNAL=	"JHEP",
	VOLUME=	"01",
	PAGES=	"018",
	YEAR=	"2017",
}

@ARTICLE{Anisha:2023xmh,
	AUTHOR=	"Anisha and Atkinson, O. and Bhardwaj, A. and Englert, C. and Naskar, W. and Stylianou, P.",
	TITLE=	"{BSM} reach of four-top production at the {LHC}",
	EPRINT=	"2302.08281",
	ARCHIVEPREFIX=	"arXiv",
	PRIMARYCLASS=	"hep-ph",
	DOI=	"10.1103/PhysRevD.108.035001",
	JOURNAL=	"Phys. Rev. D",
	VOLUME=	"108",
	PAGES=	"035001",
	YEAR=	"2023",
}

@ARTICLE{Ducu:2015fda,
	AUTHOR=	"Ducu, O. and Heurtier, L. and Maurer, J.",
	TITLE=	"{LHC} signatures of a {\PZpr} mediator between dark matter and the {SU}(3) sector",
	EPRINT=	"1509.05615",
	ARCHIVEPREFIX=	"arXiv",
	PRIMARYCLASS=	"hep-ph",
	DOI=	"10.1007/JHEP03(2016)006",
	JOURNAL=	"JHEP",
	VOLUME=	"03",
	PAGES=	"006",
	YEAR=	"2016",
}

@ARTICLE{Blasi:2023hvb,
	AUTHOR=	"Blasi, S. and Maltoni, F. and Mariotti, A. and Mimasu, K. and Pagani, D. and Tentori, S.",
	TITLE=	"Top-philic {ALP} phenomenology at the {LHC}: the elusive mass-window",
	EPRINT=	"2311.16048",
	ARCHIVEPREFIX=	"arXiv",
	PRIMARYCLASS=	"hep-ph",
	DOI=	"10.1007/JHEP06(2024)077",
	JOURNAL=	"JHEP",
	VOLUME=	"06",
	PAGES=	"077",
	YEAR=	"2024",
}

@ARTICLE{Kohda:2017fkn,
	AUTHOR=	"Kohda, M. and Modak, T. and Hou, W.-S.",
	TITLE=	"Searching for new scalar bosons via triple-top signature in {$\PQc\Pg\to\PQt\HepParticle{S}{}{0}\to\PQt\ttbar$}",
	EPRINT=	"1710.07260",
	ARCHIVEPREFIX=	"arXiv",
	PRIMARYCLASS=	"hep-ph",
	DOI=	"10.1016/j.physletb.2017.11.056",
	JOURNAL=	"Phys. Lett. B",
	VOLUME=	"776",
	PAGES=	"379",
	YEAR=	"2018",
}

@ARTICLE{Cao:2019qrb,
	AUTHOR=	"Cao, Q.-H. and Chen, S.-L. and Liu, Y. and Wang, X.-P.",
	TITLE=	"What can we learn from triple top-quark production?",
	EPRINT=	"1901.04643",
	ARCHIVEPREFIX=	"arXiv",
	PRIMARYCLASS=	"hep-ph",
	DOI=	"10.1103/PhysRevD.100.055035",
	JOURNAL=	"Phys. Rev. D",
	VOLUME=	"100",
	PAGES=	"055035",
	YEAR=	"2019",
}

@ARTICLE{Khanpour:2019qnw,
	AUTHOR=	"Khanpour, H.",
	TITLE=	"Probing top quark {FCNC} couplings in the triple-top signal at the high energy {LHC} and future circular collider",
	EPRINT=	"1909.03998",
	ARCHIVEPREFIX=	"arXiv",
	PRIMARYCLASS=	"hep-ph",
	DOI=	"10.1016/j.nuclphysb.2020.115141",
	JOURNAL=	"Nucl. Phys. B",
	VOLUME=	"958",
	PAGES=	"115141",
	YEAR=	"2020",
}

@ARTICLE{Iguro:2017ysu,
	AUTHOR=	"Iguro, S. and Tobe, K.",
	TITLE=	"${R(\HepParticle{\PD}{}{(\ast)})}$ in a general two {Higgs} doublet model",
	EPRINT=	"1708.06176",
	ARCHIVEPREFIX=	"arXiv",
	PRIMARYCLASS=	"hep-ph",
	DOI=	"10.1016/j.nuclphysb.2017.10.014",
	JOURNAL=	"Nucl. Phys. B",
	VOLUME=	"925",
	PAGES=	"560",
	YEAR=	"2017",
}

@ARTICLE{Cho:2019stk,
	AUTHOR=	"Cho, S. and Ko, P. and Lee, J. and Omura, Y. and Yu, C.",
	TITLE=	"Top {FCNC} induced by a {\PZpr} boson",
	EPRINT=	"1910.05925",
	ARCHIVEPREFIX=	"arXiv",
	PRIMARYCLASS=	"hep-ph",
	DOI=	"10.1103/PhysRevD.101.055015",
	JOURNAL=	"Phys. Rev. D",
	VOLUME=	"101",
	PAGES=	"055015",
	YEAR=	"2020",
}

@ARTICLE{Abasov:2024mwk,
	AUTHOR=	"Abasov, E. and Boos, E. and Bunichev, V. and Volkov, P. and Vorotnikov, G. and Dudko, L. and Zaborenko, A. and Iudin, E. and Markina, A. and Perfilov, M. and Savkova, N.",
	TITLE=	"Search for dark matter mediator in the production of three and four top quarks",
	EPRINT=	"2407.08308",
	ARCHIVEPREFIX=	"arXiv",
	PRIMARYCLASS=	"hep-ph",
	DOI=	"10.1134/S1063779624701764",
	JOURNAL=	"Phys. Part. Nucl.",
	VOLUME=	"56",
	PAGES=	"440",
	YEAR=	"2025",
}

@ARTICLE{Degrande:2010kt,
	AUTHOR=	"Degrande, C. and G{\'e}rard, J.-M. and Grojean, C. and Maltoni, F. and Servant, G.",
	TITLE=	"Non-resonant new physics in top pair production at hadron colliders",
	EPRINT=	"1010.6304",
	ARCHIVEPREFIX=	"arXiv",
	PRIMARYCLASS=	"hep-ph",
	DOI=	"10.1007/JHEP03(2011)125",
	JOURNAL=	"JHEP",
	VOLUME=	"03",
	PAGES=	"125",
	YEAR=	"2011",
}

@ARTICLE{Zhang:2017mls,
	AUTHOR=	"Zhang, C.",
	TITLE=	"Constraining ${\PQq\PQq\PQt\PQt}$ operators from four-top production: a case for enhanced {EFT} sensitivity",
	EPRINT=	"1708.05928",
	ARCHIVEPREFIX=	"arXiv",
	PRIMARYCLASS=	"hep-ph",
	DOI=	"10.1088/1674-1137/42/2/023104",
	JOURNAL=	"Chin. Phys. C",
	VOLUME=	"42",
	PAGES=	"023104",
	YEAR=	"2018",
}

@ARTICLE{Englert:2019zmt,
	AUTHOR=	"Englert, C. and Giudice, G. F. and Greljo, A. and McCullough, M.",
	TITLE=	"The $\widehat{\PH}$-parameter: an oblique {Higgs} view",
	EPRINT=	"1903.07725",
	ARCHIVEPREFIX=	"arXiv",
	PRIMARYCLASS=	"hep-ph",
	DOI=	"10.1007/JHEP09(2019)041",
	JOURNAL=	"JHEP",
	VOLUME=	"09",
	PAGES=	"041",
	YEAR=	"2019",
}

@ARTICLE{Banelli:2020iau,
	AUTHOR=	"Banelli, G. and Salvioni, E. and Serra, J. and Theil, T. and Weiler, A.",
	TITLE=	"The present and future of four top operators",
	EPRINT=	"2010.05915",
	ARCHIVEPREFIX=	"arXiv",
	PRIMARYCLASS=	"hep-ph",
	DOI=	"10.1007/JHEP02(2021)043",
	JOURNAL=	"JHEP",
	VOLUME=	"02",
	PAGES=	"043",
	YEAR=	"2021",
}

@ARTICLE{Darme:2021gtt,
	AUTHOR=	"Darm{\'e}, L. and Fuks, B. and Maltoni, F.",
	TITLE=	"Top-philic heavy resonances in four-top final states and their {EFT} interpretation",
	EPRINT=	"2104.09512",
	ARCHIVEPREFIX=	"arXiv",
	PRIMARYCLASS=	"hep-ph",
	DOI=	"10.1007/JHEP09(2021)143",
	JOURNAL=	"JHEP",
	VOLUME=	"09",
	PAGES=	"143",
	YEAR=	"2021",
}

@ARTICLE{Aoude:2022deh,
	AUTHOR=	"Aoude, R. and El Faham, H. and Maltoni, F. and Vryonidou, E.",
	TITLE=	"Complete {SMEFT} predictions for four top quark production at hadron colliders",
	EPRINT=	"2208.04962",
	ARCHIVEPREFIX=	"arXiv",
	PRIMARYCLASS=	"hep-ph",
	DOI=	"10.1007/JHEP10(2022)163",
	JOURNAL=	"JHEP",
	VOLUME=	"10",
	PAGES=	"163",
	YEAR=	"2022",
}

@ARTICLE{Aleshko:2023rkv,
	AUTHOR=	"Aleshko, A. and Boos, E. and Bunichev, V. and Dudko, L.",
	TITLE=	"Prospects for establishing limits on the {SMEFT} operators from the production processes of three and four top quarks in hadron collisions",
	EPRINT=	"2309.12514",
	ARCHIVEPREFIX=	"arXiv",
	PRIMARYCLASS=	"hep-ph",
	DOI=	"10.1142/S0217751X24501197",
	JOURNAL=	"Int. J. Mod. Phys. A",
	VOLUME=	"39",
	PAGES=	"2450119",
	YEAR=	"2024",
}

@ARTICLE{Aleshko:2025jua,
	AUTHOR=	"Aleshko, A. M. and Boos, E. E. and Bunichev, V. E. and Dudko, L. V.",
	TITLE=	"Sensitivity of the three top quark production process to the contribution of top-related {SMEFT} operators",
	DOI=	"10.1134/S1063779624701661",
	JOURNAL=	"Phys. Part. Nucl.",
	VOLUME=	"56",
	PAGES=	"374",
	YEAR=	"2025",
}

@ARTICLE{DiNoi:2025uhu,
	AUTHOR=	"Di Noi, S. and El Faham, H. and Gr{\"o}ber, R. and Vitti, M. and Vryonidou, E.",
	TITLE=	"Constraining four-heavy-quark operators with top-quark, {Higgs}, and electroweak precision data",
	EPRINT=	"2507.01137",
	ARCHIVEPREFIX=	"arXiv",
	PRIMARYCLASS=	"hep-ph",
	DOI=	"10.1007/JHEP01(2026)025",
	JOURNAL=	"JHEP",
	VOLUME=	"01",
	PAGES=	"025",
	YEAR=	"2026",
}

@ARTICLE{Kim:2016plm,
	AUTHOR=	"Kim, J. H. and Kong, K. and Lee, S. J. and Mohlabeng, G.",
	TITLE=	"Probing {\TeVns} scale top-philic resonances with boosted top-tagging at the high luminosity {LHC}",
	EPRINT=	"1604.07421",
	ARCHIVEPREFIX=	"arXiv",
	PRIMARYCLASS=	"hep-ph",
	DOI=	"10.1103/PhysRevD.94.035023",
	JOURNAL=	"Phys. Rev. D",
	VOLUME=	"94",
	PAGES=	"035023",
	YEAR=	"2016",
}

@ARTICLE{Darme:2025leu,
	AUTHOR=	"Darm{\'e}, L. and Fuks, B. and Li, H.-L. and Maltoni, M. and Touch{\`e}que, J.",
	TITLE=	"Searching for top-philic heavy resonances in boosted four-top final states",
	EPRINT=	"2507.05334",
	ARCHIVEPREFIX=	"arXiv",
	PRIMARYCLASS=	"hep-ph",
	DOI=	"10.1007/JHEP11(2025)091",
	JOURNAL=	"JHEP",
	VOLUME=	"11",
	PAGES=	"091",
	YEAR=	"2025",
}

@ARTICLE{Cao:2016wib,
	AUTHOR=	"Cao, Q.-H. and Chen, S.-L. and Liu, Y.",
	TITLE=	"Probing {Higgs} width and top quark {Yukawa} coupling from ${\ttbar\PH}$ and $\ttbar\ttbar$ productions",
	EPRINT=	"1602.01934",
	ARCHIVEPREFIX=	"arXiv",
	PRIMARYCLASS=	"hep-ph",
	DOI=	"10.1103/PhysRevD.95.053004",
	JOURNAL=	"Phys. Rev. D",
	VOLUME=	"95",
	PAGES=	"053004",
	YEAR=	"2017",
}

@ARTICLE{Cao:2019ygh,
	AUTHOR=	"Cao, Q.-H. and Chen, S.-L. and Liu, Y. and Zhang, R. and Zhang, Y.",
	TITLE=	"Limiting top quark-{Higgs} boson interaction and {Higgs}-boson width from multitop productions",
	EPRINT=	"1901.04567",
	ARCHIVEPREFIX=	"arXiv",
	PRIMARYCLASS=	"hep-ph",
	DOI=	"10.1103/PhysRevD.99.113003",
	JOURNAL=	"Phys. Rev. D",
	VOLUME=	"99",
	PAGES=	"113003",
	YEAR=	"2019",
}

@ARTICLE{ATLAS:Detector-2008,
	AUTHOR=	"{ATLAS Collaboration}",
	TITLE=	"The {ATLAS} experiment at the {CERN} {Large Hadron Collider}",
	DOI=	"10.1088/1748-0221/3/08/S08003",
	JOURNAL=	"JINST",
	VOLUME=	"3",
	PAGES=	"S08003",
	YEAR=	"2008",
}

@ARTICLE{ATLAS:2023dns,
	AUTHOR=	"{ATLAS Collaboration}",
	TITLE=	"The {ATLAS} experiment at the {CERN} {Large Hadron Collider}: a description of the detector configuration for \mbox{Run 3}",
	EPRINT=	"2305.16623",
	ARCHIVEPREFIX=	"arXiv",
	PRIMARYCLASS=	"physics.ins-det",
	DOI=	"10.1088/1748-0221/19/05/P05063",
	JOURNAL=	"JINST",
	VOLUME=	"19",
	PAGES=	"P05063",
	YEAR=	"2024",
}

@ARTICLE{CMS:Detector-2008,
	AUTHOR=	"{CMS Collaboration}",
	TITLE=	"The {CMS} experiment at the {CERN} {LHC}",
	DOI=	"10.1088/1748-0221/3/08/S08004",
	JOURNAL=	"JINST",
	VOLUME=	"3",
	PAGES=	"S08004",
	YEAR=	"2008",
}

@ARTICLE{CMS:PRF-21-001,
	AUTHOR=	"{CMS Collaboration}",
	TITLE=	"Development of the {CMS} detector for the {CERN} {LHC} \mbox{Run 3}",
	EPRINT=	"2309.05466",
	ARCHIVEPREFIX=	"arXiv",
	PRIMARYCLASS=	"physics.ins-det",
	DOI=	"10.1088/1748-0221/19/05/P05064",
	JOURNAL=	"JINST",
	VOLUME=	"19",
	PAGES=	"P05064",
	YEAR=	"2024",
}

@ARTICLE{CMS:SUS-16-035,
	AUTHOR=	"{CMS Collaboration}",
	TITLE=	"Search for physics beyond the standard model in events with two leptons of same sign, missing transverse momentum, and jets in proton-proton collisions at $\sqrt{s}={13\TeV}$",
	EPRINT=	"1704.07323",
	ARCHIVEPREFIX=	"arXiv",
	PRIMARYCLASS=	"hep-ex",
	DOI=	"10.1140/epjc/s10052-017-5079-z",
	JOURNAL=	"Eur. Phys. J. C",
	VOLUME=	"77",
	PAGES=	"578",
	YEAR=	"2017",
}

@ARTICLE{CMS:TOP-17-009,
	AUTHOR=	"{CMS Collaboration}",
	TITLE=	"Search for standard model production of four top quarks with same-sign and multilepton final states in proton-proton collisions at $\sqrt{s}={13\TeV}$",
	EPRINT=	"1710.10614",
	ARCHIVEPREFIX=	"arXiv",
	PRIMARYCLASS=	"hep-ex",
	DOI=	"10.1140/epjc/s10052-018-5607-5",
	JOURNAL=	"Eur. Phys. J. C",
	VOLUME=	"78",
	PAGES=	"140",
	YEAR=	"2018",
}

@ARTICLE{ATLAS:2018alq,
	AUTHOR=	"{ATLAS Collaboration}",
	TITLE=	"Search for new phenomena in events with same-charge leptons and {\PQb} jets in ${\Pp\Pp}$ collisions at $\sqrt{s}={13\TeV}$ with the {ATLAS} detector",
	EPRINT=	"1807.11883",
	ARCHIVEPREFIX=	"arXiv",
	PRIMARYCLASS=	"hep-ex",
	DOI=	"10.1007/JHEP12(2018)039",
	JOURNAL=	"JHEP",
	VOLUME=	"12",
	PAGES=	"039",
	YEAR=	"2018",
}

@ARTICLE{ATLAS:2018kxv,
	AUTHOR=	"{ATLAS Collaboration}",
	TITLE=	"Search for four-top-quark production in the single-lepton and opposite-sign dilepton final states in ${\Pp\Pp}$ collisions at $\sqrt{s}={13\TeV}$ with the {ATLAS} detector",
	EPRINT=	"1811.02305",
	ARCHIVEPREFIX=	"arXiv",
	PRIMARYCLASS=	"hep-ex",
	DOI=	"10.1103/PhysRevD.99.052009",
	JOURNAL=	"Phys. Rev. D",
	VOLUME=	"99",
	PAGES=	"052009",
	YEAR=	"2019",
}

@ARTICLE{CMS:TOP-17-019,
	AUTHOR=	"{CMS Collaboration}",
	TITLE=	"Search for the production of four top quarks in the single-lepton and opposite-sign dilepton final states in proton-proton collisions at $\sqrt{s}={13\TeV}$",
	EPRINT=	"1906.02805",
	ARCHIVEPREFIX=	"arXiv",
	PRIMARYCLASS=	"hep-ex",
	DOI=	"10.1007/JHEP11(2019)082",
	JOURNAL=	"JHEP",
	VOLUME=	"11",
	PAGES=	"082",
	YEAR=	"2019",
}

@ARTICLE{CMS:TOP-18-003,
	AUTHOR=	"{CMS Collaboration}",
	TITLE=	"Search for production of four top quarks in final states with same-sign or multiple leptons in proton-proton collisions at $\sqrt{s}={13\TeV}$",
	EPRINT=	"1908.06463",
	ARCHIVEPREFIX=	"arXiv",
	PRIMARYCLASS=	"hep-ex",
	DOI=	"10.1140/epjc/s10052-019-7593-7",
	JOURNAL=	"Eur. Phys. J. C",
	VOLUME=	"80",
	PAGES=	"75",
	YEAR=	"2020",
}

@ARTICLE{ATLAS:2020hpj,
	AUTHOR=	"{ATLAS Collaboration}",
	TITLE=	"Evidence for $\ttbar\ttbar$ production in the multilepton final state in proton-proton collisions at $\sqrt{s}={13\TeV}$ with the {ATLAS} detector",
	EPRINT=	"2007.14858",
	ARCHIVEPREFIX=	"arXiv",
	PRIMARYCLASS=	"hep-ex",
	DOI=	"10.1140/epjc/s10052-020-08509-3",
	JOURNAL=	"Eur. Phys. J. C",
	VOLUME=	"80",
	PAGES=	"1085",
	YEAR=	"2020",
}

@ARTICLE{ATLAS:2021kqb,
	AUTHOR=	"{ATLAS Collaboration}",
	TITLE=	"Measurement of the $\ttbar\ttbar$ production cross section in ${\Pp\Pp}$ collisions at $\sqrt{s}={13\TeV}$ with the {ATLAS} detector",
	EPRINT=	"2106.11683",
	ARCHIVEPREFIX=	"arXiv",
	PRIMARYCLASS=	"hep-ex",
	DOI=	"10.1007/JHEP11(2021)118",
	JOURNAL=	"JHEP",
	VOLUME=	"11",
	PAGES=	"118",
	YEAR=	"2021",
}

@ARTICLE{CMS:TOP-21-005,
	AUTHOR=	"{CMS Collaboration}",
	TITLE=	"Evidence for four-top quark production in proton-proton collisions at $\sqrt{s}={13\TeV}$",
	EPRINT=	"2303.03864",
	ARCHIVEPREFIX=	"arXiv",
	PRIMARYCLASS=	"hep-ex",
	DOI=	"10.1016/j.physletb.2023.138076",
	JOURNAL=	"Phys. Lett. B",
	VOLUME=	"844",
	PAGES=	"138076",
	YEAR=	"2023",
}

@ARTICLE{Blekman:2022jag,
	AUTHOR=	"Blekman, F. and D{\'e}liot, F. and Dutta, V. and Usai, E.",
	TITLE=	"Four-top quark physics at the {LHC}",
	EPRINT=	"2208.04085",
	ARCHIVEPREFIX=	"arXiv",
	PRIMARYCLASS=	"hep-ex",
	DOI=	"10.3390/universe8120638",
	JOURNAL=	"Universe",
	VOLUME=	"8",
	PAGES=	"638",
	YEAR=	"2022",
}

@ARTICLE{ATLAS:2023ajo,
	AUTHOR=	"{ATLAS Collaboration}",
	TITLE=	"Observation of four-top-quark production in the multilepton final state with the {ATLAS} detector",
	EPRINT=	"2303.15061",
	ARCHIVEPREFIX=	"arXiv",
	PRIMARYCLASS=	"hep-ex",
	DOI=	"10.1140/epjc/s10052-023-11573-0",
	JOURNAL=	"Eur. Phys. J. C",
	VOLUME=	"83",
	PAGES=	"496",
	YEAR=	"2023",
	NOTE=	"[Erratum: \DOI{10.1140/epjc/s10052-024-12458-6}]",
}

@ARTICLE{CMS:TOP-22-013,
	AUTHOR=	"{CMS Collaboration}",
	TITLE=	"Observation of four top quark production in proton-proton collisions at $\sqrt{s}={13\TeV}$",
	EPRINT=	"2305.13439",
	ARCHIVEPREFIX=	"arXiv",
	PRIMARYCLASS=	"hep-ex",
	DOI=	"10.1016/j.physletb.2023.138290",
	JOURNAL=	"Phys. Lett. B",
	VOLUME=	"847",
	PAGES=	"138290",
	YEAR=	"2023",
}

@ARTICLE{ATLAS:2018cye,
	AUTHOR=	"{ATLAS Collaboration}",
	TITLE=	"Search for pair production of up-type vector-like quarks and for four-top-quark events in final states with multiple {\PQb}-jets with the {ATLAS} detector",
	EPRINT=	"1803.09678",
	ARCHIVEPREFIX=	"arXiv",
	PRIMARYCLASS=	"hep-ex",
	DOI=	"10.1007/JHEP07(2018)089",
	JOURNAL=	"JHEP",
	VOLUME=	"07",
	PAGES=	"089",
	YEAR=	"2018",
}

@ARTICLE{CMS:TOP-21-003,
	AUTHOR=	"{CMS Collaboration}",
	TITLE=	"Search for new physics using effective field theory in {13\TeV} ${\Pp\Pp}$ collision events that contain a top quark pair and a boosted {\PZ} or {Higgs} boson",
	EPRINT=	"2208.12837",
	ARCHIVEPREFIX=	"arXiv",
	PRIMARYCLASS=	"hep-ex",
	DOI=	"10.1103/PhysRevD.108.032008",
	JOURNAL=	"Phys. Rev. D",
	VOLUME=	"108",
	PAGES=	"032008",
	YEAR=	"2023",
}

@ARTICLE{ATLAS:2024lyh,
	AUTHOR=	"{ATLAS Collaboration}",
	TITLE=	"Interpretations of the {ATLAS} measurements of {Higgs} boson production and decay rates and differential cross-sections in ${\Pp\Pp}$ collisions at $\sqrt{s}={13\TeV}$",
	EPRINT=	"2402.05742",
	ARCHIVEPREFIX=	"arXiv",
	PRIMARYCLASS=	"hep-ex",
	DOI=	"10.1007/JHEP11(2024)097",
	JOURNAL=	"JHEP",
	VOLUME=	"11",
	PAGES=	"097",
	YEAR=	"2024",
}

@ARTICLE{CMS:TOP-22-006,
	AUTHOR=	"{CMS Collaboration}",
	TITLE=	"Search for physics beyond the standard model in top quark production with additional leptons in the context of effective field theory",
	EPRINT=	"2307.15761",
	ARCHIVEPREFIX=	"arXiv",
	PRIMARYCLASS=	"hep-ex",
	DOI=	"10.1007/JHEP12(2023)068",
	JOURNAL=	"JHEP",
	VOLUME=	"12",
	PAGES=	"068",
	YEAR=	"2023",
}

@ARTICLE{CMS:SMP-24-003,
	AUTHOR=	"{CMS Collaboration}",
	TITLE=	"Combined effective field theory interpretation of {Higgs} boson, electroweak vector boson, top quark, and multijet measurements",
	EPRINT=	"2504.02958",
	ARCHIVEPREFIX=	"arXiv",
	PRIMARYCLASS=	"hep-ex",
	DOI=	"10.1140/epjc/s10052-025-14997-y",
	JOURNAL=	"Eur. Phys. J. C",
	VOLUME=	"86",
	PAGES=	"331",
	YEAR=	"2026",
}

@ARTICLE{Hartland:2019bjb,
	AUTHOR=	"Hartland, N. and Maltoni, F. and Nocera, E. and Rojo, J. and Slade, E. and Vryonidou, E. and Zhang, C.",
	TITLE=	"A {Monte Carlo} global analysis of the standard model effective field theory: the top quark sector",
	EPRINT=	"1901.05965",
	ARCHIVEPREFIX=	"arXiv",
	PRIMARYCLASS=	"hep-ph",
	DOI=	"10.1007/JHEP04(2019)100",
	JOURNAL=	"JHEP",
	VOLUME=	"04",
	PAGES=	"100",
	YEAR=	"2019",
}

@ARTICLE{Ethier:2021bye,
	AUTHOR=	"Ethier, J. and Magni, G. and Maltoni, F. and Mantani, L. and Nocera, E. and Rojo, J. and Slade, E. and Vryonidou, E. and Zhang, C.",
	COLLABORATION=	"SMEFiT",
	TITLE=	"Combined {SMEFT} interpretation of {Higgs}, diboson, and top quark data from the {LHC}",
	EPRINT=	"2105.00006",
	ARCHIVEPREFIX=	"arXiv",
	PRIMARYCLASS=	"hep-ph",
	DOI=	"10.1007/JHEP11(2021)089",
	JOURNAL=	"JHEP",
	VOLUME=	"11",
	PAGES=	"089",
	YEAR=	"2021",
}

@ARTICLE{Celada:2024mcf,
	AUTHOR=	"Celada, E. and Giani, T. and ter Hoeve, J. and Mantani, L. and Rojo, J. and Rossia, A. N. and Thomas, M. O. A. and Vryonidou, E.",
	TITLE=	"Mapping the {SMEFT} at high-energy colliders: from {LEP} and the {(HL-)LHC} to the {FCC}-${\Pe\Pe}$",
	EPRINT=	"2404.12809",
	ARCHIVEPREFIX=	"arXiv",
	PRIMARYCLASS=	"hep-ph",
	DOI=	"10.1007/JHEP09(2024)091",
	JOURNAL=	"JHEP",
	VOLUME=	"09",
	PAGES=	"091",
	YEAR=	"2024",
}

@ARTICLE{deBlas:2025xhe,
	AUTHOR=	"de Blas, J. and Goncalves, A. and Miralles, V. and Reina, L. and Silvestrini, L. and Valli, M.",
	TITLE=	"Constraining new physics effective interactions via a global fit of electroweak, {Drell}--{Yan}, {Higgs}, top, and flavour observables",
	EPRINT=	"2507.06191",
	ARCHIVEPREFIX=	"arXiv",
	PRIMARYCLASS=	"hep-ph",
	DOI=	"10.1007/JHEP03(2026)013",
	JOURNAL=	"JHEP",
	VOLUME=	"03",
	PAGES=	"013",
	YEAR=	"2026",
}

@ARTICLE{Ellis:2020unq,
	AUTHOR=	"Ellis, J. and Madigan, M. and Mimasu, K. and Sanz, V. and You, T.",
	TITLE=	"Top, {Higgs}, diboson and electroweak fit to the standard model effective field theory",
	EPRINT=	"2012.02779",
	ARCHIVEPREFIX=	"arXiv",
	PRIMARYCLASS=	"hep-ph",
	DOI=	"10.1007/JHEP04(2021)279",
	JOURNAL=	"JHEP",
	VOLUME=	"04",
	PAGES=	"279",
	YEAR=	"2021",
}

@ARTICLE{Miralles:2021dyw,
	AUTHOR=	"Miralles, V. and Miralles L{\'o}pez, M. and Moreno Ll{\'a}cer, M. and Pe{\~n}uelas, A. and Perell{\'o}, M. and Vos, M.",
	TITLE=	"The top quark electro-weak couplings after {LHC} \mbox{Run 2}",
	EPRINT=	"2107.13917",
	ARCHIVEPREFIX=	"arXiv",
	PRIMARYCLASS=	"hep-ph",
	DOI=	"10.1007/JHEP02(2022)032",
	JOURNAL=	"JHEP",
	VOLUME=	"02",
	PAGES=	"032",
	YEAR=	"2022",
}

@ARTICLE{ATLAS:2022rws,
	AUTHOR=	"{ATLAS Collaboration}",
	TITLE=	"Search for ${\ttbar\PH/\PSA\to\ttbar\ttbar}$ production in the multilepton final state in proton-proton collisions at $\sqrt{s}={13\TeV}$ with the {ATLAS} detector",
	EPRINT=	"2211.01136",
	ARCHIVEPREFIX=	"arXiv",
	PRIMARYCLASS=	"hep-ex",
	DOI=	"10.1007/JHEP07(2023)203",
	JOURNAL=	"JHEP",
	VOLUME=	"07",
	PAGES=	"203",
	YEAR=	"2023",
	NOTE=	"[Erratum: \DOI{10.1007/JHEP05(2025)159}]",
}

@ARTICLE{ATLAS:2023taw,
	AUTHOR=	"{ATLAS Collaboration}",
	TITLE=	"Search for top-philic heavy resonances in ${\Pp\Pp}$ collisions at $\sqrt{s}={13\TeV}$ with the {ATLAS} detector",
	EPRINT=	"2304.01678",
	ARCHIVEPREFIX=	"arXiv",
	PRIMARYCLASS=	"hep-ex",
	DOI=	"10.1140/epjc/s10052-023-12318-9",
	JOURNAL=	"Eur. Phys. J. C",
	VOLUME=	"84",
	PAGES=	"157",
	YEAR=	"2024",
}

@ARTICLE{CMS:B2G-16-015,
	AUTHOR=	"{CMS Collaboration}",
	TITLE=	"Search for \ttbar resonances in highly-boosted lepton+jets and fully hadronic final states in proton-proton collisions at $\sqrt{s}={13\TeV}$",
	EPRINT=	"1704.03366",
	ARCHIVEPREFIX=	"arXiv",
	PRIMARYCLASS=	"hep-ex",
	DOI=	"10.1007/JHEP07(2017)001",
	JOURNAL=	"JHEP",
	VOLUME=	"07",
	PAGES=	"001",
	YEAR=	"2017",
}

@ARTICLE{ATLAS:2018rvc,
	AUTHOR=	"{ATLAS Collaboration}",
	TITLE=	"Search for heavy particles decaying into top-quark pairs using lepton-plus-jets events in proton-proton collisions at $\sqrt{s}={13\TeV}$ with the {ATLAS} detector",
	EPRINT=	"1804.10823",
	ARCHIVEPREFIX=	"arXiv",
	PRIMARYCLASS=	"hep-ex",
	DOI=	"10.1140/epjc/s10052-018-5995-6",
	JOURNAL=	"Eur. Phys. J. C",
	VOLUME=	"78",
	PAGES=	"565",
	YEAR=	"2018",
}

@ARTICLE{CMS:B2G-17-017,
	AUTHOR=	"{CMS Collaboration}",
	TITLE=	"Search for resonant \ttbar production in proton-proton collisions at $\sqrt{s}={13\TeV}$",
	EPRINT=	"1810.05905",
	ARCHIVEPREFIX=	"arXiv",
	PRIMARYCLASS=	"hep-ex",
	DOI=	"10.1007/JHEP04(2019)031",
	JOURNAL=	"JHEP",
	VOLUME=	"04",
	PAGES=	"031",
	YEAR=	"2019",
}

@ARTICLE{ATLAS:2019npw,
	AUTHOR=	"{ATLAS Collaboration}",
	TITLE=	"Search for heavy particles decaying into a top-quark pair in the fully hadronic final state in ${\Pp\Pp}$ collisions at $\sqrt{s}={13\TeV}$ with the {ATLAS} detector",
	EPRINT=	"1902.10077",
	ARCHIVEPREFIX=	"arXiv",
	PRIMARYCLASS=	"hep-ex",
	DOI=	"10.1103/PhysRevD.99.092004",
	JOURNAL=	"Phys. Rev. D",
	VOLUME=	"99",
	PAGES=	"092004",
	YEAR=	"2019",
}

@ARTICLE{CMS:HIG-17-027,
	AUTHOR=	"{CMS Collaboration}",
	TITLE=	"Search for heavy {Higgs} bosons decaying to a top quark pair in proton-proton collisions at $\sqrt{s}={13\TeV}$",
	EPRINT=	"1908.01115",
	ARCHIVEPREFIX=	"arXiv",
	PRIMARYCLASS=	"hep-ex",
	DOI=	"10.1007/JHEP04(2020)171",
	JOURNAL=	"JHEP",
	VOLUME=	"04",
	PAGES=	"171",
	YEAR=	"2020",
	NOTE=	"[Erratum: \DOI{10.1007/JHEP03(2022)187}]",
}

@ARTICLE{ATLAS:2020lks,
	AUTHOR=	"{ATLAS Collaboration}",
	TITLE=	"Search for \ttbar resonances in fully hadronic final states in ${\Pp\Pp}$ collisions at $\sqrt{s}={13\TeV}$ with the {ATLAS} detector",
	EPRINT=	"2005.05138",
	ARCHIVEPREFIX=	"arXiv",
	PRIMARYCLASS=	"hep-ex",
	DOI=	"10.1007/JHEP10(2020)061",
	JOURNAL=	"JHEP",
	VOLUME=	"10",
	PAGES=	"061",
	YEAR=	"2020",
}

@ARTICLE{ATLAS:2024vxm,
	AUTHOR=	"{ATLAS Collaboration}",
	TITLE=	"Search for heavy neutral {Higgs} bosons decaying into a top quark pair in 140\fbinv of proton-proton collision data at $\sqrt{s}={13\TeV}$ with the {ATLAS} detector",
	EPRINT=	"2404.18986",
	ARCHIVEPREFIX=	"arXiv",
	PRIMARYCLASS=	"hep-ex",
	DOI=	"10.1007/JHEP08(2024)013",
	JOURNAL=	"JHEP",
	VOLUME=	"08",
	PAGES=	"013",
	YEAR=	"2024",
}

@ARTICLE{CMS:TOP-24-007,
	AUTHOR=	"{CMS Collaboration}",
	TITLE=	"Observation of a pseudoscalar excess at the top quark pair production threshold",
	EPRINT=	"2503.22382",
	ARCHIVEPREFIX=	"arXiv",
	PRIMARYCLASS=	"hep-ex",
	DOI=	"10.1088/1361-6633/adf7d3",
	JOURNAL=	"Rep. Prog. Phys.",
	VOLUME=	"88",
	PAGES=	"087801",
	YEAR=	"2025",
}

@ARTICLE{CMS:HIG-22-013,
	AUTHOR=	"{CMS Collaboration}",
	TITLE=	"Search for heavy pseudoscalar and scalar bosons decaying to a top quark pair in proton-proton collisions at $\sqrt{s}={13\TeV}$",
	EPRINT=	"2507.05119",
	ARCHIVEPREFIX=	"arXiv",
	PRIMARYCLASS=	"hep-ex",
	DOI=	"10.1088/1361-6633/ae2207",
	JOURNAL=	"Rep. Prog. Phys.",
	VOLUME=	"88",
	PAGES=	"127801",
	YEAR=	"2025",
}

@ARTICLE{ATLAS:2025kmo,
	AUTHOR=	"{ATLAS Collaboration}",
	TITLE=	"Search for \ttbar resonances in final states with exactly one or two leptons using 140\fbinv of ${\Pp\Pp}$ collision data at $\sqrt{s}={13\TeV}$ with the {ATLAS} experiment",
	EPRINT=	"2512.17856",
	ARCHIVEPREFIX=	"arXiv",
	PRIMARYCLASS=	"hep-ex",
	DOI=	"10.1007/JHEP10(2026)219",
	JOURNAL=	"JHEP",
	VOLUME=	"10",
	PAGES=	"219",
	YEAR=	"2026",
}

@ARTICLE{ATLAS:2026dbe,
	AUTHOR=	"{ATLAS Collaboration}",
	TITLE=	"Observation of a cross-section enhancement near the \ttbar production threshold in $\sqrt{s}={13\TeV}$ ${\Pp\Pp}$ collisions with the {ATLAS} detector",
	EPRINT=	"2601.11780",
	ARCHIVEPREFIX=	"arXiv",
	PRIMARYCLASS=	"hep-ex",
	DOI=	"10.1088/1361-6633/ae60a0",
	JOURNAL=	"Rep. Prog. Phys.",
	VOLUME=	"89",
	PAGES=	"057801",
	YEAR=	"2026",
}

@UNPUBLISHED{CMS:B2G-25-009,
	AUTHOR=	"{CMS Collaboration}",
	TITLE=	"Search for new particles decaying into top quark-antiquark pairs in proton-proton collisions at $\sqrt{s}={13\TeV}$",
	EPRINT=	"2603.23454",
	ARCHIVEPREFIX=	"arXiv",
	PRIMARYCLASS=	"hep-ex",
	NOTE=	"Submitted to \textit{JHEP}",
	YEAR=	"2026",
}

@ARTICLE{ATLAS:2023zkt,
	AUTHOR=	"{ATLAS Collaboration}",
	TITLE=	"Search for a ${CP}$-odd {Higgs} boson decaying into a heavy ${CP}$-even {Higgs} boson and a {\PZ} boson in the ${\Pellp\Pellm\ttbar}$ and ${\PGn\PAGn\bbbar}$ final states using 140\fbinv of data collected with the {ATLAS} detector",
	EPRINT=	"2311.04033",
	ARCHIVEPREFIX=	"arXiv",
	PRIMARYCLASS=	"hep-ex",
	DOI=	"10.1007/JHEP02(2024)197",
	JOURNAL=	"JHEP",
	VOLUME=	"02",
	PAGES=	"197",
	YEAR=	"2024",
}

@ARTICLE{CMS:B2G-23-006,
	AUTHOR=	"{CMS Collaboration}",
	TITLE=	"Search for heavy neutral {Higgs} bosons {\PSA} and {\PH} in the ${\ttbar\PZ}$ channel in proton-proton collisions at {13\TeV}",
	EPRINT=	"2412.00570",
	ARCHIVEPREFIX=	"arXiv",
	PRIMARYCLASS=	"hep-ex",
	DOI=	"10.1016/j.physletb.2025.139568",
	JOURNAL=	"Phys. Lett. B",
	VOLUME=	"866",
	PAGES=	"139568",
	YEAR=	"2025",
}

@ARTICLE{CMS:HIG-19-013,
	AUTHOR=	"{CMS Collaboration}",
	TITLE=	"Measurements of ${\ttbar\PH}$ production and the {CP} structure of the {Yukawa} interaction between the {Higgs} boson and top quark in the diphoton decay channel",
	EPRINT=	"2003.10866",
	ARCHIVEPREFIX=	"arXiv",
	PRIMARYCLASS=	"hep-ex",
	DOI=	"10.1103/PhysRevLett.125.061801",
	JOURNAL=	"Phys. Rev. Lett.",
	VOLUME=	"125",
	PAGES=	"061801",
	YEAR=	"2020",
}

@ARTICLE{ATLAS:2020ior,
	AUTHOR=	"{ATLAS Collaboration}",
	TITLE=	"${CP}$ properties of {Higgs} boson interactions with top quarks in the ${\ttbar\PH}$ and ${\PQt\PH}$ processes using ${\PH\to\PGg\PGg}$ with the {ATLAS} detector",
	EPRINT=	"2004.04545",
	ARCHIVEPREFIX=	"arXiv",
	PRIMARYCLASS=	"hep-ex",
	DOI=	"10.1103/PhysRevLett.125.061802",
	JOURNAL=	"Phys. Rev. Lett.",
	VOLUME=	"125",
	PAGES=	"061802",
	YEAR=	"2020",
}

@ARTICLE{CMS:HIG-19-008,
	AUTHOR=	"{CMS Collaboration}",
	TITLE=	"Measurement of the {Higgs} boson production rate in association with top quarks in final states with electrons, muons, and hadronically decaying tau leptons at $\sqrt{s}={13\TeV}$",
	EPRINT=	"2011.03652",
	ARCHIVEPREFIX=	"arXiv",
	PRIMARYCLASS=	"hep-ex",
	DOI=	"10.1140/epjc/s10052-021-09014-x",
	JOURNAL=	"Eur. Phys. J. C",
	VOLUME=	"81",
	PAGES=	"378",
	YEAR=	"2021",
}

@ARTICLE{CMS:HIG-21-006,
	AUTHOR=	"{CMS Collaboration}",
	TITLE=	"Search for {CP} violation in ${\ttbar\PH}$ and ${\PQt\PH}$ production in multilepton channels in proton-proton collisions at $\sqrt{s}={13\TeV}$",
	EPRINT=	"2208.02686",
	ARCHIVEPREFIX=	"arXiv",
	PRIMARYCLASS=	"hep-ex",
	DOI=	"10.1007/JHEP07(2023)092",
	JOURNAL=	"JHEP",
	VOLUME=	"07",
	PAGES=	"092",
	YEAR=	"2023",
}

@ARTICLE{ATLAS:2023cbt,
	AUTHOR=	"{ATLAS Collaboration}",
	TITLE=	"Probing the ${CP}$ nature of the top--{Higgs} {Yukawa} coupling in ${\ttbar\PH}$ and ${\PQt\PH}$ events with ${\PH\to\bbbar}$ decays using the {ATLAS} detector at the {LHC}",
	EPRINT=	"2303.05974",
	ARCHIVEPREFIX=	"arXiv",
	PRIMARYCLASS=	"hep-ex",
	DOI=	"10.1016/j.physletb.2024.138469",
	JOURNAL=	"Phys. Lett. B",
	VOLUME=	"849",
	PAGES=	"138469",
	YEAR=	"2024",
}

@ARTICLE{CMS:HIG-19-011,
	AUTHOR=	"{CMS Collaboration}",
	TITLE=	"Measurement of the ${\ttbar\PH}$ and ${\PQt\PH}$ production rates in the ${\PH\to\bbbar}$ decay channel using proton-proton collision data at $\sqrt{s}={13\TeV}$",
	EPRINT=	"2407.10896",
	ARCHIVEPREFIX=	"arXiv",
	PRIMARYCLASS=	"hep-ex",
	DOI=	"10.1007/JHEP02(2025)097",
	JOURNAL=	"JHEP",
	VOLUME=	"02",
	PAGES=	"097",
	YEAR=	"2025",
}

@ARTICLE{CMS:HIG-19-009,
	AUTHOR=	"{CMS Collaboration}",
	TITLE=	"Constraints on anomalous {Higgs} boson couplings to vector bosons and fermions in its production and decay using the four-lepton final state",
	EPRINT=	"2104.12152",
	ARCHIVEPREFIX=	"arXiv",
	PRIMARYCLASS=	"hep-ex",
	DOI=	"10.1103/PhysRevD.104.052004",
	JOURNAL=	"Phys. Rev. D",
	VOLUME=	"104",
	PAGES=	"052004",
	YEAR=	"2021",
}

@ARTICLE{CMS:HIG-22-001,
	AUTHOR=	"{CMS Collaboration}",
	TITLE=	"A portrait of the {Higgs} boson by the {CMS} experiment ten years after the discovery",
	EPRINT=	"2207.00043",
	ARCHIVEPREFIX=	"arXiv",
	PRIMARYCLASS=	"hep-ex",
	DOI=	"10.1038/s41586-022-04892-x",
	JOURNAL=	"Nature",
	VOLUME=	"607",
	PAGES=	"60",
	YEAR=	"2022",
	NOTE=	"[Author correction: \DOI{10.1038/s41586-023-06164-8}]",
}

@ARTICLE{ATLAS:2022vkf,
	AUTHOR=	"{ATLAS Collaboration}",
	TITLE=	"A detailed map of {Higgs} boson interactions by the {ATLAS} experiment ten years after the discovery",
	EPRINT=	"2207.00092",
	ARCHIVEPREFIX=	"arXiv",
	PRIMARYCLASS=	"hep-ex",
	DOI=	"10.1038/s41586-022-04893-w",
	JOURNAL=	"Nature",
	VOLUME=	"607",
	PAGES=	"52",
	YEAR=	"2022",
	NOTE=	"[Publisher correction: \DOI{10.1038/s41586-022-05581-5}, author correction: \DOI{10.1038/s41586-023-06248-5}]",
}

@ARTICLE{ATLAS:2022tnm,
	AUTHOR=	"{ATLAS Collaboration}",
	TITLE=	"Measurement of the properties of {Higgs} boson production at $\sqrt{s}={13\TeV}$ in the ${\PH\to\PGg\PGg}$ channel using 139\fbinv of ${\Pp\Pp}$ collision data with the {ATLAS} experiment",
	EPRINT=	"2207.00348",
	ARCHIVEPREFIX=	"arXiv",
	PRIMARYCLASS=	"hep-ex",
	DOI=	"10.1007/JHEP07(2023)088",
	JOURNAL=	"JHEP",
	VOLUME=	"07",
	PAGES=	"088",
	YEAR=	"2023",
}

@ARTICLE{CMS:TOP-17-004,
	AUTHOR=	"{CMS Collaboration}",
	TITLE=	"Measurement of the top quark {Yukawa} coupling from \ttbar kinematic distributions in the lepton+jets final state in proton-proton collisions at $\sqrt{s}={13\TeV}$",
	EPRINT=	"1907.01590",
	ARCHIVEPREFIX=	"arXiv",
	PRIMARYCLASS=	"hep-ex",
	DOI=	"10.1103/PhysRevD.100.072007",
	JOURNAL=	"Phys. Rev. D",
	VOLUME=	"100",
	PAGES=	"072007",
	YEAR=	"2019",
}

@ARTICLE{CMS:TOP-19-008,
	AUTHOR=	"{CMS Collaboration}",
	TITLE=	"Measurement of the top quark {Yukawa} coupling from \ttbar kinematic distributions in the dilepton final state in proton-proton collisions at $\sqrt{s}={13\TeV}$",
	EPRINT=	"2009.07123",
	ARCHIVEPREFIX=	"arXiv",
	PRIMARYCLASS=	"hep-ex",
	DOI=	"10.1103/PhysRevD.102.092013",
	JOURNAL=	"Phys. Rev. D",
	VOLUME=	"102",
	PAGES=	"092013",
	YEAR=	"2020",
}

@ARTICLE{ATLAS:2025ciy,
	AUTHOR=	"{ATLAS Collaboration}",
	TITLE=	"Measurement of the top-quark {Yukawa} coupling from \ttbar production in the lepton+jets final state using ${\Pp\Pp}$ collisions at $\sqrt{s}={13\TeV}$ with the {ATLAS} detector",
	EPRINT=	"2509.16039",
	ARCHIVEPREFIX=	"arXiv",
	PRIMARYCLASS=	"hep-ex",
	DOI=	"10.1007/JHEP01(2026)117",
	JOURNAL=	"JHEP",
	VOLUME=	"01",
	PAGES=	"117",
	YEAR=	"2026",
}

@ARTICLE{ATLAS:2024mhs,
	AUTHOR=	"{ATLAS Collaboration}",
	TITLE=	"Constraint on the total width of the {Higgs} boson from {Higgs} boson and four-top-quark measurements in ${\Pp\Pp}$ collisions at $\sqrt{s}={13\TeV}$ with the {ATLAS} detector",
	EPRINT=	"2407.10631",
	ARCHIVEPREFIX=	"arXiv",
	PRIMARYCLASS=	"hep-ex",
	DOI=	"10.1016/j.physletb.2025.139277",
	JOURNAL=	"Phys. Lett. B",
	VOLUME=	"861",
	PAGES=	"139277",
	YEAR=	"2025",
}

@ARTICLE{CMS:HIG-21-018,
	AUTHOR=	"{CMS Collaboration}",
	TITLE=	"Combined measurements and interpretations of {Higgs} boson production and decay in proton-proton collisions at $\sqrt{s}={13\TeV}$",
	EPRINT=	"2602.18611",
	ARCHIVEPREFIX=	"arXiv",
	PRIMARYCLASS=	"hep-ex",
	DOI=	"10.1088/1361-6633/ae72c4",
	JOURNAL=	"Rep. Prog. Phys.",
	VOLUME=	"89",
	PAGES=	"087801",
	YEAR=	"2026",
}

@MISC{hepdata,
	HOWPUBLISHED=	"{HEPData} record for this analysis",
	DOI=	"10.17182/hepdata.161622",
	YEAR=	"2026",
}

@ARTICLE{CMS:TRG-17-001,
	AUTHOR=	"{CMS Collaboration}",
	TITLE=	"Performance of the {CMS} {\Lone} trigger in proton-proton collisions at $\sqrt{s}={13\TeV}$",
	EPRINT=	"2006.10165",
	ARCHIVEPREFIX=	"arXiv",
	PRIMARYCLASS=	"hep-ex",
	DOI=	"10.1088/1748-0221/15/10/P10017",
	JOURNAL=	"JINST",
	VOLUME=	"15",
	PAGES=	"P10017",
	YEAR=	"2020",
}

@ARTICLE{CMS:TRG-12-001,
	AUTHOR=	"{CMS Collaboration}",
	TITLE=	"The {CMS} trigger system",
	EPRINT=	"1609.02366",
	ARCHIVEPREFIX=	"arXiv",
	PRIMARYCLASS=	"physics.ins-det",
	DOI=	"10.1088/1748-0221/12/01/P01020",
	JOURNAL=	"JINST",
	VOLUME=	"12",
	PAGES=	"P01020",
	YEAR=	"2017",
}

@ARTICLE{CMS:TRG-19-001,
	AUTHOR=	"{CMS Collaboration}",
	TITLE=	"Performance of the {CMS} high-level trigger during {LHC} \mbox{Run 2}",
	EPRINT=	"2410.17038",
	ARCHIVEPREFIX=	"arXiv",
	PRIMARYCLASS=	"physics.ins-det",
	DOI=	"10.1088/1748-0221/19/11/P11021",
	JOURNAL=	"JINST",
	VOLUME=	"19",
	PAGES=	"P11021",
	YEAR=	"2024",
}

@ARTICLE{CMS:EGM-17-001,
	AUTHOR=	"{CMS Collaboration}",
	TITLE=	"Electron and photon reconstruction and identification with the {CMS} experiment at the {CERN} {LHC}",
	EPRINT=	"2012.06888",
	ARCHIVEPREFIX=	"arXiv",
	PRIMARYCLASS=	"hep-ex",
	DOI=	"10.1088/1748-0221/16/05/P05014",
	JOURNAL=	"JINST",
	VOLUME=	"16",
	PAGES=	"P05014",
	YEAR=	"2021",
}

@ARTICLE{CMS:MUO-16-001,
	AUTHOR=	"{CMS Collaboration}",
	TITLE=	"Performance of the {CMS} muon detector and muon reconstruction with proton-proton collisions at $\sqrt{s}={13\TeV}$",
	EPRINT=	"1804.04528",
	ARCHIVEPREFIX=	"arXiv",
	PRIMARYCLASS=	"physics.ins-det",
	DOI=	"10.1088/1748-0221/13/06/P06015",
	JOURNAL=	"JINST",
	VOLUME=	"13",
	PAGES=	"P06015",
	YEAR=	"2018",
}

@ARTICLE{CMS:TRK-11-001,
	AUTHOR=	"{CMS Collaboration}",
	TITLE=	"Description and performance of track and primary-vertex reconstruction with the {CMS} tracker",
	EPRINT=	"1405.6569",
	ARCHIVEPREFIX=	"arXiv",
	PRIMARYCLASS=	"physics.ins-det",
	DOI=	"10.1088/1748-0221/9/10/P10009",
	JOURNAL=	"JINST",
	VOLUME=	"9",
	PAGES=	"P10009",
	YEAR=	"2014",
}

@ARTICLE{CMS:PRF-14-001,
	AUTHOR=	"{CMS Collaboration}",
	TITLE=	"Particle-flow reconstruction and global event description with the {CMS} detector",
	EPRINT=	"1706.04965",
	ARCHIVEPREFIX=	"arXiv",
	PRIMARYCLASS=	"physics.ins-det",
	DOI=	"10.1088/1748-0221/12/10/P10003",
	JOURNAL=	"JINST",
	VOLUME=	"12",
	PAGES=	"P10003",
	YEAR=	"2017",
}

@ARTICLE{CMS:TAU-16-003,
	AUTHOR=	"{CMS Collaboration}",
	TITLE=	"Performance of reconstruction and identification of {\PGt} leptons decaying to hadrons and {\PGnGt} in ${\Pp\Pp}$ collisions at $\sqrt{s}={13\TeV}$",
	EPRINT=	"1809.02816",
	ARCHIVEPREFIX=	"arXiv",
	PRIMARYCLASS=	"hep-ex",
	DOI=	"10.1088/1748-0221/13/10/P10005",
	JOURNAL=	"JINST",
	VOLUME=	"13",
	PAGES=	"P10005",
	YEAR=	"2018",
}

@ARTICLE{CMS:JME-13-004,
	AUTHOR=	"{CMS Collaboration}",
	TITLE=	"Jet energy scale and resolution in the {CMS} experiment in ${\Pp\Pp}$ collisions at {8\TeV}",
	EPRINT=	"1607.03663",
	ARCHIVEPREFIX=	"arXiv",
	PRIMARYCLASS=	"hep-ex",
	DOI=	"10.1088/1748-0221/12/02/P02014",
	JOURNAL=	"JINST",
	VOLUME=	"12",
	PAGES=	"P02014",
	YEAR=	"2017",
}

@ARTICLE{CMS:JME-17-001,
	AUTHOR=	"{CMS Collaboration}",
	TITLE=	"Performance of missing transverse momentum reconstruction in proton-proton collisions at $\sqrt{s}={13\TeV}$ using the {CMS} detector",
	EPRINT=	"1903.06078",
	ARCHIVEPREFIX=	"arXiv",
	PRIMARYCLASS=	"hep-ex",
	DOI=	"10.1088/1748-0221/14/07/P07004",
	JOURNAL=	"JINST",
	VOLUME=	"14",
	PAGES=	"P07004",
	YEAR=	"2019",
}

@ARTICLE{CMS:JME-18-001,
	AUTHOR=	"{CMS Collaboration}",
	TITLE=	"Pileup mitigation at {CMS} in {13\TeV} data",
	EPRINT=	"2003.00503",
	ARCHIVEPREFIX=	"arXiv",
	PRIMARYCLASS=	"hep-ex",
	DOI=	"10.1088/1748-0221/15/09/P09018",
	JOURNAL=	"JINST",
	VOLUME=	"15",
	PAGES=	"P09018",
	YEAR=	"2020",
}

@ARTICLE{Cacciari:2008gp,
	AUTHOR=	"Cacciari, M. and Salam, G. P. and Soyez, G.",
	TITLE=	"The anti-\kt jet clustering algorithm",
	EPRINT=	"0802.1189",
	ARCHIVEPREFIX=	"arXiv",
	PRIMARYCLASS=	"hep-ph",
	DOI=	"10.1088/1126-6708/2008/04/063",
	JOURNAL=	"JHEP",
	VOLUME=	"04",
	PAGES=	"063",
	YEAR=	"2008",
}

@ARTICLE{Cacciari:2011ma,
	AUTHOR=	"Cacciari, M. and Salam, G. P. and Soyez, G.",
	TITLE=	"{\FASTJET} user manual",
	EPRINT=	"1111.6097",
	ARCHIVEPREFIX=	"arXiv",
	PRIMARYCLASS=	"hep-ph",
	DOI=	"10.1140/epjc/s10052-012-1896-2",
	JOURNAL=	"Eur. Phys. J. C",
	VOLUME=	"72",
	PAGES=	"1896",
	YEAR=	"2012",
}

@TECHREPORT{CMS:DP-2021-033,
	AUTHOR=	"{CMS Collaboration}",
	TITLE=	"Jet energy scale and resolution measurements with legacy \mbox{Run 2} data collected by {CMS} at {13\TeV}",
	TYPE=	"CMS Detector Performance Note",
	NUMBER=	"CMS-DP-2021-033",
	YEAR=	"2021",
	URL=	"https://cds.cern.ch/record/2792322",
}

@ARTICLE{CMS:BTV-16-002,
	AUTHOR=	"{CMS Collaboration}",
	TITLE=	"Identification of heavy-flavour jets with the {CMS} detector in ${\Pp\Pp}$ collisions at {13\TeV}",
	EPRINT=	"1712.07158",
	ARCHIVEPREFIX=	"arXiv",
	PRIMARYCLASS=	"physics.ins-det",
	DOI=	"10.1088/1748-0221/13/05/P05011",
	JOURNAL=	"JINST",
	VOLUME=	"13",
	PAGES=	"P05011",
	YEAR=	"2018",
}

@ARTICLE{Bols:2020bkb,
	AUTHOR=	"Bols, E. and Kieseler, J. and Verzetti, M. and Stoye, M. and Stakia, A.",
	TITLE=	"Jet flavour classification using {DeepJet}",
	EPRINT=	"2008.10519",
	ARCHIVEPREFIX=	"arXiv",
	PRIMARYCLASS=	"hep-ex",
	DOI=	"10.1088/1748-0221/15/12/P12012",
	JOURNAL=	"JINST",
	VOLUME=	"15",
	PAGES=	"P12012",
	YEAR=	"2020",
}

@TECHREPORT{CMS:DP-2023-005,
	AUTHOR=	"{CMS Collaboration}",
	TITLE=	"Performance summary of {AK4} jet {\PQb} tagging with data from proton-proton collisions at {13\TeV} with the {CMS} detector",
	TYPE=	"CMS Detector Performance Note",
	NUMBER=	"CMS-DP-2023-005",
	YEAR=	"2023",
	URL=	"https://cds.cern.ch/record/2854609",
}

@TECHREPORT{CMS:DP-2020-021,
	AUTHOR=	"{CMS Collaboration}",
	TITLE=	"{ECAL} 2016 refined calibration and \mbox{Run 2} summary plots",
	TYPE=	"CMS Detector Performance Note",
	NUMBER=	"CMS-DP-2020-021",
	YEAR=	"2020",
	URL=	"https://cds.cern.ch/record/2717925",
}

@ARTICLE{CMS:EGM-13-001,
	AUTHOR=	"{CMS Collaboration}",
	TITLE=	"Performance of electron reconstruction and selection with the {CMS} detector in proton-proton collisions at $\sqrt{s}={8\TeV}$",
	EPRINT=	"1502.02701",
	ARCHIVEPREFIX=	"arXiv",
	PRIMARYCLASS=	"physics.ins-det",
	DOI=	"10.1088/1748-0221/10/06/P06005",
	JOURNAL=	"JINST",
	VOLUME=	"10",
	PAGES=	"P06005",
	YEAR=	"2015",
}

@ARTICLE{CMS:CFT-09-014,
	AUTHOR=	"{CMS Collaboration}",
	TITLE=	"Performance of {CMS} muon reconstruction in cosmic-ray events",
	EPRINT=	"0911.4994",
	ARCHIVEPREFIX=	"arXiv",
	PRIMARYCLASS=	"physics.ins-det",
	DOI=	"10.1088/1748-0221/5/03/T03022",
	JOURNAL=	"JINST",
	VOLUME=	"5",
	PAGES=	"T03022",
	YEAR=	"2010",
}

@ARTICLE{CMS:MUO-17-001,
	AUTHOR=	"{CMS Collaboration}",
	TITLE=	"Performance of the reconstruction and identification of high-momentum muons in proton-proton collisions at $\sqrt{s}={13\TeV}$",
	EPRINT=	"1912.03516",
	ARCHIVEPREFIX=	"arXiv",
	PRIMARYCLASS=	"physics.ins-det",
	DOI=	"10.1088/1748-0221/15/02/P02027",
	JOURNAL=	"JINST",
	VOLUME=	"15",
	PAGES=	"P02027",
	YEAR=	"2020",
}

@ARTICLE{CMS:HIG-17-018,
	AUTHOR=	"{CMS Collaboration}",
	TITLE=	"Evidence for associated production of a {Higgs} boson with a top quark pair in final states with electrons, muons, and hadronically decaying {\PGt} leptons at $\sqrt{s}={13\TeV}$",
	EPRINT=	"1803.05485",
	ARCHIVEPREFIX=	"arXiv",
	PRIMARYCLASS=	"hep-ex",
	DOI=	"10.1007/JHEP08(2018)066",
	JOURNAL=	"JHEP",
	VOLUME=	"08",
	PAGES=	"066",
	YEAR=	"2018",
}

@ARTICLE{CMS:TOP-18-008,
	AUTHOR=	"{CMS Collaboration}",
	TITLE=	"Observation of single top quark production in association with a {\PZ} boson in proton-proton collisions at $\sqrt{s}={13\TeV}$",
	EPRINT=	"1812.05900",
	ARCHIVEPREFIX=	"arXiv",
	PRIMARYCLASS=	"hep-ex",
	DOI=	"10.1103/PhysRevLett.122.132003",
	JOURNAL=	"Phys. Rev. Lett.",
	VOLUME=	"122",
	PAGES=	"132003",
	YEAR=	"2019",
}

@ARTICLE{CMS:SUS-19-012,
	AUTHOR=	"{CMS Collaboration}",
	TITLE=	"Search for electroweak production of charginos and neutralinos in proton-proton collisions at $\sqrt{s}={13\TeV}$",
	EPRINT=	"2106.14246",
	ARCHIVEPREFIX=	"arXiv",
	PRIMARYCLASS=	"hep-ex",
	DOI=	"10.1007/JHEP04(2022)147",
	JOURNAL=	"JHEP",
	VOLUME=	"04",
	PAGES=	"147",
	YEAR=	"2022",
}

@ARTICLE{CMS:SMP-20-012,
	AUTHOR=	"{CMS Collaboration}",
	TITLE=	"Measurements of the electroweak diboson production cross sections in proton-proton collisions at $\sqrt{s}={5.02\TeV}$ using leptonic decays",
	EPRINT=	"2107.01137",
	ARCHIVEPREFIX=	"arXiv",
	PRIMARYCLASS=	"hep-ex",
	DOI=	"10.1103/PhysRevLett.127.191801",
	JOURNAL=	"Phys. Rev. Lett.",
	VOLUME=	"127",
	PAGES=	"191801",
	YEAR=	"2021",
}

@ARTICLE{CMS:TOP-20-010,
	AUTHOR=	"{CMS Collaboration}",
	TITLE=	"Inclusive and differential cross section measurements of single top quark production in association with a {\PZ} boson in proton-proton collisions at $\sqrt{s}={13\TeV}$",
	EPRINT=	"2111.02860",
	ARCHIVEPREFIX=	"arXiv",
	PRIMARYCLASS=	"hep-ex",
	DOI=	"10.1007/JHEP02(2022)107",
	JOURNAL=	"JHEP",
	VOLUME=	"02",
	PAGES=	"107",
	YEAR=	"2022",
}

@ARTICLE{CMS:MUO-22-001,
	AUTHOR=	"{CMS Collaboration}",
	TITLE=	"Muon identification using multivariate techniques in the {CMS} experiment in proton-proton collisions at $\sqrt{s}={13\TeV}$",
	EPRINT=	"2310.03844",
	ARCHIVEPREFIX=	"arXiv",
	PRIMARYCLASS=	"hep-ex",
	DOI=	"10.1088/1748-0221/19/02/P02031",
	JOURNAL=	"JINST",
	VOLUME=	"19",
	PAGES=	"P02031",
	YEAR=	"2024",
}

@ARTICLE{CMS:LUM-17-003,
	AUTHOR=	"{CMS Collaboration}",
	TITLE=	"Precision luminosity measurement in proton-proton collisions at $\sqrt{s}={13\TeV}$ in 2015 and 2016 at {CMS}",
	EPRINT=	"2104.01927",
	ARCHIVEPREFIX=	"arXiv",
	PRIMARYCLASS=	"hep-ex",
	DOI=	"10.1140/epjc/s10052-021-09538-2",
	JOURNAL=	"Eur. Phys. J. C",
	VOLUME=	"81",
	PAGES=	"800",
	YEAR=	"2021",
}

@TECHREPORT{CMS:LUM-17-004,
	AUTHOR=	"{CMS Collaboration}",
	TITLE=	"{CMS} luminosity measurement for the 2017 data-taking period at $\sqrt{s}={13\TeV}$",
	TYPE=	"CMS Physics Analysis Summary",
	NUMBER=	"CMS-PAS-LUM-17-004",
	YEAR=	"2018",
	URL=	"https://cds.cern.ch/record/2621960",
}

@TECHREPORT{CMS:LUM-18-002,
	AUTHOR=	"{CMS Collaboration}",
	TITLE=	"{CMS} luminosity measurement for the 2018 data-taking period at $\sqrt{s}={13\TeV}$",
	TYPE=	"CMS Physics Analysis Summary",
	NUMBER=	"CMS-PAS-LUM-18-002",
	YEAR=	"2019",
	URL=	"https://cds.cern.ch/record/2676164",
}

@ARTICLE{NNPDF:2017mvq,
	AUTHOR=	"Ball, R. D. and Bertone, V. and Carrazza, S. and Del Debbio, L. and Forte, S. and Groth-Merrild, P. and Guffanti, A. and Hartland, N. P. and Kassabov, Z. and Latorre, J. I. and Nocera, E. R. and Rojo, J. and Rottoli, L. and Slade, E. and Ubiali, M.",
	COLLABORATION=	"NNPDF",
	TITLE=	"Parton distributions from high-precision collider data",
	EPRINT=	"1706.00428",
	ARCHIVEPREFIX=	"arXiv",
	PRIMARYCLASS=	"hep-ph",
	DOI=	"10.1140/epjc/s10052-017-5199-5",
	JOURNAL=	"Eur. Phys. J. C",
	VOLUME=	"77",
	PAGES=	"663",
	YEAR=	"2017",
}

@ARTICLE{Sjostrand:2014zea,
	AUTHOR=	"Sj{\"o}strand, T. and Ask, S. and Christiansen, J. R. and Corke, R. and Desai, N. and Ilten, P. and Mrenna, S. and Prestel, S. and Rasmussen, C. O. and Skands, P. Z.",
	TITLE=	"An introduction to {\PYTHIA8.2}",
	EPRINT=	"1410.3012",
	ARCHIVEPREFIX=	"arXiv",
	PRIMARYCLASS=	"hep-ph",
	DOI=	"10.1016/j.cpc.2015.01.024",
	JOURNAL=	"Comput. Phys. Commun.",
	VOLUME=	"191",
	PAGES=	"159",
	YEAR=	"2015",
}

@ARTICLE{CMS:GEN-17-001,
	AUTHOR=	"{CMS Collaboration}",
	TITLE=	"Extraction and validation of a new set of {CMS} {\PYTHIA8} tunes from underlying-event measurements",
	EPRINT=	"1903.12179",
	ARCHIVEPREFIX=	"arXiv",
	PRIMARYCLASS=	"hep-ex",
	DOI=	"10.1140/epjc/s10052-019-7499-4",
	JOURNAL=	"Eur. Phys. J. C",
	VOLUME=	"80",
	PAGES=	"4",
	YEAR=	"2020",
}

@ARTICLE{GEANT4:2002zbu,
	AUTHOR=	"Agostinelli, S. and others",
	COLLABORATION=	"GEANT4",
	TITLE=	"{\GEANTfour}---a simulation toolkit",
	DOI=	"10.1016/S0168-9002(03)01368-8",
	JOURNAL=	"Nucl. Instrum. Meth. A",
	VOLUME=	"506",
	PAGES=	"250",
	YEAR=	"2003",
}

@TECHREPORT{CMS:DP-2020-045,
	AUTHOR=	"{CMS Collaboration}",
	TITLE=	"Simulation of the silicon strip tracker pre-amplifier in early 2016 data",
	TYPE=	"CMS Detector Performance Note",
	NUMBER=	"CMS-DP-2020-045",
	YEAR=	"2020",
	URL=	"https://cds.cern.ch/record/2740688",
}

@ARTICLE{Alwall:2014hca,
	AUTHOR=	"Alwall, J. and Frederix, R. and Frixione, S. and Hirschi, V. and Maltoni, F. and Mattelaer, O. and Shao, H.-S. and Stelzer, T. and Torrielli, P. and Zaro, M.",
	TITLE=	"The automated computation of tree-level and next-to-leading order differential cross sections, and their matching to parton shower simulations",
	EPRINT=	"1405.0301",
	ARCHIVEPREFIX=	"arXiv",
	PRIMARYCLASS=	"hep-ph",
	DOI=	"10.1007/JHEP07(2014)079",
	JOURNAL=	"JHEP",
	VOLUME=	"07",
	PAGES=	"079",
	YEAR=	"2014",
}

@ARTICLE{Artoisenet:2012st,
	AUTHOR=	"Artoisenet, P. and Frederix, R. and Mattelaer, O. and Rietkerk, R.",
	TITLE=	"Automatic spin-entangled decays of heavy resonances in {Monte Carlo} simulations",
	EPRINT=	"1212.3460",
	ARCHIVEPREFIX=	"arXiv",
	PRIMARYCLASS=	"hep-ph",
	DOI=	"10.1007/JHEP03(2013)015",
	JOURNAL=	"JHEP",
	VOLUME=	"03",
	PAGES=	"015",
	YEAR=	"2013",
}

@ARTICLE{Bevilacqua:2012em,
	AUTHOR=	"Bevilacqua, G. and Worek, M.",
	TITLE=	"Constraining {BSM} physics at the {LHC}: Four top final states with {NLO} accuracy in perturbative {QCD}",
	EPRINT=	"1206.3064",
	ARCHIVEPREFIX=	"arXiv",
	PRIMARYCLASS=	"hep-ph",
	DOI=	"10.1007/JHEP07(2012)111",
	JOURNAL=	"JHEP",
	VOLUME=	"07",
	PAGES=	"111",
	YEAR=	"2012",
}

@ARTICLE{Maltoni:2015ena,
	AUTHOR=	"Maltoni, F. and Pagani, D. and Tsinikos, I.",
	TITLE=	"Associated production of a top-quark pair with vector bosons at {NLO} in {QCD}: impact on ${\ttbar\PH}$ searches at the {LHC}",
	EPRINT=	"1507.05640",
	ARCHIVEPREFIX=	"arXiv",
	PRIMARYCLASS=	"hep-ph",
	DOI=	"10.1007/JHEP02(2016)113",
	JOURNAL=	"JHEP",
	VOLUME=	"02",
	PAGES=	"113",
	YEAR=	"2016",
}

@ARTICLE{Frederix:2017wme,
	AUTHOR=	"Frederix, R. and Pagani, D. and Zaro, M.",
	TITLE=	"Large {NLO} corrections in ${\ttbar\PWpm}$ and $\ttbar\ttbar$ hadroproduction from supposedly subleading {EW} contributions",
	EPRINT=	"1711.02116",
	ARCHIVEPREFIX=	"arXiv",
	PRIMARYCLASS=	"hep-ph",
	DOI=	"10.1007/JHEP02(2018)031",
	JOURNAL=	"JHEP",
	VOLUME=	"02",
	PAGES=	"031",
	YEAR=	"2018",
}

@ARTICLE{Jezo:2021smh,
	AUTHOR=	"Je{\v{z}}o, T. and Kraus, M.",
	TITLE=	"Hadroproduction of four top quarks in the {\POWHEG} \textsc{box}",
	EPRINT=	"2110.15159",
	ARCHIVEPREFIX=	"arXiv",
	PRIMARYCLASS=	"hep-ph",
	DOI=	"10.1103/PhysRevD.105.114024",
	JOURNAL=	"Phys. Rev. D",
	VOLUME=	"105",
	PAGES=	"114024",
	YEAR=	"2022",
}

@ARTICLE{Dimitrakopoulos:2024qib,
	AUTHOR=	"Dimitrakopoulos, N. and Worek, M.",
	TITLE=	"Four top final states with {NLO} accuracy in perturbative {QCD}: 4 lepton channel",
	EPRINT=	"2401.10678",
	ARCHIVEPREFIX=	"arXiv",
	PRIMARYCLASS=	"hep-ph",
	DOI=	"10.1007/JHEP06(2024)129",
	JOURNAL=	"JHEP",
	VOLUME=	"06",
	PAGES=	"129",
	YEAR=	"2024",
}

@ARTICLE{vanBeekveld:2022hty,
	AUTHOR=	"van Beekveld, M. and Kulesza, A. and Moreno Valero, L.",
	TITLE=	"Threshold resummation for the production of four top quarks at the {LHC}",
	EPRINT=	"2212.03259",
	ARCHIVEPREFIX=	"arXiv",
	PRIMARYCLASS=	"hep-ph",
	DOI=	"10.1103/PhysRevLett.131.211901",
	JOURNAL=	"Phys. Rev. Lett.",
	VOLUME=	"131",
	PAGES=	"211901",
	YEAR=	"2023",
}

@ARTICLE{vanBeekveld:2025ghw,
	AUTHOR=	"van Beekveld, M. and Kulesza, A. and Lupattelli, M. and Saracco, T.",
	TITLE=	"Invariant-mass threshold resummation for the production of four top quarks at the {LHC}",
	EPRINT=	"2505.10381",
	ARCHIVEPREFIX=	"arXiv",
	PRIMARYCLASS=	"hep-ph",
	DOI=	"10.1007/JHEP10(2025)209",
	JOURNAL=	"JHEP",
	VOLUME=	"10",
	PAGES=	"209",
	YEAR=	"2025",
}

@ARTICLE{Frixione:2008yi,
	AUTHOR=	"Frixione, S. and Laenen, E. and Motylinski, P. and Webber, B. R. and White, C. D.",
	TITLE=	"Single-top hadroproduction in association with a {\PW} boson",
	EPRINT=	"0805.3067",
	ARCHIVEPREFIX=	"arXiv",
	PRIMARYCLASS=	"hep-ph",
	DOI=	"10.1088/1126-6708/2008/07/029",
	JOURNAL=	"JHEP",
	VOLUME=	"07",
	PAGES=	"029",
	YEAR=	"2008",
}

@ARTICLE{Demartin:2016axk,
	AUTHOR=	"Demartin, F. and Maier, B. and Maltoni, F. and Mawatari, K. and Zaro, M.",
	TITLE=	"${\PQt\PW\PH}$ associated production at the {LHC}",
	EPRINT=	"1607.05862",
	ARCHIVEPREFIX=	"arXiv",
	PRIMARYCLASS=	"hep-ph",
	DOI=	"10.1140/epjc/s10052-017-4601-7",
	JOURNAL=	"Eur. Phys. J. C",
	VOLUME=	"77",
	PAGES=	"34",
	YEAR=	"2017",
}

@MISC{Durieux:2023ttt,
	AUTHOR=	"Durieux, G.",
	TITLE=	"Triple top-quark production at {NLO} in {QCD}",
	DOI=	"10.5281/zenodo.7679328",
	YEAR=	"2023",
	HOWPUBLISHED=	"Zenodo",
}

@UNPUBLISHED{Durieux:2026kus,
	AUTHOR=	"Durieux, G. and El Faham, H. and Frederix, R. and Pagani, D. and Zaro, M.",
	TITLE=	"The inseparable three and four tops",
	EPRINT=	"2607.27323",
	ARCHIVEPREFIX=	"arXiv",
	PRIMARYCLASS=	"hep-ph",
	YEAR=	"2026",
}

@ARTICLE{Barger:2010uw,
	AUTHOR=	"Barger, V. and Keung, W.-Y. and Yencho, B.",
	TITLE=	"Triple-top signal of new physics at the {LHC}",
	EPRINT=	"1001.0221",
	ARCHIVEPREFIX=	"arXiv",
	PRIMARYCLASS=	"hep-ph",
	DOI=	"10.1016/j.physletb.2010.03.001",
	JOURNAL=	"Phys. Lett. B",
	VOLUME=	"687",
	PAGES=	"70",
	YEAR=	"2010",
}

@ARTICLE{Chen:2014ewl,
	AUTHOR=	"Chen, C.-R.",
	TITLE=	"Searching for new physics with triple-top signal at the {LHC}",
	DOI=	"10.1016/j.physletb.2014.07.041",
	JOURNAL=	"Phys. Lett. B",
	VOLUME=	"736",
	PAGES=	"321",
	YEAR=	"2014",
}

@ARTICLE{Malekhosseini:2018fgp,
	AUTHOR=	"Malekhosseini, M. and Ghominejad, M. and Khanpour, H. and Mohammadi Najafabadi, M.",
	TITLE=	"Constraining top quark flavor violation and dipole moments through three and four top quark productions at the {LHC}",
	EPRINT=	"1804.05598",
	ARCHIVEPREFIX=	"arXiv",
	PRIMARYCLASS=	"hep-ph",
	DOI=	"10.1103/PhysRevD.98.095001",
	JOURNAL=	"Phys. Rev. D",
	VOLUME=	"98",
	PAGES=	"095001",
	YEAR=	"2018",
}

@ARTICLE{Boos:2021yat,
	AUTHOR=	"Boos, E. and Dudko, L.",
	TITLE=	"Triple top quark production in standard model",
	EPRINT=	"2107.07629",
	ARCHIVEPREFIX=	"arXiv",
	PRIMARYCLASS=	"hep-ph",
	DOI=	"10.1142/S0217751X22500233",
	JOURNAL=	"Int. J. Mod. Phys. A",
	VOLUME=	"37",
	PAGES=	"2250023",
	YEAR=	"2022",
}

@ARTICLE{CMS:TOP-21-011,
	AUTHOR=	"{CMS Collaboration}",
	TITLE=	"Measurement of the cross section of top quark-antiquark pair production in association with a {\PW} boson in proton-proton collisions at $\sqrt{s}={13\TeV}$",
	EPRINT=	"2208.06485",
	ARCHIVEPREFIX=	"arXiv",
	PRIMARYCLASS=	"hep-ex",
	DOI=	"10.1007/JHEP07(2023)219",
	JOURNAL=	"JHEP",
	VOLUME=	"07",
	PAGES=	"219",
	YEAR=	"2023",
}

@INPROCEEDINGS{Voss:2007jxm,
	AUTHOR=	"Voss, H. and H{\"o}cker, A. and Stelzer, J. and Tegenfeldt, F.",
	TITLE=	"\textsc{tmva}, the toolkit for multivariate data analysis with \textsc{root}",
	BOOKTITLE=	"{Proc. 11th International Workshop on Advanced Computing and Analysis Techniques in Physics Research (ACAT 2007): Amsterdam, The Netherlands, April 23--27, 2007}",
	EPRINT=	"physics/0703039",
	ARCHIVEPREFIX=	"arXiv",
	PRIMARYCLASS=	"physics.data-an",
	DOI=	"10.22323/1.050.0040",
	YEAR=	"2007",
	NOTE=	"[PoS (ACAT2007) 040]",
}

@ARTICLE{CMS:SUS-15-008,
	AUTHOR=	"{CMS Collaboration}",
	TITLE=	"Search for new physics in same-sign dilepton events in proton-proton collisions at $\sqrt{s}={13\TeV}$",
	EPRINT=	"1605.03171",
	ARCHIVEPREFIX=	"arXiv",
	PRIMARYCLASS=	"hep-ex",
	DOI=	"10.1140/epjc/s10052-016-4261-z",
	JOURNAL=	"Eur. Phys. J. C",
	VOLUME=	"76",
	PAGES=	"439",
	YEAR=	"2016",
}

@ARTICLE{Butterworth:2015oua,
	AUTHOR=	"Butterworth, J. and others",
	TITLE=	"{PDF4LHC} recommendations for {LHC} \mbox{Run 2}",
	EPRINT=	"1510.03865",
	ARCHIVEPREFIX=	"arXiv",
	PRIMARYCLASS=	"hep-ph",
	DOI=	"10.1088/0954-3899/43/2/023001",
	JOURNAL=	"J. Phys. G",
	VOLUME=	"43",
	PAGES=	"023001",
	YEAR=	"2016",
}

@ARTICLE{CMS:TOP-18-002,
	AUTHOR=	"{CMS Collaboration}",
	TITLE=	"Measurement of the cross section for \ttbar production with additional jets and {\PQb} jets in ${\Pp\Pp}$ collisions at $\sqrt{s}={13\TeV}$",
	EPRINT=	"2003.06467",
	ARCHIVEPREFIX=	"arXiv",
	PRIMARYCLASS=	"hep-ex",
	DOI=	"10.1007/JHEP07(2020)125",
	JOURNAL=	"JHEP",
	VOLUME=	"07",
	PAGES=	"125",
	YEAR=	"2020",
}

@ARTICLE{CMS:TOP-22-009,
	AUTHOR=	"{CMS Collaboration}",
	TITLE=	"Inclusive and differential cross section measurements of $\ttbar\bbbar$ production in the lepton+jets channel at $\sqrt{s}={13\TeV}$",
	EPRINT=	"2309.14442",
	ARCHIVEPREFIX=	"arXiv",
	PRIMARYCLASS=	"hep-ex",
	DOI=	"10.1007/JHEP05(2024)042",
	JOURNAL=	"JHEP",
	VOLUME=	"05",
	PAGES=	"042",
	YEAR=	"2024",
}

@ARTICLE{Frederix:2021agh,
	AUTHOR=	"Frederix, R. and Tsinikos, I.",
	TITLE=	"On improving {NLO} merging for ${\ttbar\PW}$ production",
	EPRINT=	"2108.07826",
	ARCHIVEPREFIX=	"arXiv",
	PRIMARYCLASS=	"hep-ph",
	DOI=	"10.1007/JHEP11(2021)029",
	JOURNAL=	"JHEP",
	VOLUME=	"11",
	PAGES=	"029",
	YEAR=	"2021",
}

@TECHREPORT{CMS:NOTE-2011-005,
	AUTHOR=	"{ATLAS and CMS Collaborations, and LHC Higgs Combination Group}",
	TITLE=	"Procedure for the {LHC} {Higgs} boson search combination in {Summer} 2011",
	NUMBER=	"CMS-NOTE-2011-005, ATL-PHYS-PUB-2011-11",
	YEAR=	"2011",
	URL=	"https://cds.cern.ch/record/1379837",
}

@ARTICLE{Barlow:1993dm,
	AUTHOR=	"Barlow, R. and Beeston, C.",
	TITLE=	"Fitting using finite {Monte Carlo} samples",
	DOI=	"10.1016/0010-4655(93)90005-W",
	JOURNAL=	"Comput. Phys. Commun.",
	VOLUME=	"77",
	PAGES=	"219",
	YEAR=	"1993",
}

@INPROCEEDINGS{Conway:2011in,
	AUTHOR=	"Conway, J. S.",
	TITLE=	"Incorporating nuisance parameters in likelihoods for multisource spectra",
	BOOKTITLE=	"{Proc. 2011 Workshop on Statistical Issues Related to Discovery Claims in Search Experiments and Unfolding (PHYSTAT 2011): Geneva, Switzerland, January 17--20, 2011}",
	EPRINT=	"1103.0354",
	ARCHIVEPREFIX=	"arXiv",
	PRIMARYCLASS=	"physics.data-an",
	DOI=	"10.5170/CERN-2011-006.115",
	YEAR=	"2011",
}

@ARTICLE{CMS:CAT-23-001,
	AUTHOR=	"{CMS Collaboration}",
	TITLE=	"The {CMS} statistical analysis and combination tool: \textsc{combine}",
	EPRINT=	"2404.06614",
	ARCHIVEPREFIX=	"arXiv",
	PRIMARYCLASS=	"physics.data-an",
	DOI=	"10.1007/s41781-024-00121-4",
	JOURNAL=	"Comput. Softw. Big Sci.",
	VOLUME=	"8",
	PAGES=	"19",
	YEAR=	"2024",
}

@INPROCEEDINGS{Verkerke:2003ir,
	AUTHOR=	"Verkerke, W. and Kirkby, D.",
	TITLE=	"The \textsc{RooFit} toolkit for data modeling",
	BOOKTITLE=	"{Proc. 13th International Conference on Computing in High Energy and Nuclear Physics (CHEP 2003): La Jolla CA, United States, March 24--28, 2003}",
	EPRINT=	"physics/0306116",
	ARCHIVEPREFIX=	"arXiv",
	PRIMARYCLASS=	"physics.data-an",
	YEAR=	"2003",
	NOTE=	"[eConf C0303241 (2003) MOLT007]",
	URL=	"https://www.slac.stanford.edu/econf/C0303241/proc/papers/MOLT007.PDF",
}

@INPROCEEDINGS{Moneta:2010pm,
	AUTHOR=	"Moneta, L. and Belasco, K. and Cranmer, K. S. and Kreiss, S. and Lazzaro, A. and Piparo, D. and Schott, G. and Verkerke, W. and Wolf, M.",
	TITLE=	"The \textsc{RooStats} project",
	BOOKTITLE=	"{Proc. 13th International Workshop on Advanced Computing and Analysis Techniques in Physics Research (ACAT 2010): Jaipur, India, February 22--27, 2010}",
	EPRINT=	"1009.1003",
	ARCHIVEPREFIX=	"arXiv",
	PRIMARYCLASS=	"physics.data-an",
	DOI=	"10.22323/1.093.0057",
	YEAR=	"2010",
	NOTE=	"[PoS (ACAT2010) 057]",
}

@ARTICLE{Cowan:2010js,
	AUTHOR=	"Cowan, G. and Cranmer, K. and Gross, E. and Vitells, O.",
	TITLE=	"Asymptotic formulae for likelihood-based tests of new physics",
	EPRINT=	"1007.1727",
	ARCHIVEPREFIX=	"arXiv",
	PRIMARYCLASS=	"physics.data-an",
	DOI=	"10.1140/epjc/s10052-011-1554-0",
	JOURNAL=	"Eur. Phys. J. C",
	VOLUME=	"71",
	PAGES=	"1554",
	YEAR=	"2011",
	NOTE=	"[Erratum: \DOI{10.1140/epjc/s10052-013-2501-z}]",
}

@ARTICLE{Wilks:1938dza,
	AUTHOR=	"Wilks, S. S.",
	TITLE=	"The large-sample distribution of the likelihood ratio for testing composite hypotheses",
	DOI=	"10.1214/aoms/1177732360",
	JOURNAL=	"Annals Math. Statist.",
	VOLUME=	"9",
	PAGES=	"60",
	YEAR=	"1938",
}

@ARTICLE{Bernlochner:2022oiw,
	AUTHOR=	"Bernlochner, F. U. and Fry, D. C. and Menary, S. B. and Persson, E.",
	TITLE=	"Cover your bases: asymptotic distributions of the profile likelihood ratio when constraining effective field theories in high-energy physics",
	EPRINT=	"2207.01350",
	ARCHIVEPREFIX=	"arXiv",
	PRIMARYCLASS=	"physics.data-an",
	DOI=	"10.21468/SciPostPhysCore.6.1.013",
	JOURNAL=	"SciPost Phys. Core",
	VOLUME=	"6",
	PAGES=	"013",
	YEAR=	"2023",
}

@ARTICLE{Degrande:2012wf,
	AUTHOR=	"Degrande, C. and Greiner, N. and Kilian, W. and Mattelaer, O. and Mebane, H. and Stelzer, T. and Willenbrock, S. and Zhang, C.",
	TITLE=	"Effective field theory: A modern approach to anomalous couplings",
	EPRINT=	"1205.4231",
	ARCHIVEPREFIX=	"arXiv",
	PRIMARYCLASS=	"hep-ph",
	DOI=	"10.1016/j.aop.2013.04.016",
	JOURNAL=	"Annals Phys.",
	VOLUME=	"335",
	PAGES=	"21",
	YEAR=	"2013",
}

@ARTICLE{Brivio:2017btx,
	AUTHOR=	"Brivio, I. and Jiang, Y. and Trott, M.",
	TITLE=	"The {SMEFTsim} package, theory and tools",
	EPRINT=	"1709.06492",
	ARCHIVEPREFIX=	"arXiv",
	PRIMARYCLASS=	"hep-ph",
	DOI=	"10.1007/JHEP12(2017)070",
	JOURNAL=	"JHEP",
	VOLUME=	"12",
	PAGES=	"070",
	YEAR=	"2017",
}

@ARTICLE{Brivio:2020onw,
	AUTHOR=	"Brivio, I.",
	TITLE=	"{SMEFTsim} 3.0---a practical guide",
	EPRINT=	"2012.11343",
	ARCHIVEPREFIX=	"arXiv",
	PRIMARYCLASS=	"hep-ph",
	DOI=	"10.1007/JHEP04(2021)073",
	JOURNAL=	"JHEP",
	VOLUME=	"04",
	PAGES=	"073",
	YEAR=	"2021",
}

@ARTICLE{Mattelaer:2016gcx,
	AUTHOR=	"Mattelaer, O.",
	TITLE=	"On the maximal use of {Monte Carlo} samples: re-weighting events at {NLO} accuracy",
	EPRINT=	"1607.00763",
	ARCHIVEPREFIX=	"arXiv",
	PRIMARYCLASS=	"hep-ph",
	DOI=	"10.1140/epjc/s10052-016-4533-7",
	JOURNAL=	"Eur. Phys. J. C",
	VOLUME=	"76",
	PAGES=	"674",
	YEAR=	"2016",
}

@ARTICLE{CMS:TOP-19-001,
	AUTHOR=	"{CMS Collaboration}",
	TITLE=	"Search for new physics in top quark production with additional leptons in proton-proton collisions at $\sqrt{s}={13\TeV}$ using effective field theory",
	EPRINT=	"2012.04120",
	ARCHIVEPREFIX=	"arXiv",
	PRIMARYCLASS=	"hep-ex",
	DOI=	"10.1007/JHEP03(2021)095",
	JOURNAL=	"JHEP",
	VOLUME=	"03",
	PAGES=	"095",
	YEAR=	"2021",
}

@ARTICLE{Degrande:2020evl,
	AUTHOR=	"Degrande, C. and Durieux, G. and Maltoni, F. and Mimasu, K. and Vryonidou, E. and Zhang, C.",
	TITLE=	"Automated one-loop computations in the standard model effective field theory",
	EPRINT=	"2008.11743",
	ARCHIVEPREFIX=	"arXiv",
	PRIMARYCLASS=	"hep-ph",
	DOI=	"10.1103/PhysRevD.103.096024",
	JOURNAL=	"Phys. Rev. D",
	VOLUME=	"103",
	PAGES=	"096024",
	YEAR=	"2021",
}

@ARTICLE{Gritsan:2016hjl,
	AUTHOR=	"Gritsan, A. V. and R{\"o}ntsch, R. and Schulze, M. and Xiao, M.",
	TITLE=	"Constraining anomalous {Higgs} boson couplings to the heavy-flavor fermions using matrix element techniques",
	EPRINT=	"1606.03107",
	ARCHIVEPREFIX=	"arXiv",
	PRIMARYCLASS=	"hep-ph",
	DOI=	"10.1103/PhysRevD.94.055023",
	JOURNAL=	"Phys. Rev. D",
	VOLUME=	"94",
	PAGES=	"055023",
	YEAR=	"2016",
}

@ARTICLE{Artoisenet:2013puc,
	AUTHOR=	"Artoisenet, P. and de Aquino, P. and Demartin, F. and Frederix, R. and Frixione, S. and Maltoni, F. and Mandal, M. K. and Mathews, P. and Mawatari, K. and Ravindran, V. and Seth, S. and Torrielli, P. and Zaro, M.",
	TITLE=	"A framework for {Higgs} characterisation",
	EPRINT=	"1306.6464",
	ARCHIVEPREFIX=	"arXiv",
	PRIMARYCLASS=	"hep-ph",
	DOI=	"10.1007/JHEP11(2013)043",
	JOURNAL=	"JHEP",
	VOLUME=	"11",
	PAGES=	"043",
	YEAR=	"2013",
}

@ARTICLE{Demartin:2014fia,
	AUTHOR=	"Demartin, F. and Maltoni, F. and Mawatari, K. and Page, B. and Zaro, M.",
	TITLE=	"Higgs characterisation at {NLO} in {QCD}: ${CP}$ properties of the top-quark {Yukawa} interaction",
	EPRINT=	"1407.5089",
	ARCHIVEPREFIX=	"arXiv",
	PRIMARYCLASS=	"hep-ph",
	DOI=	"10.1140/epjc/s10052-014-3065-2",
	JOURNAL=	"Eur. Phys. J. C",
	VOLUME=	"74",
	PAGES=	"3065",
	YEAR=	"2014",
}

@ARTICLE{Brod:2022bww,
	AUTHOR=	"Brod, J. and Cornell, J. M. and Skodras, D. and Stamou, E.",
	TITLE=	"Global constraints on {Yukawa} operators in the standard model effective theory",
	EPRINT=	"2203.03736",
	ARCHIVEPREFIX=	"arXiv",
	PRIMARYCLASS=	"hep-ph",
	DOI=	"10.1007/JHEP08(2022)294",
	JOURNAL=	"JHEP",
	VOLUME=	"08",
	PAGES=	"294",
	YEAR=	"2022",
}

@ARTICLE{CMS:HIG-17-031,
	AUTHOR=	"{CMS Collaboration}",
	TITLE=	"Combined measurements of {Higgs} boson couplings in proton-proton collisions at $\sqrt{s}={13\TeV}$",
	EPRINT=	"1809.10733",
	ARCHIVEPREFIX=	"arXiv",
	PRIMARYCLASS=	"hep-ex",
	DOI=	"10.1140/epjc/s10052-019-6909-y",
	JOURNAL=	"Eur. Phys. J. C",
	VOLUME=	"79",
	PAGES=	"421",
	YEAR=	"2019",
}
\cleardoublepage \appendix\section{The CMS Collaboration \label{app:collab}}\begin{sloppypar}\hyphenpenalty=5000\widowpenalty=500\clubpenalty=5000\cmsinstitute{Yerevan Physics Institute, Yerevan, Armenia}
{\tolerance=6000
A.~Hayrapetyan, V.~Makarenko\cmsorcid{0000-0002-8406-8605}, A.~Tumasyan\cmsAuthorMark{1}\cmsorcid{0009-0000-0684-6742}
\par}
\cmsinstitute{Institut f\"{u}r Hochenergiephysik, Vienna, Austria}
{\tolerance=6000
W.~Adam\cmsorcid{0000-0001-9099-4341}, L.~Benato\cmsorcid{0000-0001-5135-7489}, T.~Bergauer\cmsorcid{0000-0002-5786-0293}, M.~Dragicevic\cmsorcid{0000-0003-1967-6783}, C.~Giordano\cmsorcid{0000-0001-6317-2481}, P.S.~Hussain\cmsorcid{0000-0002-4825-5278}, M.~Jeitler\cmsAuthorMark{2}\cmsorcid{0000-0002-5141-9560}, N.~Krammer\cmsorcid{0000-0002-0548-0985}, A.~Li\cmsorcid{0000-0002-4547-116X}, D.~Liko\cmsorcid{0000-0002-3380-473X}, M.~Matthewman, J.~Schieck\cmsAuthorMark{2}\cmsorcid{0000-0002-1058-8093}, R.~Sch\"{o}fbeck\cmsAuthorMark{2}\cmsorcid{0000-0002-2332-8784}, M.~Shooshtari\cmsorcid{0009-0004-8882-4887}, M.~Sonawane\cmsorcid{0000-0003-0510-7010}, W.~Waltenberger\cmsorcid{0000-0002-6215-7228}, C.E.~Wulz\cmsAuthorMark{2}\cmsorcid{0000-0001-9226-5812}
% ...
\par}
$^{1}$Also at Yerevan State University, Yerevan, Armenia\\
$^{2}$Also at Technische Universit\"{a}t Wien, Vienna, Austria\\
% ...
\end{sloppypar}
%%% END EDITABLE REGION %%%
% skeleton_end
\end{document}